\documentclass[a4paper,fleqn,usenatbib]{mnras}

\usepackage{newtxtext,newtxmath}

\usepackage[T1]{fontenc}
\usepackage{ae,aecompl}

\usepackage{graphicx}	
\usepackage{amsmath}	
\usepackage{amssymb}	
\newcommand\numberthis{\addtocounter{equation}{1}\tag{\theequation}}
\usepackage{cancel}
\usepackage{pdflscape}	
\usepackage[T1]{fontenc}
\usepackage{ae,aecompl}
\usepackage{glossaries}
\usepackage{amsmath, bm}
\usepackage{svg}
\usepackage{float}
\usepackage{xcolor}

\title[Linear amplification of Langmuir waves]{Pulsar radio emission mechanism I : On the amplification of Langmuir waves in the linear regime}

\author[Rahaman et al.]{
Sk. Minhajur Rahaman,$^{1}$\thanks{E-mail: rahaman@ncra.tifr.res.in}
Dipanjan Mitra,$^{1,2}$
George I. Melikidze$^{2,3}$
\\
$^{1}$,National Centre for Radio Astrophysics,Tata Institute of Fundamental Research, Post Bag 3, Ganeshkind,Pune-411007,INDIA\\
$^{2}$Janusz Gil Institute of Astronomy, University of Zielona G\'ora, ul Szafrana 2, 65-516 Zielana G\'ora, Poland \\
$^{3}$ Evgeni Kharadze Georgian National Astrophysical Observatory, 0301, Abastumani, Georgia }

\date{Accepted XXX. Received YYY; in original form ZZZ}

\pubyear{2015}

\begin{document}
\label{firstpage}
\pagerange{\pageref{firstpage}--\pageref{lastpage}}
\maketitle

\begin{abstract}
Observations suggest that in normal period radio pulsars, coherent curvature radiation is excited within 10$\%$ of the light cylinder. The coherence is attributed to Langmuir mode instability in a relativistically streaming one-dimensional plasma flow along the open magnetic field lines. In this work, we use a hot plasma treatment to solve the hydrodynamic dispersion relation of Langmuir mode for realistic pulsar parameters. The solution involves three scenarios of two-stream instability viz., driven by high energy beams, due to longitudinal drift that leads to a separation of electron-positron distribution functions in the secondary plasma and due to cloud-cloud interaction causing spatial overlap of two successive secondary plasma
clouds. We find that sufficient amplification can be obtained only for the latter two scenarios. Our analysis shows that longitudinal drift is characterized by high growth rates only for certain multi-polar surface field
geometry. For these configurations, very high growth rates are obtained starting from a few tens of km from the neutron star surface, which then falls monotonically with increasing distance. For cloud-cloud overlap,
growth rates become high starting only after a few hundred km from the
surface, which first increases and then decreases with increasing
distance. A spatial window of up to around a 1000 km above the neutron
star surface has been found where large amplitude Langmuir waves can be excited
while the pair plasma is dense enough to account for high brightness
temperature.
\end{abstract}




\section{Introduction}
\label{introduction}
Observations of radio emission from normal period pulsars (with periods $P$
longer than $\sim$ 0.1 seconds) suggest that: a) The radio emission
has exceedingly high brightness temperature $T_\mathrm{b} \sim 10^{25} -
10^{27} $ K, which is at least 12 orders of magnitude higher than the
incoherent synchrotron limit of $10^{12}$ K (see
\citealt{1969ApJ...155L..71K} ). This necessarily requires a coherent
radio emission mechanism 
(e.g. \citealt{1969Ap&SS...4..464G,
1975ARA&A..13..511G,1979SSRv...24..567C,
1993ppl..conf..105M,2017JApA...38...52M}) ; 
b) The radio emission is highly polarized, which is consistent with
coherent curvature radiation ( hereafter CCR ) (e.g. 
\citealt{2009ApJ...696L.141M,2014ApJ...794..105M}) ; c)
The radio emission detaches from the pulsar magnetosphere a few
hundred km away from the surface (e.g. \citealt{1993ApJ...405..285R,
2017JApA...38...52M}).

These observations require plasma processes where stable charge
bunches can form and excite CCR in relativistically streaming pair
plasma, which can eventually escape from the plasma to reach the
observer (see \citealt{2014ApJ...794..105M,
2004ApJ...600..872G,2009ApJ...696L.141M}). For a
general non-zero angle between the propagation vector and the ambient
magnetic field, the pulsar pair plasma consists of two eigen-modes
viz., the purely transverse X-mode and the quasi-transverse O-mode (see
\citealt{1986ApJ...302..120A}). The quasi-transverse O-mode has a
sub-Lumininal Alfven branch and super-Luminal LO branch.  A number of
works show that cyclotron instabilities of the X and O modes can be
excited close to the light-cylinder (see for e.g.
\citealt{1991MNRAS.253..377K,1999JPlPh..62...65L}).  However, several works
have shown that close to neutron star surface, where the radio emission originates, the excitation of the Alfv\'en branch
is inefficient(e.g. \citealt{1986FizPl..12.1233L}; \citealt{1991MNRAS.253..377K} ; \citealt{2000MNRAS.315...31L}).  For the special case when the angle between the
propagation vector and the ambient magnetic field is zero, the O-mode
becomes purely longitudinal and is referred to as the Langmuir mode
(see fig. 2 of \citealt{1986ApJ...302..120A}). It has been shown in
several studies (e.g. \citealt{2002nsps.conf..240U}) that closer to the
neutron star surface this longitudional Langmuir mode can become
unstable.  Langmuir mode instability is a popular candidate for these
CCR charge bunches. Theoretically, a combination of linear and
non-linear plasma theory is needed to form stable charge bunch (see
\citealt{2000ApJ...544.1081M}). The linear part of the theory involves
development of two stream instability in the plasma that leads to the
growth of the amplitude of the longitudinal and electrostatic Langmuir
wave mode.  While the oscillating electric field of the Langmuir mode
can form longitudinal concentrations of charges, it is well known that
such linear Langmuir bunches are not capable of radiating coherently
(see e.g. \citealt{1986FizPl..12.1233L,
2000ApJ...544.1081M}). Analytical studies show that under
certain approximations stable bunches viz.
relativistic Langmuir charge soliton can form when non-linear effects
are taken into account (see
\citealt{1980Ap&SS..68...49P,2000ApJ...544.1081M}). Recent numerical
analysis have also found such stable charge bunches, when all
non-linear interactions are properly taken into account (see
\citealt{2018MNRAS.480.4526L}).  However there are several gaps in the
theory that remains to be addressed. The non-linear theory requires a
priori very large amplitude for the electrostatic waves.  The
  crucial question of quantitative estimates of linear growth rates of
  Langmuir waves for realistic pulsar plasma parameters, and if the
  growth rate is sufficient to drive the system beyond the linear
  regime requires thorough investigation.

The radio emission is excited in relativistically streaming pulsar
plasma that consists of a dense secondary positron-electron ($e^{+}e^{-}$) 
pair plasma, a tenuous high energy primary positron or electron 
($e^{+}/e^{-}$) beam and a
tenuous high energy ion beam. The growth of Langmuir instability
requires a two-stream condition to be established in this plasma.
Some early works on Langmuir mode in pulsar plasma in fact concluded
that Langmuir mode cannot become unstable (e.g.
\citealt{1975Afz....11..305S} ). However \cite{1979JETP...49..483L},
discussed that in relativistic plasma, particles close to the
velocity of light can be in resonance with the Langmuir mode. The
authors also discussed two regimes of growth viz., the kinetic and the
hydrodynamic regime.  There are three ways (referred to as case {\em
  C}1, {\em C}2, {\em C}3 hereafter) by which the two-stream instability can
develop in this flow: first for {\em C}1 between the high energy beams
and secondary plasma system, second for {\em C}2 between the electrons
and positrons in the secondary plasma itself due to longitudinal
drift, and third for {\em C}3 between the overlapping fast and slow
particles overlap of successive secondary plasma clouds due to
intermittent discharges at the polar gap.

Previous studies of the growth of Langmuir wave in pulsar plasma for the
three aforementioned cases of two-stream instability can be briefly
summarized as follows:

\noindent {\em C}1: Initial studies of pulsar radio emission mechanism
(e.g. \citealt{1975ApJ...196...51R} hereafter RS75) appealed to a
two-stream instability driven by high energy cold $e^{+}/e^{-}$
beam. Subsequent works(e.g. \citealt{1977MNRAS.179..189B}) found very
small growth rates for such $e^{+}/e^{-}$ cold
beam. \citet{1983Afz....19..753E} presented a hot plasma treatment of
the high energy $e^{+}/e^{-}$ beam and showed that
kinetic regime is suppressed and only the
    hydrodynamic regime survives. \citet{2002MNRAS.337..422G}
explored beam-driven hydrodynamic instability of a low frequency
longitudinal beam mode rather than the high-frequency Langmuir
mode. Most of the subsequent works 
(see \citealt{1999JPlPh..62...65L,1999ApJ...521..351M,2019JPlPh..85f9003R})
have focussed on this $e^{+}/e^{-}$ beam and
    found the growth rates to be negligible.\par

\noindent {\em C}2: The study by 
\citet[hereafter CR77]{1977ApJ...212..800C}
showed that as the
secondary pair plasma moves along the curved magnetic field line,
longitudinal drift causes the electron and positron distribution
function to separate. This can lead to two-stream instability in the
secondary plasma. However, they did not consider a hot plasma treatment
of the secondary plasma and obtained order of magnitude estimates of
growth rate using simple 
assumptions. \citet[hereafter AM98]{1998MNRAS.301...59A}
revisited the problem
where they presented a hot plasma treatment of the shifted
electron-positron distribution function within the secondary plasma cloud.

\noindent {\em C}3: \citet{1987ApJ...320..333U} showed that in
non-stationary plasma flow models, slow and fast moving particles of
two successive plasma clouds can overlap within a few hundred km from the
surface leading to the development of a two-stream instability in the
overlapping region. \citet{1988Ap&SS.140..325U} revisited the problem
and tried to estimate growth rates by approximating the distribution
function of the fast and slow particles in the overlapping region by
delta-function. AM98 extended and presented a
more realistic analytical way of constructing the form of the hot plasma
distribution function in the overlapping region.\par

The aforementioned studies suggest that large amplitude Langmuir wave cannot exist due to $e^{+}/e^{-}$ beam in {\em C}1. AM98 showed that two-stream instability in {\em C}2 and {\em C}3
  can result in high growth rates of Langmuir wave. They also found that the growth rate for {\em C}3 to be significantly larger than {\em C}2. Thus AM98 provided the necessary justification, that in principle large amplitude Langmuir waves can be triggered for both {\em C}2 and {\em C}3.

However AM98 obtained growth rates in the
    hydrodynamic regime for {\em C}2 and {\em C}3 using many
    simplifying assumptions. For example, they assumed the surface magnetic field to be dipolar while observations suggest the existence of a strong multipolar magnetic field at the surface. Also, they estimated growth rates as a function of the distance from the neutron star,
    using coarse spatial ( and temporal ) resolution. In their numerical scheme, AM98 did not obtain the complete solution of the dispersion relation at a given height and estimated growth rates only for some representative wave numbers.
    The coarse resolution in their analysis can wash away many important features of the evolution of the growth rate as a function of the distance from the neutron star. Hence it is necessary to undertake an updated study of AM98 where these shortcomings should be addressed appropriately. This is the primary focus of this work. Further
    keeping in view that a tenous high energy beam of ions can exist,
    we study the effect of the same on Langmuir mode instability in {\em C}1 and
    compare it with the $e^{+}/e^{-}$ beam.

It must be noted that the amplification of the Langmuir wave for a given
frequency $\omega$ depends on the gain `G' which is a product of the
growth rate ($\omega_\mathrm{I}$) and the time available for growth
($\Delta t$) as the amplitude is $\propto e^{G = \omega_\mathrm{I}
  \Delta t}$. If time $\Delta t$ is small, even with a high
$\omega_\mathrm{I}$, the amplification factor $G$ will be small and
one cannot use these waves to participate in the coherent emission
mechanism. In this work, we go beyond just the
    estimation of growth rate and present a method that employs the
    complete bandwidth of the growing waves to estimate $\Delta t$ and
    thereby the maximum gain possible for a given frequency. 
We present an exhaustive treatment of Langmuir
    mode instability for {\em C}1, {\em C}2 and {\em C}3 to examine
    the existence of large amplitude Langmuir waves in the pulsar
    radio emission region.
The outline of the paper is as follows: In sections \ref{Inputs to the plasma processes} we discuss physical constraints for the hot
plasma description and models of plasma flow. In sections \ref{Analysis
  of Langmuir instability} and \ref{Estimation of growth rates} we
describe the analysis method for the linear Langmuir instability and
study the growth rates and gain factors for these cases. In sections
\ref{Discussions and results} and \ref{conclusion} we discuss the
results and state our conclusions.

\section{ Inputs to the pulsar plasma parameters}\label{Inputs to the plasma processes}

\subsection{Constraints from radio emission height} \label{Implications from radio emission height}

A number of studies : \citet{1991ApJ...370..643B},
\citet{1997A&AS..126..121V} , \citet{2004A&A...421..215M} ,
\citet{2011ApJ...727...92M} , \citet{2008MNRAS.391.1210W} has
consistently found the emission region to be below 10 $\%$ of LC
across pulsar period (see fig. 3 of \citealt{2017JApA...38...52M}).  As
discussed by \citet{2004A&A...421..215M}, the various methods employed
to find radio emission heights can be affected due to measurement as
well as systematic errors, however for normal pulsars average
estimates of a few hundred kilometers above the neutron star surface
is considered reasonable. Specific studies
(e.g. \citealt{2002ApJ...577..322M}) also, focus on estimating the
range of emission heights as a function of frequency, and it is found
that a certain radius to frequency mapping exists in pulsars where
progressively higher frequencies arise closer and closer to the
neutron star surface.  These studies reveal that the broad-band pulsar
emission range from about few ten to hundred km at the highest
frequency $\sim 5$ GHz and to several hundred km at the lowest
frequency $\sim 100$ MHz. \citet{1991MNRAS.253..377K} showed that
cyclotron resonances can be excited only near the light cylinder. At
the radio emission heights all cyclotron resonances are suppressed and
only the Cherenk\'ov resonance condition can operate.

\subsection{Signature of Coherent Curvature Radiation} \label{Signature of curvature radiation}

Several lines of evidence (see \citealt{2001ApJ...549.1111L,
2005MNRAS.364.1397J, 2007ApJ...664..443R,
2012MNRAS.423.2736N,2013MNRAS.430.2281N,
2015MNRAS.453.4485F}) have revealed that the polarization of
the emergent pulsar radiation are directed either perpendicular or
parallel to the magnetic field line plane.  
These polarization modes are commonly
referred to as the extraordinary and Ordinary mode respectively which
are defined with their electric field vector being perpendicular and
parallel to the magnetic field plane respectively.  The eigenmodes of
the pulsar plasma viz., the X-mode and the O-mode is perpendicular and
parallel to the $\vec{k}-\vec{B}$ plane, where $\vec{B}$ is the 
ambient magnetic field and $\vec{k}$ is the propagation vector
of the wave.  
If the underlying excitation
mechanism is due to CCR, then these two planes need to be co-incident. For
any other form of excitation, these two planes can maintain arbitrary
orientation to each other. This implies that the
polarization of emergent radiation carries information about the
underlying excitation mechanism. This idea was applied by
\citet{2009ApJ...696L.141M} to a sample of nearly 100 $\%$ linearly
polarized single pulses which established CCR as the underlying
emission mechanism.

\subsection{Multi-polar surface magnetic fields and particle flows}\label{Evidence for multipolar field}
It is well known (see e.g. \citealt{2004A&A...421..215M}) that at a
few hundred kilometers above the neutron star surface, from regions
where the radio emission originates, the underlying magnetic field
structure is dipolar. However, in recent years there are
several pieces of evidence for the presence of surface multipolar fields. For
example \citet{2001ApJ...550..383G} and \citet{2020MNRAS.492.2468M}
suggested that the radio-loud nature of the extremal long period 8.5 s
pulsar J2144-3933 (\citealt{1999Natur.400..848Y}) can only be
explained if surface magnetic fields have a radius of curvature
$\rho_\mathrm{c} \sim 10^{5}$ cm at the surface, which is only possible
due to presence of strong multipolar surface magnetic field. 
The X-ray observations have also confirmed the
presence of multipolar fields on the surface (see e.g.
\citealt{2019MNRAS.489.4589A}).

The presence of multipolar surface magnetic field significantly affects the description
of the plasma. At the polar cap magnetically induced pair creation
processes are triggered.  The presence of multipolar surface magnetic field decreases
the radius of curvature at the surface thereby increasing 
the efficiency of the pair creation process.  As a result the number
density of the pair plasma exceeds the co-rotational value
$n_\mathrm{GJ}$ by a multiplicity factor $\kappa_\mathrm{GJ}$.
Observations of PWNe has revealed $\kappa_\mathrm{GJ} \sim 10^{4} -
10^{5}$ (see \citealt{2007ApJ...658.1177D} ;
\citealt{2011ASSP...21..624B} ).  To get this high value of
$\kappa_\mathrm{GJ}$ estimations show that to get $\kappa_\mathrm{GJ}
\sim 10^{4} - 10^{5}$ multi-polar fields are required (see
\citealt{2010MNRAS.406.1379M} ; \citealt{2015MNRAS.447.2295S} ;
\citealt{2019ApJ...871...12T}) whereas for purely dipolar fields
$\kappa_\mathrm{GJ}$ is about the order of a few tens to hundred (see
\citealt{2001ApJ...560..871H}; \citealt{2002ApJ...581..451A})

\subsubsection{Need for multipolar surface magnetic field for CCR}\label{Need for a dense plasma}
The limiting brightness temperature for incoherent curvature radiation
is $T^\mathrm{ICR}_\mathrm{lim} \approx 10^{13} \;$K$ $ (see
\citealt{1978ApJ...225..557M}).  In the Rayleigh-Jeans regime, the
brightness temperature is proportional to power. CCR is an `$N^2$'
process meaning if `$N$' particles are involved, the power is boosted
by a factor `$N$' compared to what would be achieved if the charged
particles were emitting independently (or incoherently). The number of
particles participating in CCR to explain the observed high brightness
temperature is given by $ N_\mathrm{CCR} \;=\;
{T_\mathrm{obs}}/{T^\mathrm{ICR}_\mathrm{lim}} \;\approx\; 10^{12}
$. Radio emission from pulsars are received from $10$ MHz to $10$
GHz. The length of the bunch should satisfy the constraint $L \ll
{c}/{\nu_\mathrm{High}} \sim 3$ cm for coherence to be maintained for
all frequencies. Assuming $L \sim 1 $ cm, the corresponding number
density required $n_\mathrm{CCR} \sim 10^{12} \; $cm$^{-3} $.  At an
emission height of $r_\mathrm{em} \;=\; 50 \; R_\mathrm{NS}$, the
Goldreich-Julian value is given by $ n_\mathrm{GJ} \;=\; 5.52 \times
10^{8} \; (1 \; $sec$ \;/\; P)\;(B \;/\; 10^{12} \; $Gauss$ )\;
(r_\mathrm{NS} / r_\mathrm{em})^3 $ cm$^{-3}$ (see
\citealt{1969ApJ...157..869G}). Thus, CCR requires number density in
excess of the Goldreich-Julian value by a factor of $10^{4}$.

\subsubsection{Description of particle flow and secondary plasma distribution functions} 

The models of plasma flow can be divided into two classes: a) The
steady flow model (also known as SCLF model) given by
\citet{1979ApJ...231..854A} where when condition above the polar cap
is such that $\vec{\Omega}_\mathrm{Rot} \cdot \vec{B} > 0 $ 
 (here 
$\vec{\Omega}_\mathrm{Rot} = 2 \pi/ P$ is the pulsar rotational 
frequency),
the electrons can be easily pulled out from the star and a stationary
flow of electron beam-plasma can be maintained and; b) The non-stationary
spark discharge model (also referred to Inner acceleration gap model
or the pure vacuum gap model) by RS75 for pulsars with
$\vec{\Omega}_\mathrm{Rot} \cdot \vec{B} < 0 $ giving rise to
an intermittent plasma flow due to sparking discharges at the polar
gap.  In both these models the beam-plasma system is established.

The vacuum gap model of RS75 is more successful in explaining
pulsar radio observations like sub-pulse drift phenomenon, however the original
model required certain modifications.
\citet{2003A&A...407..315G} noticed that the sub-pulse drift rates and 
the temperature of the thermal X-ray emitting polar cap are both lower than that 
predicted by the pure vacuum gap model of RS75. 
They suggested that the pure vacuum gap is untenable and must be
partially screened such that the potential is $\Delta V_\mathrm{vac}$
across the gap is replaced by $ \eta \; \Delta V_\mathrm{vac}$ where
$\eta$ is the screening factor. For a pulsar of period 1 second and
dipolar magnetic strength of $10^{12}$ gauss, the maximum potential
drop available in vacuum is $ \Delta V_\mathrm{vac} = 6 \times
10^{12}$ volts. The authors constrained $\eta = 0.1$ , which gives the
Lorentz factor of the high energy primary beams of $e^{+}/e^{-}$ and
ions to be given by $\gamma_\mathrm{b,e^{+}/e^{-}} \sim 10^{6}$ and
$\gamma_\mathrm{b,ions} \sim 10^{3}$ respectively. Assuming CCR we can
find an order of magnitude estimate of the bulk Lorentz factor of the
secondary pair plasma.  Most of the power in CCR for charge bunch with
Lorentz factor $\gamma$ is concentrated near the critical frequency $
\omega_\mathrm{c} \;=\; 1.5 \; {\gamma^3 c}\;/\;{\rho_\mathrm{c}}$,
(see \citealt{1962clel.book.....J},  where $c$ is the velocity of
light). Assuming observing frequency
$\nu_\mathrm{obs} = 1.4$ GHz to be close to the critical frequency at
$r_\mathrm{em} \;=\; 50\; R_\mathrm{NS} $ where $\rho_\mathrm{c} \;
\approx \; 10^{8}$ cm, we have $\gamma \approx 200-300$.

For our work we assume the distribution functions of all the species
to be relativistically streaming gaussians. For secondary plasma, the
mean and the width are assumed to be $\sim 200-300$ and $\sim 40-60$
respectively. Note that the two stream-condition can be established in
non-stationary flow by all three cases of {\em C}1, {\em C}2 and {\em
  C}3 whereas for stationary flow only the cases {\em C}1 and {\em
  C}2.

To summarize CCR needs to be excited by large amplitude Langmuir waves
in a hot relativistically streaming dense secondary pair plasma
. At the radio emission heights, the wave-particle interaction is
mediated by the Cherenk\'ov resonance condition. In subsequent
sections we address how large amplitude Langmuir waves can be
triggered for the three cases, {\em C}1,{\em C}2, {\em C}3 of two-stream
instability discussed earlier.

\section{Analysis of Langmuir instability}\label{Analysis of Langmuir instability}
 
In the following subsections, we establish the methodology for studying
  Langmuir instability. To do this we define a threshold gain for a wave of a
particular frequency that can be used as a proxy for the breakdown of
the linear theory. This, in turn, is achieved by solving the complex
frequencies using the appropriate dispersion relation. For this
analysis, the following aspects need to be considered.

\subsection{The Dispersion relation in the observer frame of reference} 
The dispersion relation of the Langmuir mode for a strictly
one-dimensional relativistic flow in the observer frame of reference
is given by (see section 4 of AM98)
\begin{eqnarray}
&\ \epsilon(\omega, k)  = k c + \sum_{\alpha} \omega_\mathrm{p,\alpha}^2 \int_{-\infty}^{+\infty} \; dp_{\alpha}\; \frac{\partial f_{\alpha}^\mathrm{(0)}}{\partial p_{\alpha}} \;\frac{1}{\left(\omega - \beta_{\alpha} \; k c\right)} = 0
\label{Dispersion relation for the Langmuir mode}
\end{eqnarray}
where $\omega^2_\mathrm{p,\alpha}= 4\pi n_{\alpha} q^2_{\alpha} /
m_{\alpha}; \gamma = \sqrt{1 + p^2_{\alpha}} ; \beta_{\alpha} =
p_{\alpha}/\sqrt{1 + p^2_{\alpha}}$. Here $n_{\alpha}$, $q_{\alpha}$,
$m_{\alpha}$, $p_{\alpha}$ and $f^\mathrm{(0)}_{\alpha}$ is the number
density, charge, mass, dimensionless momenta and the equilibrium
distribution function of the $\alpha$-th species in the plasma such
that $n_{\alpha} = \kappa_\mathrm{GJ,\alpha} \; n_\mathrm{GJ}$,
$p_{\alpha} = \gamma m_{\alpha} v / m_{\alpha} c = \gamma \beta $ and
$\int^{+\infty}_{-\infty} \; dp_{\alpha} \; f^{(0)}_{\alpha} = 1$. We
assume $f^{(0)}_{\alpha}$ = ($1/(\sqrt{2 \pi \sigma^2_{\alpha}})\;$
exp $[-(p_{\alpha} - \mu_{\alpha})^2/ 2 \sigma^2]$), with mean
$\mu_{\alpha}$ and width $\sigma_{\alpha}$ for all $\alpha$
species. In the super-Luminal region the Cherenk\'ov resonance
condition cannot be satisfied and hence there is no singularity in the
integral of Eq. \ref{Dispersion relation for the Langmuir mode}. The
integral can be integrated by parts to give the dispersion relation as
\begin{equation}\label{super-Luminal region}
1 - \sum_{\alpha} \omega_\mathrm{p,\alpha}^2 \int_{-\infty}^{+\infty}
dp_{\alpha} \; \frac{1}{\gamma^3} \frac{f_{\alpha}^{(0)}}{(\omega -
  \beta_{\alpha}\kappa c)^2} = 0
\end{equation}

At $k =0 $ the cut-off $\omega_\mathrm{0}$ is given by $\omega_0^2
\;=\; \sum_{\alpha} \omega_\mathrm{p,\alpha}^2
\int_{-\infty}^{+\infty} \; dp_{\alpha} \;f_{\alpha}^{(0)} / \gamma^3$
while the frequency $\omega_1$ at which the Langmuir mode touches the
$\omega = \kappa c$ line is given by
\begin{equation}\label{Electrostatic transition}
\omega_\mathrm{1}^2 \;=\; \sum_{\alpha} \omega_\mathrm{p,\alpha}^2
\int_{-\infty}^{+\infty}\; dp_{\alpha} \; \frac{1}{\gamma^3}
\frac{f_{\alpha}^{(0)}}{(1 - \beta_{\alpha})^2}
\end{equation}

The dispersion relation can be cast in the dimensionless form using
$\omega_1$ as a scaling factor to give
\begin{eqnarray}
&\ \epsilon(\Omega, K) = K + \sum_{\alpha} \chi_{\alpha}
  \int_{-\infty}^{+\infty} \; dp_{\alpha}\; \frac{\partial
    f_{\alpha}^\mathrm{(0)}}{\partial p_{\alpha}}
  \;\frac{1}{\left(\Omega - \beta_{\alpha} \; K\right)} = 0
\label{dimensionless dispersion relation for the Langmuir mode}
\end{eqnarray}
such that $ \Omega = \omega/\omega_1 ; K = k c / \omega_1 ;
\chi_{\alpha} = \omega_\mathrm{p,\alpha}^2/\omega_1^2$.  All
integration using the distribution function for the species ``
$\alpha$ '' will be denoted by
$\left\langle(...)\right\rangle_{\alpha} = \int_{-\infty}^{+\infty}
dp_{\alpha}\; f_{\alpha}^{(0)} \;(...)$.

\subsubsection{Growth rates in the sub-luminal regime.}\label{Expression for growth rate in kinetic and hydrodynamic regime }

The Cherenk\'ov resonance condition 
$ \omega - \beta_{\alpha} kc $ in the denominator of
  Eq. \ref{Dispersion relation for the Langmuir mode} is satisfied in
  the sub-Luminal regime, and produces a singularity in the integral
  for Langmuir wave frequencies ($ \geq \omega_1$). The pole
$p_\mathrm{pole} = \omega/\sqrt{(\kappa c)^2 - \omega^2}$ of the
dispersion function needs to be treated using Landau prescription. For
the growth of Langmuir waves, the Landau prescription allows for two
regimes of growth (see Appendix A for discussion) viz., the kinetic
regime, and the hydrodynamic regime.

In the kinetic regime the pole lies very close to the real axis
contour such that the Landau contour has to be analytically continued
to the lower half plane. In this regime the dispersion relation is
broken into a principal value integral and a residue at the pole. The
dimensionless growth rate in the kinetic regime is given by (see
Eq. A13 of Appendix A )
\begin{equation}\label{kinetic growth rate}
\Gamma_\mathrm{kin} = \frac{\pi}{2K^2}\frac{ \chi_\mathrm{b}
  \left(\frac{\partial f_\mathrm{b}^{(0)}}{\partial p_\mathrm{b}}
  \gamma^3\right)\Big\rvert_{p_\mathrm{b}
    \;=\;p_\mathrm{b,res}}}{\chi_\mathrm{s} \left\langle
  \gamma^3(1+\beta_\mathrm{s})^3 \right\rangle_\mathrm{s}}
\end{equation}  
such that $\Gamma_\mathrm{kin} = \omega_\mathrm{I,kin}/\omega_1$ and
$\langle (...)\rangle_{\alpha} = \int^{+\infty}_{-\infty} \;
dp_{\alpha} \; (...) \; f^{(0)}_{\alpha}$.

Here subscript $\mathrm{b}$ and $\mathrm{s}$ correspond to the 
beam and secondary plasma respectively. Note that the distribution
having the pole correspnd to $\mathrm{b}$ and the distribution
function away from the pole correspond to $\mathrm{s}$. 
The kinetic growth rate is a local description as it requires
  only the derivative of the distribution functions at
  $p_\mathrm{\alpha,res}$, and is referred to be of resonant type where
  only the set of particles at and around $p_\mathrm{\alpha,res}$
  contribute to the growth. It must also be noted that the expression
for kinetic growth rate has been derived under the assumption that the
slopes are gentle viz., $\sigma_{\alpha}$ is broad and the distribution
functions have no discontinuity.

In the hydrodynamic regime the pole lies above the Landau contour. In
this regime the dispersion relation can be integrated by parts along
the real axis for complex frequency $\omega = \omega_\mathrm{R} + i
\omega_\mathrm{I}$ where $\omega_\mathrm{I} > 0$.The real and
imaginary part of the dimensionless dispersion relation (see Eq. A18
of Appendix A) in the hydrodynamic regime are given by
\begin{align*}\label{Hydrodynamic equations}
&\ 1 \;-\; \sum_{\alpha} \chi_{\alpha} \int_{-\infty}^{+\infty}
  dp_{\alpha}\; \frac{f_{\alpha}^{(0)}}{ \gamma^3}
  \;\frac{\left\lbrace\left(\Omega_\mathrm{R} -
    \beta_{\alpha}K\right)^2 -
    \Omega_\mathrm{I}^2\right\rbrace}{\left[\left(\Omega_\mathrm{R} -
      \beta_{\alpha}K\right)^2 + \Omega_\mathrm{I}^2\right]^2} \;=\; 0
  \\ &\ - i\;2\;\Omega_\mathrm{I}\;\; \sum_{\alpha} \chi_{\alpha}
  \int_{-\infty}^{+\infty} dp_{\alpha}\; \frac{f_{\alpha}^{(0)}}{
    \gamma^3} \;\frac{(\Omega_\mathrm{R} -
    \beta_{\alpha}K)}{\left[\left(\Omega_\mathrm{R} -
      \beta_{\alpha}K\right)^2 + \Omega_\mathrm{I}^2\right]^2} \;=\; 0
  \numberthis
\end{align*}
where $\Omega_\mathrm{I} = \omega_\mathrm{I}/\omega_1 ;
\Omega_\mathrm{R} = \omega_\mathrm{R}/\omega_1 $.  The above set of
equations have to be solved simultaneously to get a solution for
the dimensionless quantities
$\Omega_\mathrm{R}$ and
$\Omega_\mathrm{I}$. The dimensional growth
    rate is a product of the dimensionlesss growth
    rate($\Omega_\mathrm{I}$) and the scaling factor $\omega_1$.
Quantities in the dimensional form will have the following notation $
\omega_\mathrm{R} = $Re$(\omega)$ [in rad/s]; $\omega_\mathrm{I} =
$Im$(\omega)$[in s$^{-1}$]. Unlike the kinetic regime, the growth
rates in the hydrodynamic regime require a complete description of the
distribution function for all the species involved. In this sense the
hydrodynamic regime represents a non-resonant type of growth where all
the particles contribute to the growth. It must be noted that the
growth rates in the hydrodynamic regime are necessarily greater than
that in the kinetic regime.

Next, we define the equivalent distribution function (hereafter EDF) as the
number density weighted summation of the distribution functions of the
species that constitute the system. The expression for growth rates
in both regimes requires that the EDF satisfy the relativistic
generalization of Gardner's theorem \footnote{ see
  \citet{1963PhFl....6..839G} for the original version stated for a
  non-relativistic plasma system.} states that if the EDF of a plasma
system is single-humped then such a system cannot support a growing
set of waves ( see Appendix B for proof). Thus two-stream instability
cannot be satisfied for a single-humped EDF.

If the distribution functions in the EDF are given by gaussians and
the mean of the gaussians are well-separated, the hydrodynamic
growth-rate has to satisfy the condition that
\begin{align*}\label{separation of the regimes}
\Omega_\mathrm{I} \geq \frac{\Delta \gamma_{T}}{\bar{\gamma_{>}}^3} \;      \numberthis
\end{align*}
where the quantities $\Delta \gamma_{T}$ and $\bar{\gamma_{>}}$ refer
to the width and mean of the gaussian distribution function with the
higher mean 
(see e.g. Eq. 49 of AM98). The condition Eq.~\ref{separation of
    the regimes} can be used as a separation between the hydrodynamic
  and the kinetic regime, where the condition is reversed for the
  kinetic regime. However, if the means of the gaussians are
not-sufficiently separated this threshold is much lower.

In this work apart from a brief discussion of the kinetic regime in {\em C}1, we focus exclusively on the hydrodynamic regime for all three scenarios. The algorithm for
    solving the hydrodynamic equations are presented in Appendix D.

\subsection{Constraint on $\omega_1$} \label{Constraint on omega1}

The solution of the dispersion relation must have the character that
Re$ \; (\omega) \geq \omega_1$. Combining the number density
constraint as shown in \ref{Need for a dense plasma}, the
corresponding value for the scaling $\omega_1$ using
Eq. \ref{Electrostatic transition} is given by
\begin{align*}\label{omega_1_threshold}
\omega^\mathrm{Th}_1 \sim  \sqrt{\gamma} \sqrt{n_\mathrm{CCR}} \times 10^{4.5} \; \mathrm{rad/s} \geq 10^{11} \; \mathrm{rad/s}  \numberthis
\end{align*}

\subsection{Maximum gain factor for a particular Re($\omega$) } \label{Maximum gain factor for a particular frequency}
 Following \citet{2002MNRAS.337..422G} we
    introduce a method to estimate the maximum amplification for a
    given frequency Re($\omega$) using the bandwidth of growing waves
    as a proxy for the time available ($\Delta t$) for growth. Let
us consider the dispersion relation at two points `$A$' and `$B$'
along a given open magnetic field line. The ratio of the scaling
frequencies at these two points is given by ${\omega^\mathrm{B}_1}/{\omega^\mathrm{A}_1}
=\left( {r_\mathrm{A}}/{r_\mathrm{B}}\right)^{3/2}$. The same frequency corresponds
to the frequency $\omega^\mathrm{B}_1 + \Delta \omega_\mathrm{B}$ at point `$B$' where
$\Delta \omega_\mathrm{B}$ is the bandwidth of growing waves at `$B$'. Then we
have
\begin{align*}\label{delta t}
&\ \frac{\Delta \omega^\mathrm{B}}{\omega_1^\mathrm{B}} = \left( \frac{\omega^\mathrm{A}}{\omega^\mathrm{A}_1}\right) \left( \frac{r_\mathrm{B}}{r_\mathrm{A}}\right)^{3/2} - 1 \\
\Rightarrow &\  \Delta r = r_\mathrm{B} - r_\mathrm{A} = r_\mathrm{A} \left[ \left( \frac{\Delta \Omega^\mathrm{B} + 1}{\Omega^\mathrm{A}}\right)^{2/3} - 1 \right]  \numberthis
\end{align*}

The time for which the growth rate for frequency $\omega_\mathrm{A}$ can be
sustained is given by $ \Delta t \;=\; {\Delta r}/{c} $. Assuming that
the growth rate remains constant and using Eq. \ref{delta t} the
maximum gain for frequency $\omega^\mathrm{A}$ is given by
\begin{align*} \label{Maximum gain }
\mathrm{G}_\mathrm{max} \;=\; \Gamma_{\omega^\mathrm{A}} \Delta t =\Gamma_\mathrm{\omega^\mathrm{A}} \frac{\Delta r}{c} \numberthis
\end{align*}
This will be used for the estimation of the maximum gain following the
numerical solution of the dispersion relations to get the growth rate
($\Gamma_{\omega^{\mathrm{A}}}$) and the bandwidth for cases {\em C2} and {\em
  C3}.

\subsection{Criterion for breakdown of the linear theory} 
\label{Criterion for breakdown }

The solution of the dispersion relation does not carry information
about the amplitude of the Langmuir wave i.e, we can only calculate
$e^\mathrm{G}$ given the growth rate and the time available for growth. The
amplitude is given by $E(t) \;=\; E(t=0)\; e^\mathrm{G}$. The initial
amplitude $E(t=0)$ of the wave at a particular frequency has to be
obtained from a different treatment of the dispersion function as is
done in subsection C1 of Appendix C. However, since the Langmuir wave
grows at the expense of the particles in the plasma, the maximum
energy that the wave can gain is equal to the total energy of all the
particles in the plasma. Thus although the linear theory can predict
arbitrary gain, in reality, there exists a gain threshold which cannot
be exceeded. The next paragraph describes how to get 
an estimate of this threshold from consideration of maximum energy
available in the plasma. If the linear theory predicts a gain 
close to or higher than this threshold, then it must be taken as a
definitive indicator of the breakdown of the linear theory.

To indicate the breakdown of the linear theory we propose the
following hypothetical situation: The growth rates are sufficient for
breakdown of linear theory if the linear theory predicts the energy
density in the field to be equal to the total energy density.  Let us
consider the dispersion relation for a wave of frequency
$\omega^\mathrm{A}_1$ at two points `$A$' and `$B$' with point `$B$'
higher up along a given field line. Assuming a constant growth rate
the field energy density at point `$B$' for $\omega^\mathrm{A}_1$
should satisfy the condition $|E_\mathrm{\omega^\mathrm{A}_1}|_\mathrm{B}^2 /  8\pi
\approx (|E_\mathrm{\omega^\mathrm{A}_1}|_\mathrm{A}^2 / 8\pi) \; e^\mathrm{2
  G_{max}} = W_\mathrm{B}$ where $W_\mathrm{A}$ and $W_\mathrm{B}$ are
the total energy density at points ``$A$'' and ``$B$''. Using Eq. C9
from Appendix C we obtain a threshold gain indicating the breakdown of
the linear regime viz., $ G^\mathrm{Th}_\mathrm{max} \approx 
\mathrm{ln} \left[ \sum_{\alpha} \gamma^2_{\alpha} \right] / 
2$. For high energy beam driven instability this threshold is dictated
by the Lorentz factor of the high energy beams. The gain threshold for
instability in case {\em C}1 driven by $e^{\pm}$ beam and the ion beam
comes out to be 12 and 6 respectively. The gain threshold for cases
{\em C}2 and {\em C}3 involving only the species in the secondary
plasma is given by 5. Thus, in all three cases {\em C}1,{\em C}2, {\em C}3 a
representative threshold of gain to indicate the breakdown of linearity
can be taken as
\begin{align*}\label{Threshold for non-linear regime}
    \mathrm{G}^\mathrm{Th}_\mathrm{max} \;=\; 5  \numberthis
\end{align*}

\section{Estimation of growth rates and gain}
\label{Estimation of growth rates}
In what follows all analyses will be done along the last open field
line of an aligned rotator with period $ P = 1 $ second and global
dipolar strength $B_\mathrm{d} = 10^{12}$ Gauss. For case {\em C1} we get an
analytical estimate of the maximum gain. For cases {\em C2} and {\em
  C3} the hydrodynamic equations Eq. \ref{Hydrodynamic equations} are
solved numerically ( see Appendix D ) to obtain growth rates and
maximum gain as a function of $r/R_\mathrm{NS}$.

\subsection{{\em C}1: Beam-driven Growth}\label{Beam driven growth}
The beam distribution function is given by
\begin{equation}\label{beam distribution function}
    f_\mathrm{b}^\mathrm{(0)} = \frac{1} {\sqrt{\pi} p_\mathrm{T_b}}  e^{  - (p_\mathrm{b} -\bar{p}_\mathrm{b} )^2/{p^2_\mathrm{T_b}}}
\end{equation}
such that $\mu_\mathrm{b} = \bar{p}$ and $\sigma_\mathrm{b} \;=\; p_\mathrm{T_b} / \sqrt{2}$. Further let us introduce the width to mean ratio given by    
\begin{align*}\label{width to mean ratio}
     x_\mathrm{b} \;=\; \frac{p_\mathrm{Tb} }{\bar{\gamma_\mathrm{b} }}       \numberthis
\end{align*}

From subsection \ref{Maximum gain factor for a particular frequency} the
maximum gain for a particular frequency $\omega_\mathrm{A}$ is given by  
\begin{align*}
G^\mathrm{b}_\mathrm{max} = \Gamma_\mathrm{\omega^\mathrm{A}} \left( \frac {r_\mathrm{A}} 
{c} \right) \left[ \left( \frac{\Delta \Omega^\mathrm{B} + 1 } {\Omega^\mathrm{A}} \right)^{2/3} - 1  
  \right] 
\end{align*} 
Combining this with expressions for bandwidth of growing
waves (from subsection C2 of
Appendix C ) and the threshold  Eq. \ref{separation of the regimes}
we obtain the expression for the maximum gain for
Langmuir waves of frequency $\omega^\mathrm{A}_1$ due to beam-driven instability as
\begin{equation}\label{Gain for a beam}
G^\mathrm{b}_\mathrm{max} \approx x_\mathrm{b}  \frac{p_\mathrm{T_{b}}}{\bar{\gamma_\mathrm{b}}^2}  \frac{r_\mathrm{A}}{c} \left[ \frac{8 x_\mathrm{b}}{3} \left( \frac{\gamma_\mathrm{s}}{\bar{\gamma}_\mathrm{b}}\right)^2 \right] \omega^\mathrm{A}_1   
\end{equation}
\subsubsection{High energy positron/electron beam with $\gamma_b \sim 10^{6}$}

\citet{1983Afz....19..753E} demonstrated that
    only the hydrodynamic regime exists for the high energy
    $e^{+}/e^{-}$ beam even for a broad distribution function or large
    $x_\mathrm{b}$.  To estimate the gain we chose a representative
    value $x_\mathrm{b,e^{+}/e^{-}} = 0.3 $ in Eq. \ref{Gain for a
        beam} and get G$_\mathrm{max,e^{+}/e^{-}}$, which is plotted
    as solid blue curve in Fig.\ref{Gain high energy beam} as a
    function of $r/R_\mathrm{NS}$.

\subsubsection{High energy ion beam with $\gamma_{ion} \sim 10^{3}$}
Unlike the $e^{+}/e^{-}$ beam case, there is
    no apriori information indicating for what value of
    $x_\mathrm{ion}$ will the hydrodynamic regime exist exclusively. Thus to
    proceed we get an estimate of `$x_\mathrm{min,ion}$' for which the
    kinetic regime gets suppressed completely.  We assume the ion beam
    to be composed of iron ions such that $ {n_\mathrm{s} }
/{n_\mathrm{ion}} = \kappa_\mathrm{GJ} \sim 10^{4}$ ,
${\gamma_\mathrm{ion} }/{\gamma_\mathrm{s}} \sim 10$ , $
{m_\mathrm{ion}}/{m_\mathrm{e}} \sim 56 $ and
${Q_\mathrm{ion}}/{Q_\mathrm{e}} \sim 26 $. By substituting
Eq. \ref{beam distribution function} in Eq. \ref{kinetic growth
    rate} and estimating it at $p = \sqrt{2} p_\mathrm{T_b}$ , we
have the maximum dimensionless growth rate in the kinetic regime $
\Gamma^\mathrm{max}_\mathrm{kin} \approx \sqrt{{\pi}/{2}} (
      {\chi_\mathrm{ion} \gamma^{3}_\mathrm{ion}} / { 8
        p^2_\mathrm{T,ion} e^2 \chi_\mathrm{s} \gamma^3_\mathrm{s} })
      $. $\Gamma^\mathrm{max}_\mathrm{kin}$ so obtained must follow
      the constraint $ \Gamma^\mathrm{max}_\mathrm{kin} \leq
      {p_\mathrm{T,ion}} / {\gamma^3_\mathrm{ion}}$. This gives $
      x_\mathrm{b,ion} \geq [ \sqrt{ {\pi}/{2}} ( {n_\mathrm{ion}
          Q^2_\mathrm{ion} m_\mathrm{e} \gamma^3_\mathrm{ion} } / { 8
          e^2 2 n_\mathrm{s} Q^2_\mathrm{e} m_\mathrm{ion} \;
          \gamma^3_\mathrm{s} } )]^{1/3} \approx 0.01 $. Substituting
      $x_\mathrm{min,ion} = 0.01$ in Eq. \ref{Gain for a beam} we get
      G$_\mathrm{max,ion}$ plotted as dashed red line in Fig.\ref{Gain
        high energy beam} as a function of $r/R_\mathrm{NS}$.  As
      evident in the figure, the maximum gain for the ion is larger
      than the $e^{+}/e^{-}$ beam, however still significantly smaller
      than the gain threshold given by Eq. \ref{Threshold for
          non-linear regime}.

\begin{figure}
\centering
\includegraphics[width = \columnwidth]{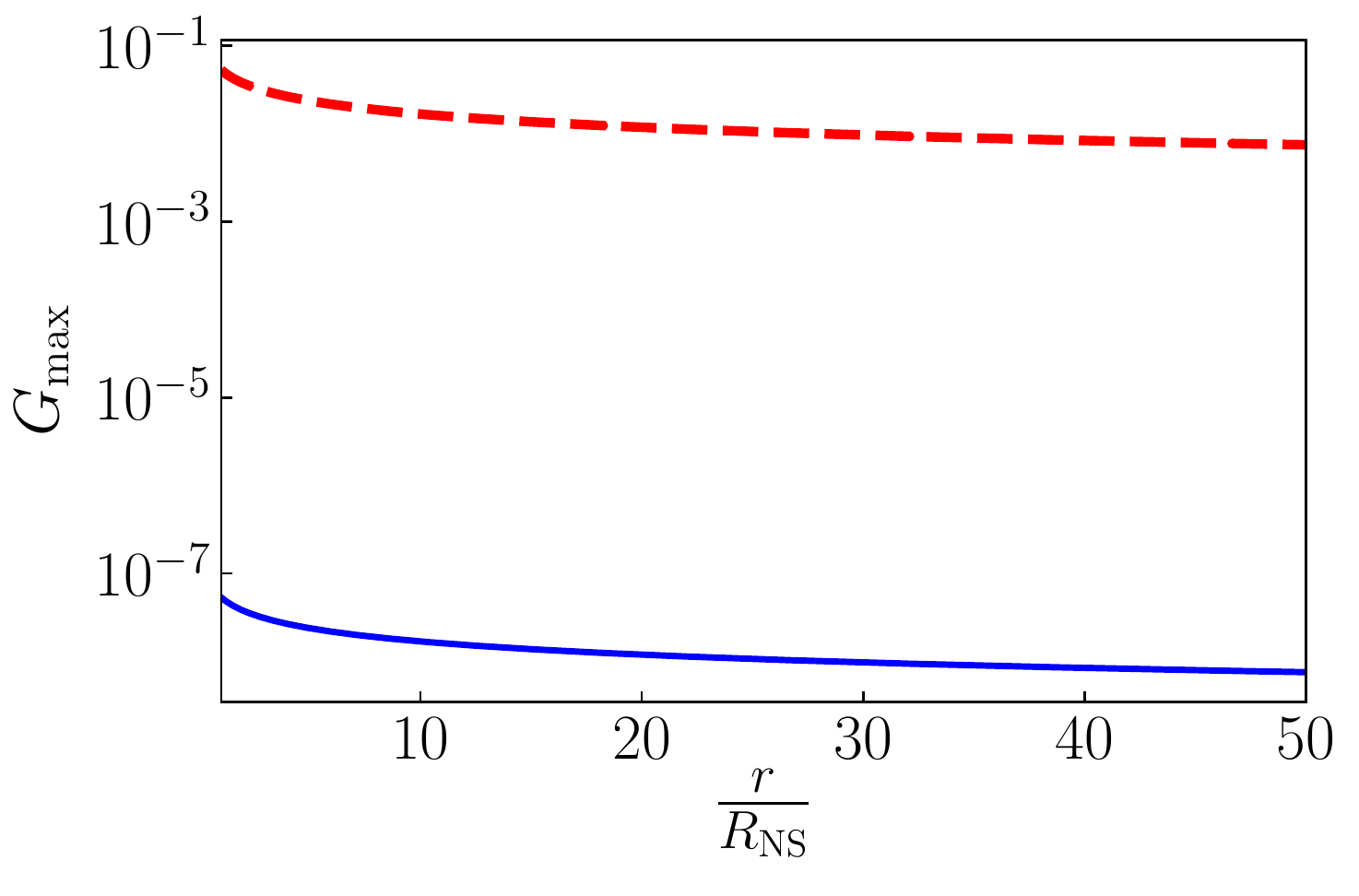}
\caption{Plot for maximum gain that can be obtained for high energy
  beam driven instability. The dashed red line and the solid blue line
  gives the maximum gain for an ion beam and positron/electron beam
  with $x_\mathrm{min,ion} = 0.01$ and $x_\mathrm{e^{+}/e^{-}} = 0.3$
  . A multiplicity factor of $\kappa_\mathrm{GJ} = 10^{4}$ and
  $\gamma_\mathrm{s} = 200$ has been used.}
\label{Gain high energy beam}
\end{figure}

It can be seen from Fig. \ref{Gain high energy beam} that none of the high energy beams 
can exceed the gain threshold given by Eq. \ref{Threshold for non-linear regime}.

\subsection{ {\em C}2: Growth due to longitudinal drift }\label{Longitudinal drift}

CR77 suggested that due to the motion of the
combined system of ``beam + secondary plasma'' along curved magnetic
field lines , the electron-positron distribution function in secondary
plasma has to separate to provide a steady state current dictated by
the local Goldreich-Julian value and the solenoidal nature of current
flow.  The separation of the bulk velocity $\Delta \beta$ of the
species in the secondary plasma at a distance of $r_\mathrm{A}$  from the
neutron star surface is given by
\begin{align*}
\label{longdrift}
|\Delta \beta|_\mathrm{A} \;\approx\; \left| \left(\frac{\rho_\mathrm{b} }{\rho_{s}} \right)_\mathrm{o} \;\left[ \frac{(\hat{\Omega}_\mathrm{Rot} \cdot \hat{B} \; f_\mathrm{Rot})_\mathrm{A}}{(\hat{\Omega}_\mathrm{Rot} \cdot \hat{B}\; f_\mathrm{Rot})_\mathrm{o}} \;-\; 1\right]\right| \numberthis
\end{align*}

where $\rho_\mathrm{b}$ and $\rho_\mathrm{s}$ correspond to the
  charge density of the beam and secondary plasma respectively.  The
  ratio $(\rho_\mathrm{b}/\rho_\mathrm{s}) = 1/\kappa_{GJ}$, which we
call as the density term.  Here the reference point `` $O$'' is taken
at $r = 1.02 \; R_\mathrm{NS}$ where pair creation cascades ceases,
a. The correction `$f_\mathrm{Rot} \approx 1 + \mathrm{\bm{O}}
({\Omega^2r^2}/{c^2})$' due to rotation can be taken to be 1, as the
higher order term $\mathrm{\bm{O}} ({\Omega^2r^2}/{c^2}) \sim 0.01$ at
$r = 50 R_\mathrm{NS} $ for a pulsar with $P = $ 1 second. It can be
seen that the separation of the electron-positron distribution
function is a product of two terms viz., the density term
$({\rho_b}/{\rho_{s}})_{o}$ and the geometrical term $
[{(\hat{\Omega}_\mathrm{Rot}\cdot \hat{B}
    )_\mathrm{A}}/{(\hat{\Omega}_\mathrm{Rot}\cdot\hat{B} )_\mathrm{o}} - 1 ]$.
The geometrical factor is zero only for very
    straight magnetic field lines.  Thus curved magnetic field line is
    a necessary requirement of longitudinal drift/ separation of
    $e^{\pm}$ distribution in a secondary plasma.

{ \textbf{Simulating EDF for {\em C}2: } }
Eq. \ref{longdrift} just gives the difference between
$\beta_\mathrm{(+)}$ and $\beta_\mathrm{(-)}$. To solve for
$\beta_\mathrm{(+)}$ and $\beta_\mathrm{(-)}$ we need an additional
constraint. We make the simplifying
    assumptions that [i] longitudinal drift affects only the mean of
    the distribution functions i.e, the separation of the bulk
    velocity is equal to the separation of the mean of the $e^{\pm}$
    distribution functions; and [ii] The $e^{\pm}$ distribution
    functions are co-incident at ``$O$'' with mean lorentz factor $
    \gamma^\mathrm{O}_\mathrm{(\pm)}$ and at any point $r_\mathrm{A}$
    the mean of the distribution functions separate to attain values
    that are symmetrical about $ \gamma^\mathrm{O}_\mathrm{(\pm)}$.

The requirement of symmetry translates to the condition that for any other point `$A$' along
the field line
\begin{align*}\label{Deltagamma}
\left| \Delta \gamma_\mathrm{(+)} \right| \;=\; \left| \Delta \gamma_\mathrm{(-)} \right| \;=\; \left| \Delta \gamma \right| \;=\; \left|  \gamma^\mathrm{A}_\mathrm{(\pm)} - \gamma^\mathrm{O}_\mathrm{(\pm)} \right|  \numberthis 
\end{align*}
where $ \gamma^\mathrm{O}_\mathrm{(\pm)}$ is the mean of the
overlapped distribution function at point ``$O$'' and $
\gamma^\mathrm{A}_\mathrm{(\pm)}$ is the mean of the electron-positron
distribution function at ``$A$''.

Let the beta value corresponding to the bulk velocity of both the
electrons and positrons at `$O$' be denoted by $\beta_\mathrm{o}$. For
any other point `$A$', let the beta factor corresponding to the bulk
velocity of the positrons and electrons be denoted by $\beta_{(+)}$
and $\beta_{(-)}$, then $\beta_\mathrm{(+)} - \beta_\mathrm{(-)} \;=\;
[ \beta_\mathrm{0} + \Delta \beta_\mathrm{(+)}] - [ \beta_\mathrm{0} -
  \Delta \beta_\mathrm{(-)}] $ is given by
\begin{align*}\label{Deltabeta}
\Delta \beta \;=\; \Delta \beta_\mathrm{(+)} + \Delta \beta_\mathrm{(-)} \numberthis
\end{align*}

Let us introduce the factor `$f_\mathrm{ratio}$' given by $ \Delta
\beta_\mathrm{(+)} = f_\mathrm{ratio} \;\Delta \beta_\mathrm{(-)}$ and
perform the following steps to get the separation ($2\Delta \gamma $)
at any distance $r_\mathrm{A}$. For a given
    mean lorentz factor $ \gamma^\mathrm{O}_\mathrm{(\pm)}$ at `$O$',
    and the density term($= 1/\kappa_\mathrm{GJ}$) and geometrical
    factor at ``$A$'', Eq. \ref{Deltabeta}
    is solved for $f_\mathrm{ratio}$ so as to
    satisfy the symmetry constraint given by Eq. \ref{Deltagamma}.
    Once $f_\mathrm{ratio}$ is obtained, the bulk
    velocity for the separated distribution functions can be estimated
    as $\beta_{(+)} = \beta_\mathrm{o} + \Delta \beta f_\mathrm{ratio}
    / \left( 1 + f_\mathrm{ratio} \right)$ and $\beta_{(-)} =
    \beta_\mathrm{o} - \Delta \beta / \left( 1 + f_\mathrm{ratio}
    \right)$. The bulk velocity so obtained are transformed to the
    mean lorentz factors $\gamma_{(+)}$ and $\gamma_{(-)}$ via the
    transformation $\gamma_{(\pm)} = 1 /\sqrt{1 - \beta^2_{\pm}}$. The
    corresponding momenta is given by $ p_{(\pm)} \approx
    \gamma_{{(\pm)}}$. 

In this case the EDF consists of the
    summation of the shifted gaussian distribution functions with mean
    $\gamma_{(+)}$ and $\gamma_{(-)}$.
    After getting the EDF we follow the steps
    outlined in Appendix D to solve the hydrodynamic Eq.
    \ref{Hydrodynamic equations}.

We proceed to solve growth rates for two surface magnetic field
configuration viz., a purely dipolar one and multi polar
field. For both field configurations we
    consider the last open field line for an aligned rotator and
    assume secondary plasma distribution function to be a gaussian (
    with mean $\mu = 250$, width $\sigma = 40$) at $r/R_\mathrm{NS} =
    1.02$.

\subsubsection{Simple Dipolar Geometry}\label{Simple Dipolar Geometry}

\begin{figure*}
\begin{tabular}{cc}
  \includegraphics[width=70mm]{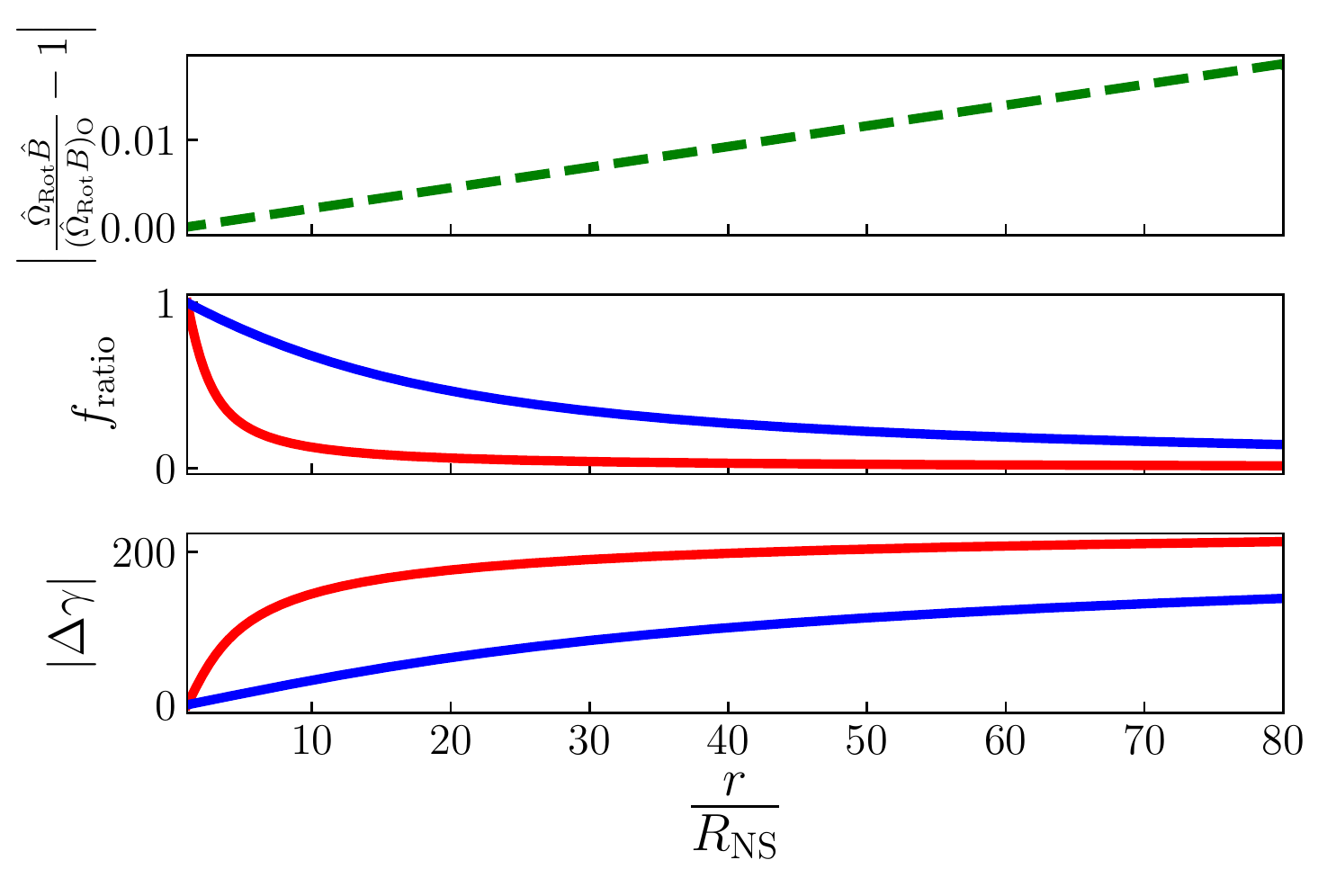} & \includegraphics[width=70mm]{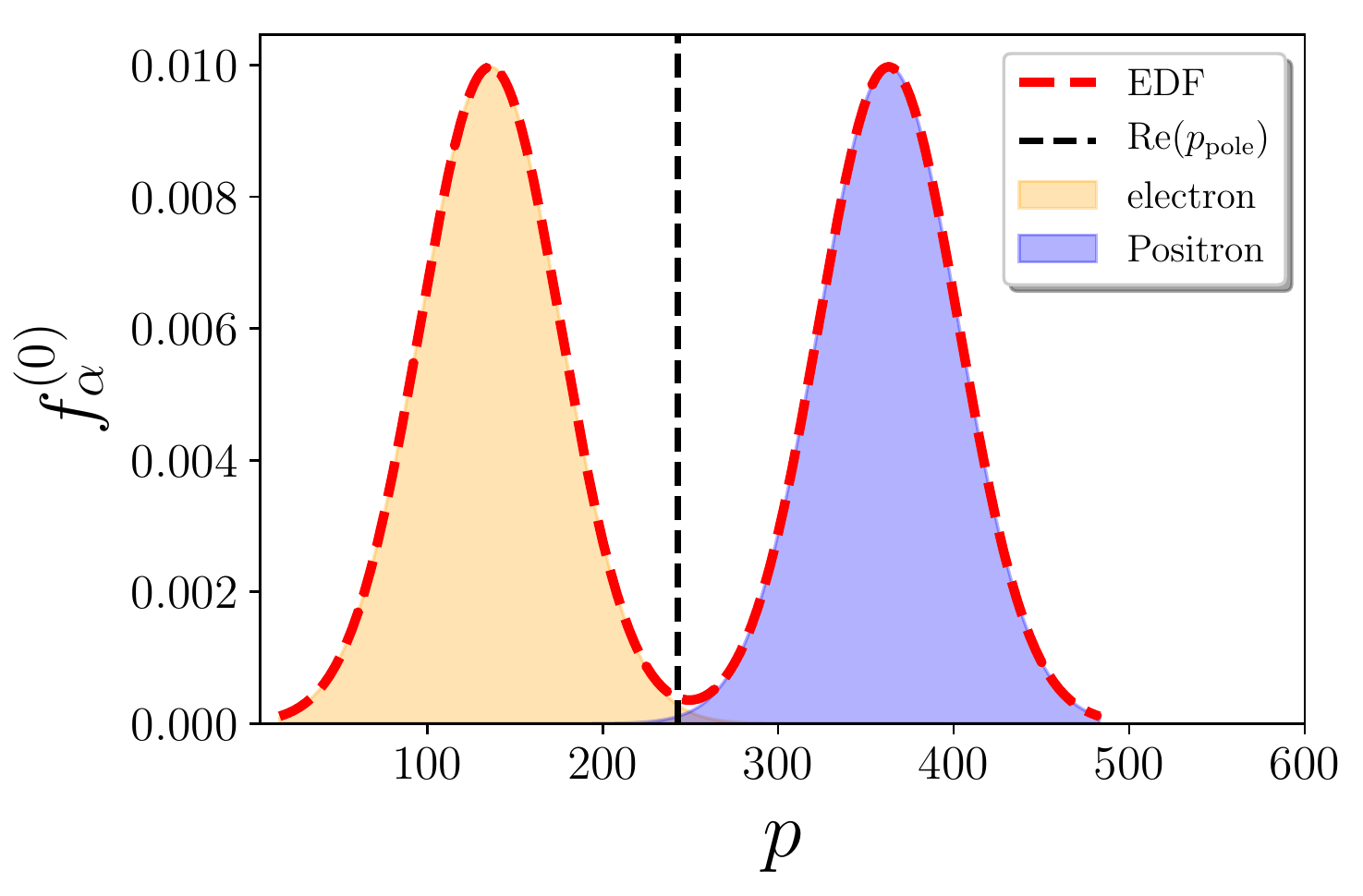} \\
(A) Splitting of the distribution function in dipolar geometry & (B) EDF at $r/R_\mathrm{NS} = $50 for  $\kappa_\mathrm{GJ} = 500 $.  \\[4pt]
 \includegraphics[width=70mm]{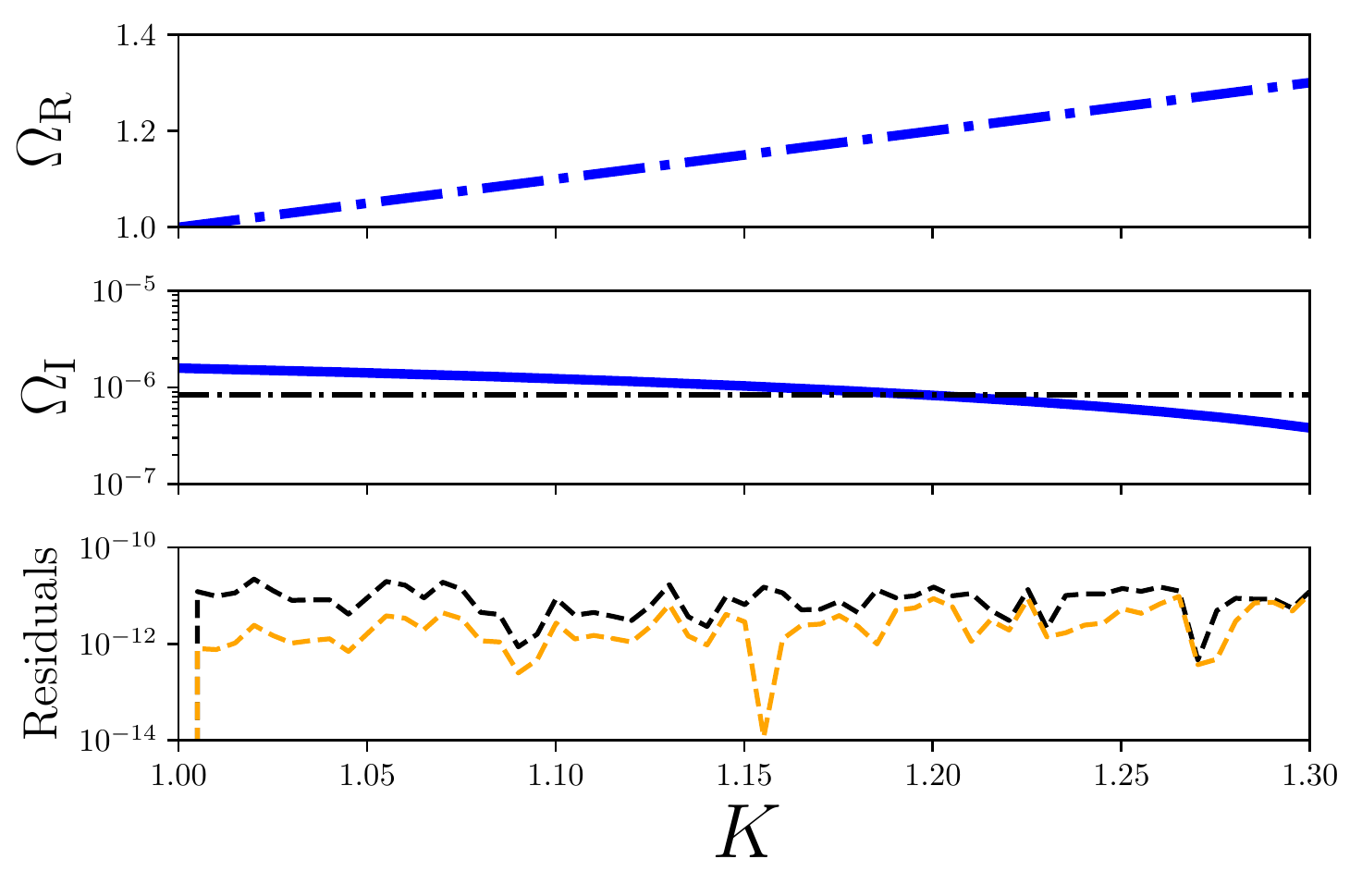} &   \includegraphics[width=70mm]{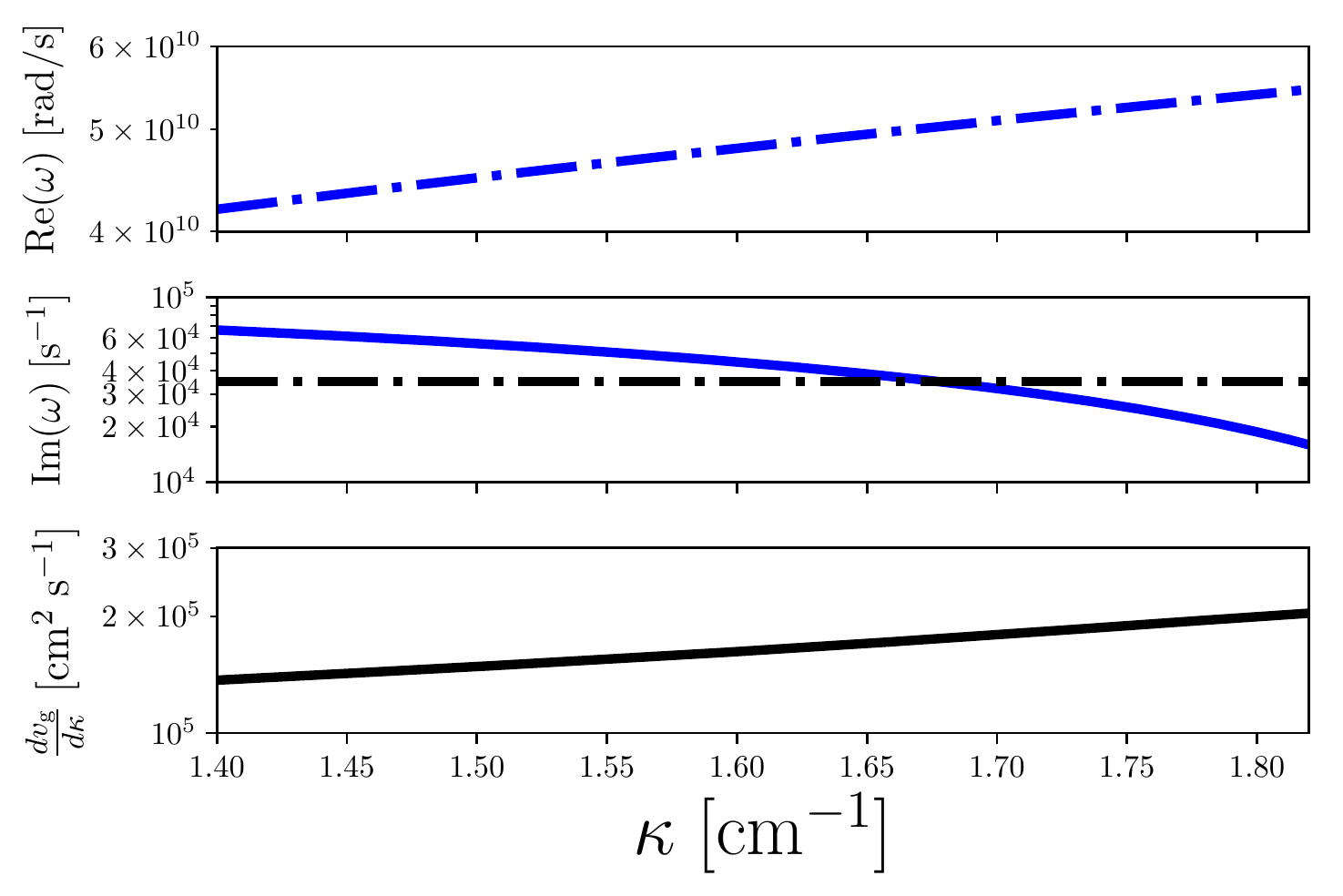} \\
(C) Dimensionless dispersion relation at $r/R_\mathrm{NS} = $ 50 for $\kappa_\mathrm{GJ} = 500$. & (D) Dimensional dispersion relation at $r/R_\mathrm{NS} = $ 50 for $\kappa_\mathrm{GJ} = 500$. \\[4pt]
 \includegraphics[width=70mm]{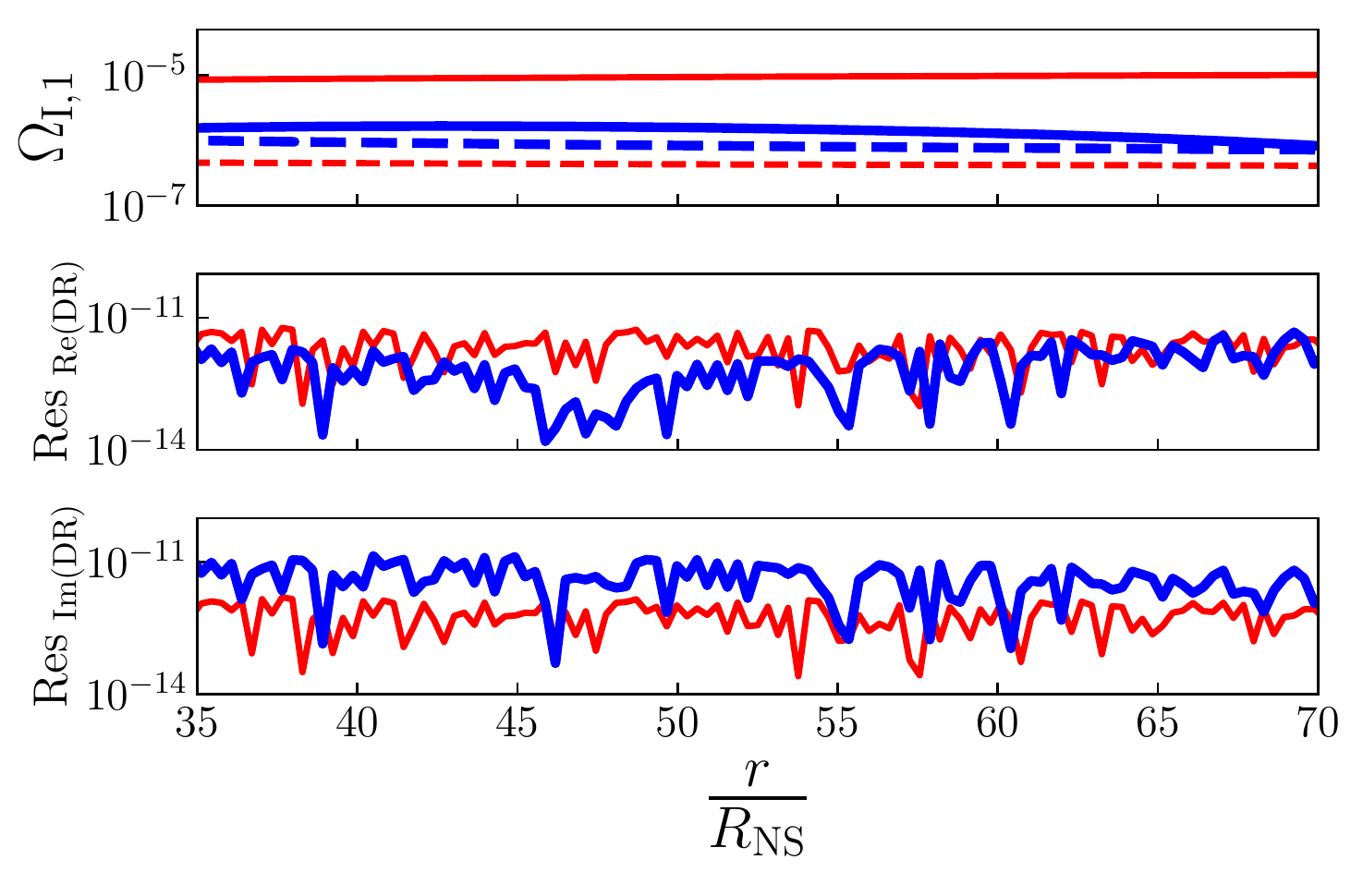} &   \includegraphics[width=70mm]{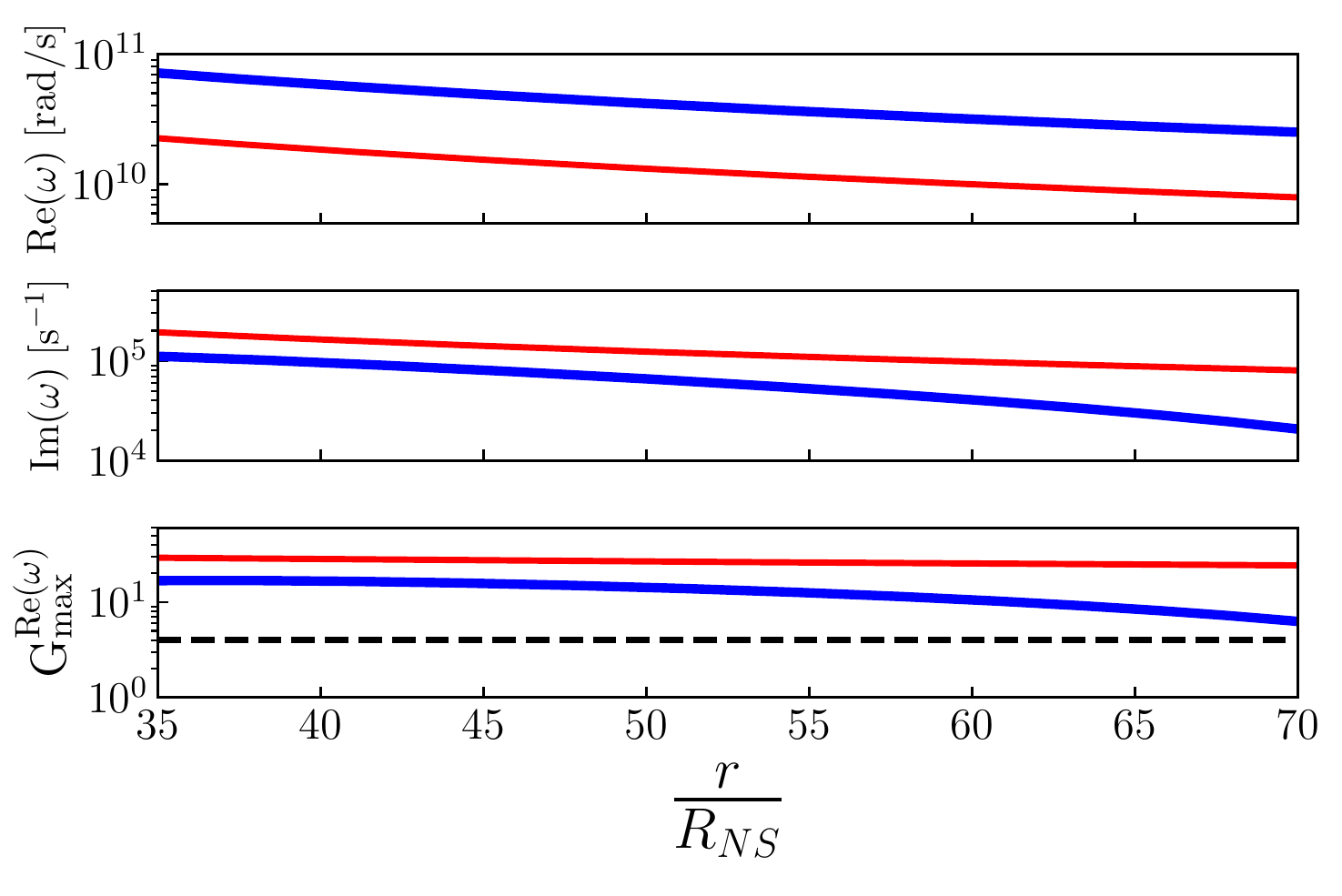} \\
(E) Dimensionless growth rate as a function of $r/R_\mathrm{NS}$.  & (F) Dimensional growth rate as a function of $r/R_\mathrm{NS}$. \\[4pt]

\end{tabular}
\caption{Plots for {\em C}2 longitudinal drift for a purely dipolar
geometry as discussed in sections \ref{Longitudinal drift} and
\ref{Simple Dipolar Geometry}.  Left top panel (A) has three subplots,
and from top to bottom shows the geometrical factor, the
$f_\mathrm{ratio}$ and $\Delta \gamma$ as a
function of $r/R_\mathrm{NS}$. The red and blue line corresponds to
$\kappa_\mathrm{GJ} = 50$ and $\kappa_\mathrm{GJ} = 500$
respectively. The top right panel (B) shows the EDF at
$r/R_\mathrm{NS} =50$. The vertical black dot-dashed line shows Re$(p_\mathrm{pole})$.  The middle left panel (C), top and
middle subplot shows the dimensionless real (in blue dash-dot line)
and imaginary parts (in solid blue line) of the dispersion relation as
a function of the dimensionless wavenumber $K$ corresponding to the
EDF shown in (B) and the black dash-dot line corresponds to the
analytical threshold given by Eq. \ref{separation of the regimes}. The
third subplot shows the residuals of the real and imaginary parts of
the dispersion relation by dashed black and orange line
respectively. The middle right panel (D) top and middle subplot is similar to
that of panel C and correspond to the dimensional dispersion relation for
$\kappa_\mathrm{GJ} = 500 $.  The bottom
subplot of (D) shows the group velocity dispersion as a function of
wavenumber $k$.  The lower left panel (E) top, middle, and bottom subplot shows the maximum dimensionless growth rate
and the residuals of the real and imaginary part of the dispersion
relation as a function of $r/R_\mathrm{NS}$. The red and blue lines
correspond to multiplicity factors $\kappa_\mathrm{GJ} = 50 $ 
and $\kappa_\mathrm{GJ} = 500 $, and the dashed red and blue lines refer to the threshold given by Eq. \ref{separation of the regimes}. 
The lower right panel (F), the top and middle subplot shows
the maximum growth rate for
$\kappa_\mathrm{GJ} = 50$ and $\kappa_\mathrm{GJ} = 500$ shown as
solid red and blue line respectively as a function of
$r/R_\mathrm{NS}$. The third subplot shows the maximum gain calculated
from Eq. \ref{Maximum gain } with the dashed black line showing the
threshold Eq. \ref{Threshold for non-linear regime}.}
\label{Longitudinal drift for dipolar geometry}
\end{figure*}

The results are shown in Fig. \ref{Longitudinal drift for dipolar
  geometry}. The top panel shows (A) the separation of the
distribution functions as a function of $r/R_\mathrm{NS}$ for two
multiplicity factors ($\kappa_\mathrm{GJ} \sim 50 \;,\; 500$) and (B)
shows the EDF for $\kappa_\mathrm{GJ} \;=\; 500$ at a distance of 500
km from the neutron star surface.  In the middle panel (C) the
solution of the dispersion relation along with the residuals in 
the dimensionless form is shown while (D) shows the dimensional growth
rate along with the group velocity dispersion at $r/R_\mathrm{NS} =
50$ for $\kappa_\mathrm{GJ} = 500$ . In the last panel (E) the
dimensionless and (F) the dimensional growth rate along with maximum
gain for an unstable wave of a given frequency is shown as a function
of $r/R_\mathrm{NS}$. It must be noted that the $\mathrm{Re} \;
(\omega)$ does not satisfy the constraint given in section(\ref{Need
  for a dense plasma}).

\subsubsection{Multi-polar Geometry}\label{Multipolar geometry}

\begin{figure}
\centering
\includegraphics[width = \columnwidth]{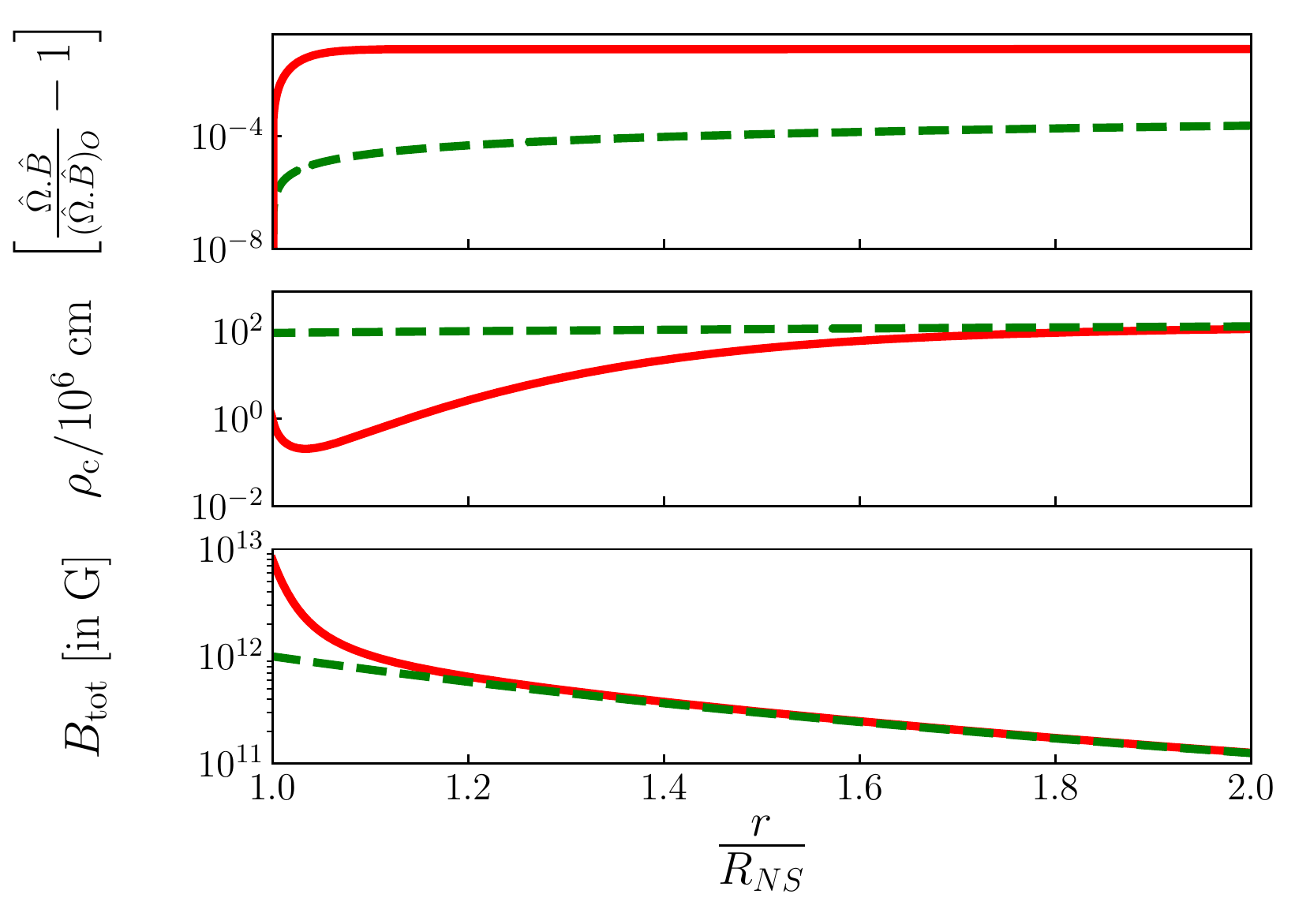}
\caption{Plot of the geometrical factor , radius of curvature and
  magnetic field strength as a function of $r/R_\mathrm{NS}$ in the
  first, second and third panel respectively, for the last open field
  line for a pulsar of period $P = 1$ seconds and global dipolar field
  of $B_\mathrm{d} = 10^{12}$ gauss. The local crust-anchored surface
  field has the parameters $ b = 10 \;,\; \theta_\mathrm{m} = - 0.01
  \; $radians$ \;, \theta_\mathrm{r} = 0.08 \; $radians$ $ such that
  $\left| m/d \right| = 0.0125 $. The solid red line and the dashed
  green line shows the variation of the aforementioned quantities for
  a multipolar configuration and a purely dipolar configuration
  respectively.}
\label{Comparison of dipolar and multipolar configuration}  
\end{figure}

{ \textbf{Simulating the multipolar field
    configuration:}} As discussed in section
    \ref{Evidence for multipolar field} any multipolar field
    configuration must satisfy the following two conditions for CCR :
    [i] At the radio emission heights the pulsar magnetic field must
    have a purely dipolar character ; [ii] The neutron surface must
    have a much smaller radius of curvature $\rho_\mathrm{c}$ compared
    to a purely dipolar field.  As a model for surface multipolar
    magnetic field we employ the prescription by
    \cite{2002A&A...388..235G}. In this model the magnetic field
    configuration is a superposition of two dipoles viz., a star
    centred global dipole with strength $B_\mathrm{d}$ and a
    crust-anchored local dipole embedded within $\Delta R \;=\; 0.05
    R_\mathrm{NS}$ from the surface with dipole strength $B_\mathrm{s}
    \;= \; b B_\mathrm{d} $. This local component is situated at the
    co-ordinates ($\theta_\mathrm{m}, \theta_\mathrm{r}$) with respect
    to the global dipole field (see fig.1 of
    \citealt{2002A&A...388..235G}). The strengths of the magnetic
    moments of the global dipolar field and the local crustal field is
    given by $ |\vec{d}| \;=\; 0.5 \; B_\mathrm{d} R^3_\mathrm{NS} $
    and $ |\vec{m}| \;=\; 0.5 \; B_\mathrm{m} (0.05R_\mathrm{NS})^3 $
    respectively. The boundary condition is chosen such that at the
    radio emission region $r/R_{NS} = 10$ the composite magnetic
    configuration should satisfy condition[i]. The middle panel of
    Fig. \ref{Comparison of dipolar and multipolar configuration}
    shows $\rho_\mathrm{c}$ as a function of $r/R_\mathrm{NS}$ for a
    purely dipole field ( shown in dashed green line ) and a composite
    configuration (shown as a solid red line) for certain model
    parameters are given in the caption. It can be seen that
    $\rho_\mathrm{c}$ due to the multi-polar configuration satisfies
    condition [ii] at the surface.  Both $\rho_{c}$ and the strength
    of the magnetic field $B_\mathrm{tot}$ shown in the middle and
    the lower panel of Fig. \ref{Comparison of dipolar and multipolar
      configuration} resembles that of a purely dipolar configuration
    within 10 km from the surface. The magnetic field strength for the
    multipolar configuration differs from that of a purely dipolar
    configuration by less than $0.8\%$ at $r/R_\mathrm{NS} \;=\;
    2$. This means that the superposed field configuration is
    insensitive to any change in the boundary condition beyond few
    tens of km from the surface. Since the multi-polar configuration
    has $\rho_\mathrm{c} \sim 10^{5}$ cm, we can justifiably use high
    $\kappa_\mathrm{GJ}$. It must be noted that for $r/R_\mathrm{NS}
    \geq 2$ both $\rho_\mathrm{c}$ and $B_\mathrm{tot}$ attain a
    purely dipolar character while the geometrical factor quickly
    attains a boosted steady value compared to a purely dipolar
    surface geometry as shown in the upper panel of
    Fig. \ref{Comparison of dipolar and multipolar
      configuration}. This simulated geometrical factor and high
    $\kappa_\mathrm{GJ}$ are then used as inputs for simulating the
    EDF.

The results of our analysis are shown in Fig.~\ref{Longitudinal drift
  for multipolar geometry} and the plot description are similar to Fig
\ref{Longitudinal drift for dipolar geometry} and the parameters for
the simulations are described in the caption to the figure. It is
important to note that in this case, unlike the dipolar example above,
$\mathrm{Re} \; (\omega)$ satisfy the constraint given in
section(\ref{Need for a dense plasma}).

\begin{figure*}
\begin{tabular}{cc}
  \includegraphics[width=70mm]{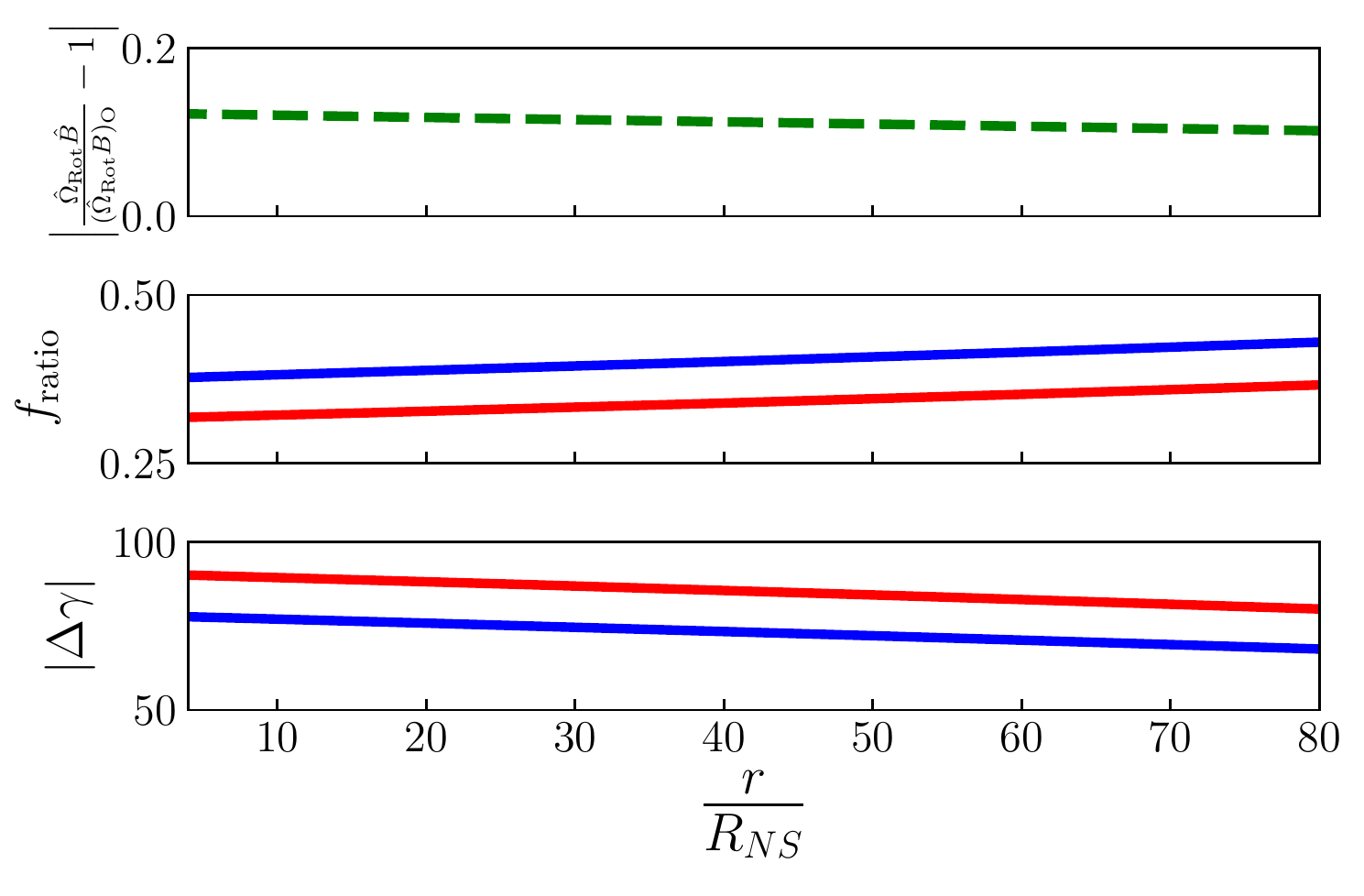} & \includegraphics[width=70mm]{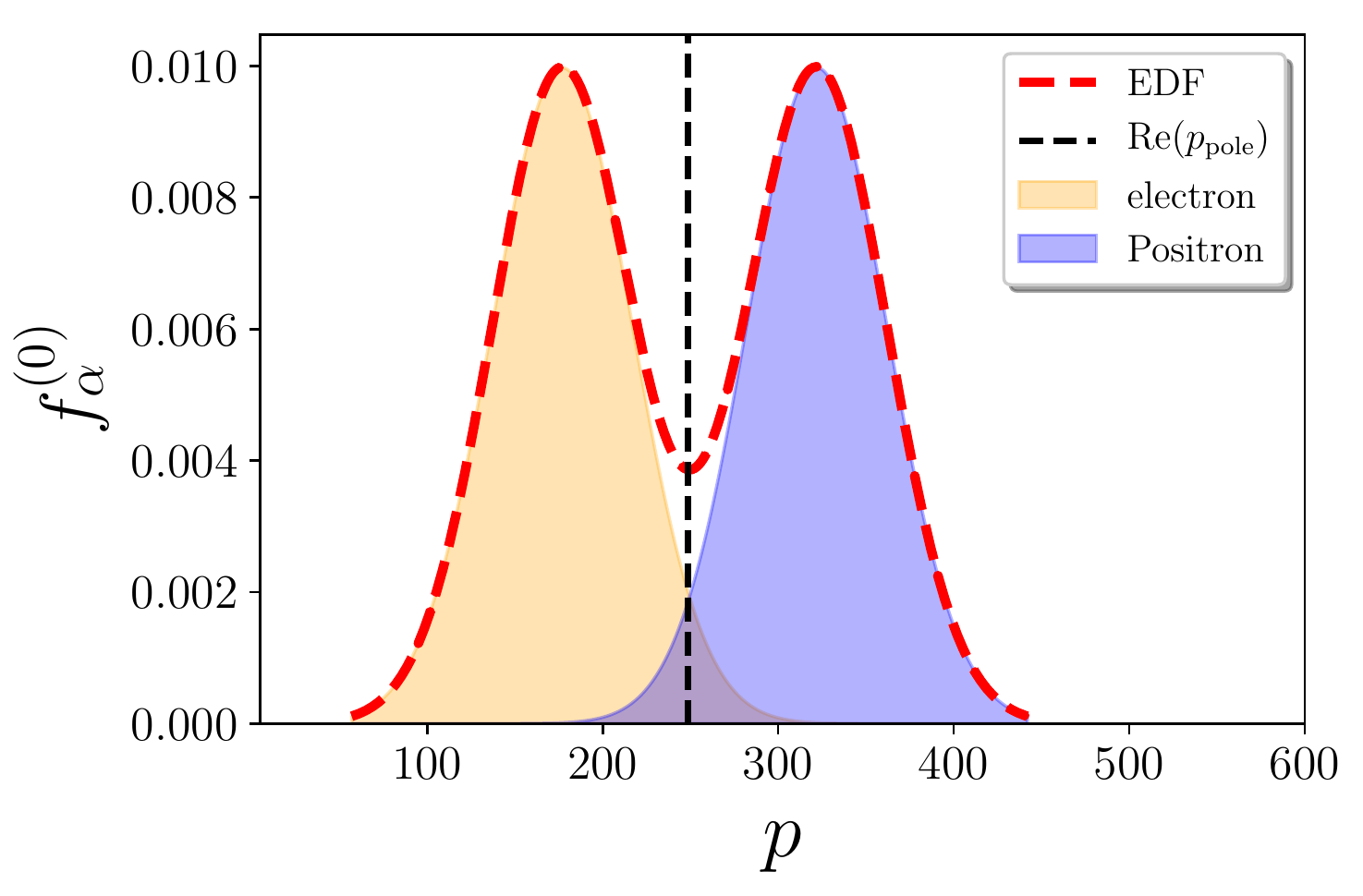} \\
(A) Splitting of the distribution function in multipolar geometry. & (B) EDF at $r/R_\mathrm{NS} = $ 50 for $\kappa_\mathrm{GJ} = 10^{4}$.  \\[4pt]
 \includegraphics[width=70mm]{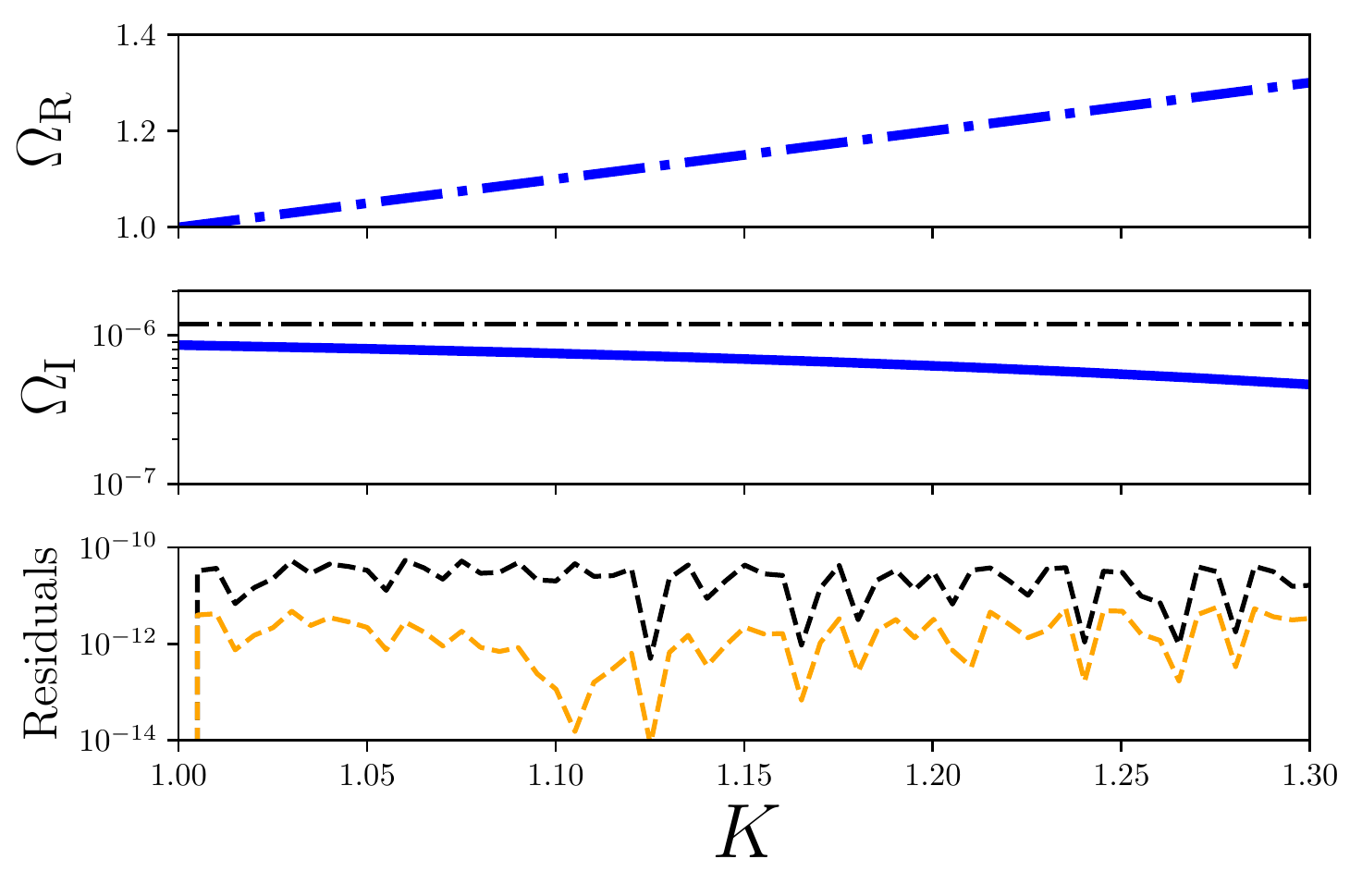} &   \includegraphics[width=70mm]{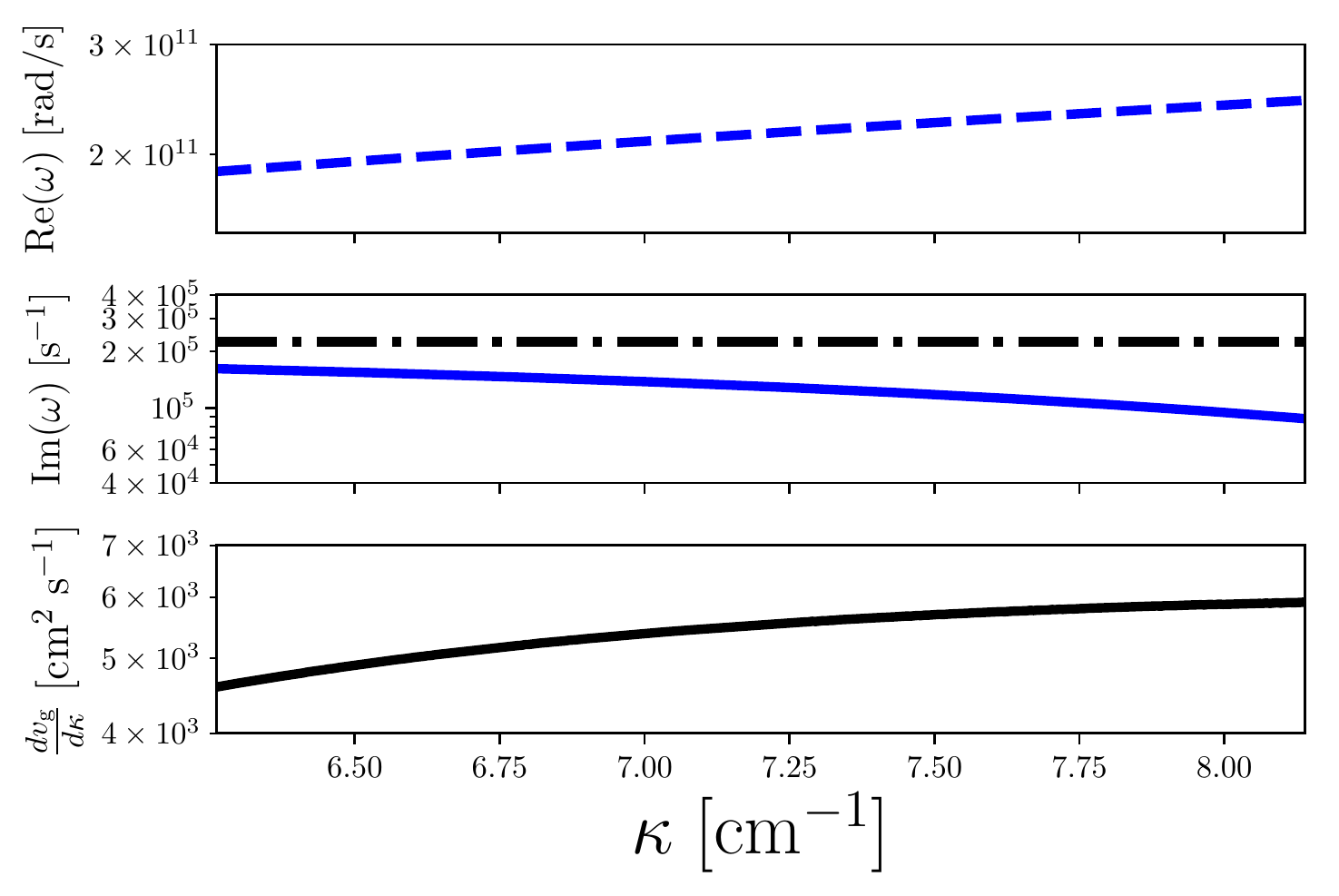} \\
(C) Dimensionless dispersion relation at 500 km for $\kappa_\mathrm{GJ} = 10^{4}$. & (D) Dimensional dispersion relation at $r/R_\mathrm{NS} = $ 50 for $\kappa_\mathrm{GJ} = 10^{4}$. \\[4pt]
 \includegraphics[width=70mm]{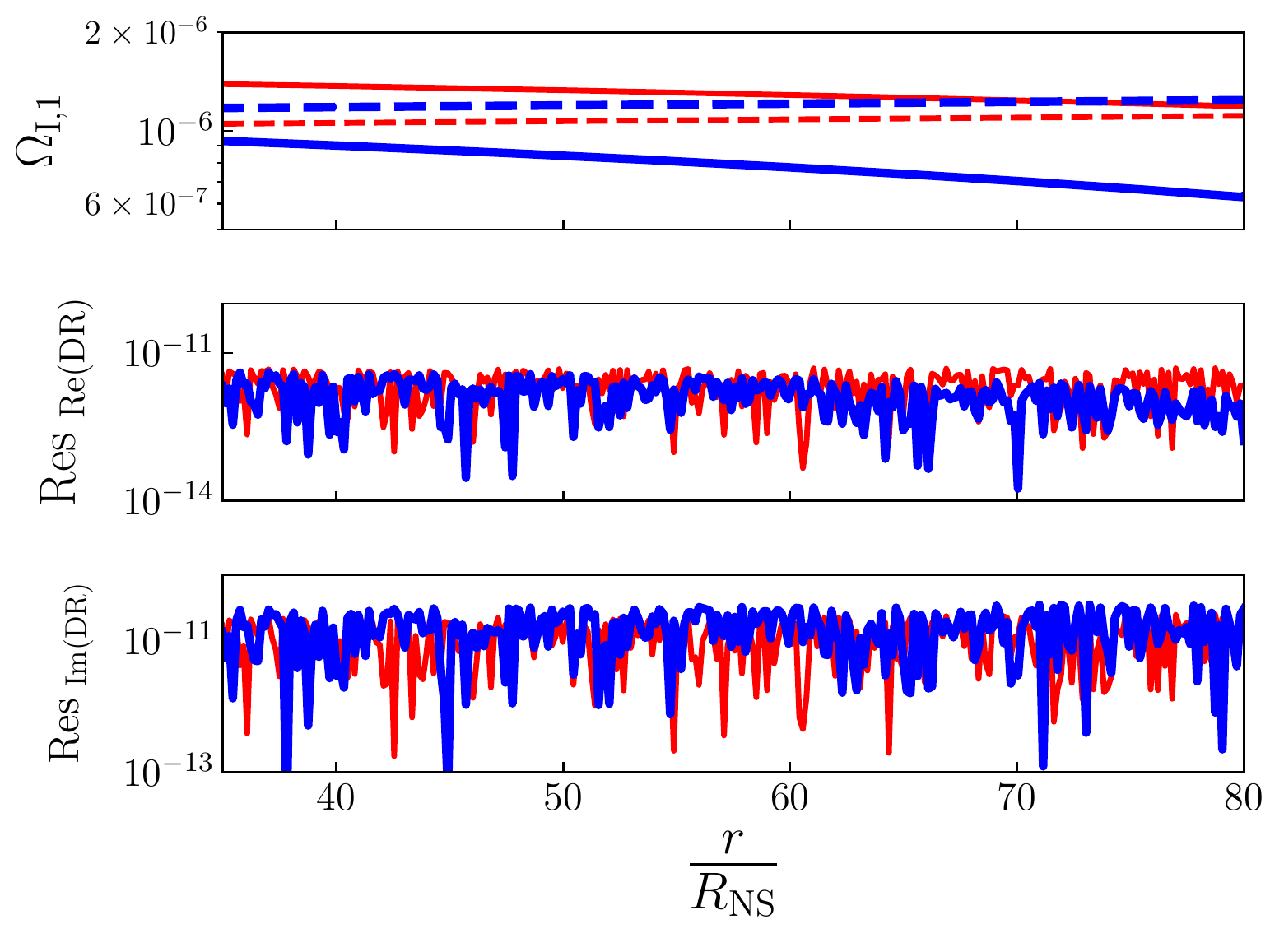} &   \includegraphics[width=70mm]{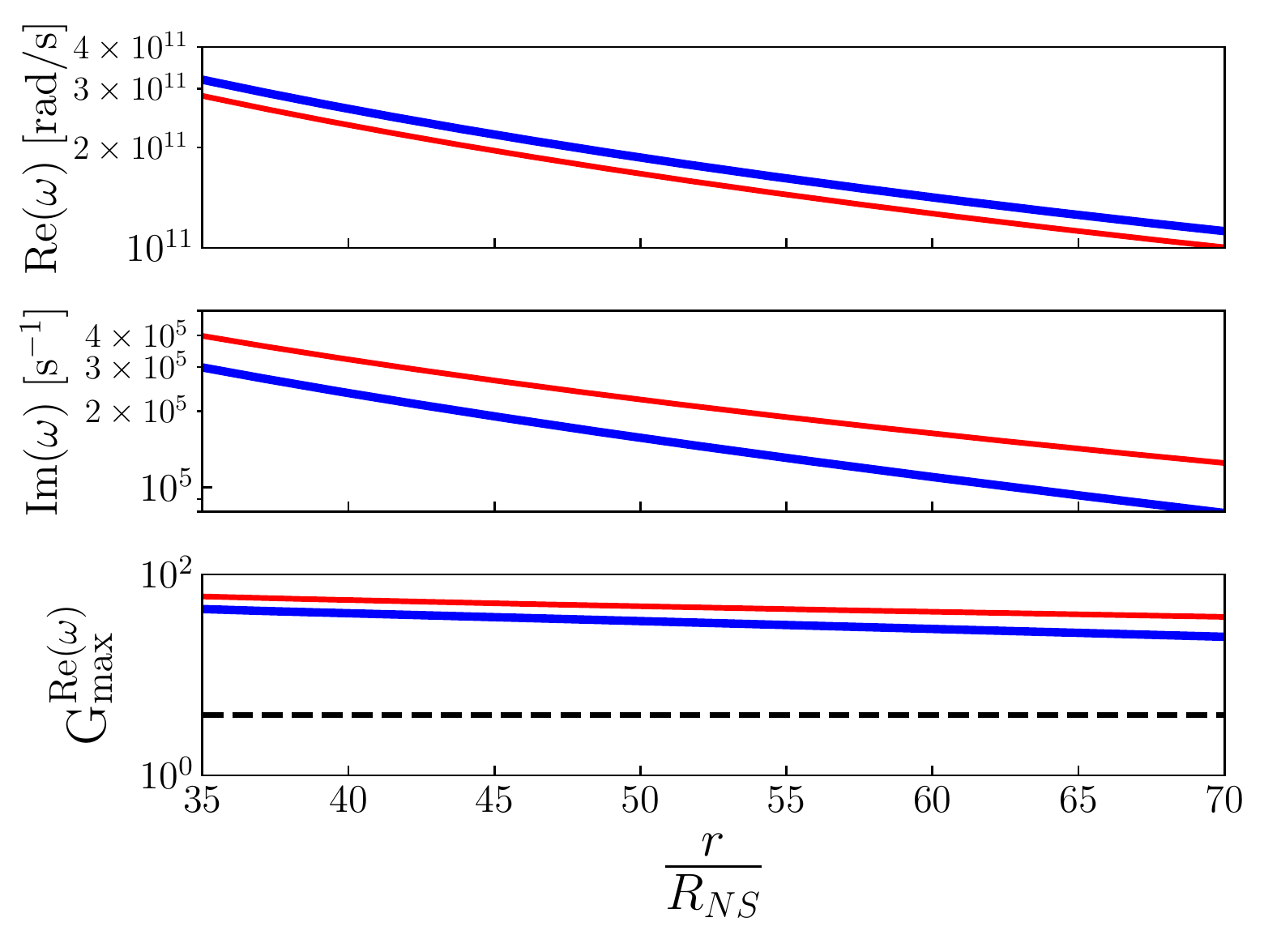} \\
(E) Dimensionless growth rate as a function of $r/R_\mathrm{NS}$.  & (F) Dimensional growth rate as a function of $r/R_\mathrm{NS}$. \\[4pt]
\end{tabular}
\caption{Plots for {\em C}2 along the last open field line
  for a  multi-polar field configuration parameters (see
  Fig. \ref{Comparison of dipolar and multipolar configuration} ) as discussed in section \ref{Multipolar geometry}. The plot description is the same as for Fig.\ref{Longitudinal drift for dipolar geometry} with the red and blue lines representing $\kappa_\mathrm{GJ} = $ 8$\times 10^{3}$ and 10$^{4}$ respectively.}
\label{Longitudinal drift for multipolar  geometry}
\end{figure*}

To summarize as seen in the bottom panel of figures \ref{Simple
  Dipolar Geometry} and \ref{Multipolar geometry} sufficient growth
rates exceeding the threshold limit Eq. \ref{Threshold for non-linear
  regime} can be obtained for both dipolar and multipolar field
configuration.

\subsection{{\em C}3: Growth due to cloud-cloud overlap}\label{Section on cloud-cloud overlap}

This model by \citet{1987ApJ...320..333U}  later developed by \citet{1988Ap&SS.140..325U} is based on the non-steady sparking discharge model (also referred to Inner acceleration gap model or the pure vacuum gap model) by RS75. In the
  RS75 model for pulsars with $\vec{\Omega}_\mathrm{Rot}
  \cdot \vec{B} < 0 $, positive charges are needed to screen
  the co-rotational electric field above the polar cap. However due to the high binding energy of the ions supply to positive charges are inhibited, and a vacuum gap with a strong electric field develops above the polar cap. The gap initially grows, however, once it reaches a height $h\approx 60-100$ m, it discharges via magnetic pair creation. Due to the strong electric field in the gap, the electrons are accelerated towards the stellar surface, while the positron streams relativistically away from the stellar surface. The upstreaming positron has sufficient energy to produce pair cascade,
  thus creating the secondary plasma cloud.  This process continues until the electric field in the gap is screened, and hence for the gap emptying time $h$ is a time $\tau = h/c \sim $ which is about a
  few hundreds of nanoseconds, the sparking process stops. Once the gap empties, the electric field grows and the sparking process starts again. Hence during steady-state, a non-stationary flow of secondary plasma cloud is generated, with each cloud having a spread in particle velocity.  In the original model of
  \citet{1988Ap&SS.140..325U} the overlap of the fastest and slowest
  particles of these successive secondary plasma clouds leads to
  two-stream instability.
  
AM98 extended the cloud-cloud overlap formalism of \citet{1988Ap&SS.140..325U} by categorizing the particles in each cloud of the secondary plasma into fast, slow and intermediate particles based on their speeds $v$. The authors
    presented an analytical expression for EDF in the overlapped
    region using $\Psi = x - vt $. The integral of motion $\Psi$ kept track of the position of these three categories of particles in each secondary plasma cloud. The distribution function for each
    cloud is given by F(p, $\Psi$) = F(p) F($\Psi$). The phase
    function F($\Psi$) modulates the shape of the distribution
    function F(p) as a function of $r/R_\mathrm{NS}$. In our scheme of constructing the EDF, We assume that F($\Psi$) can be ignored within a single secondary plasma cloud. We justify this assumption based on two considerations.  Firstly, in the hydrodynamic regime, the dip in the EDF containing Re (p$_\mathrm{pole}$) is of paramount importance. Since the hydrodynamic equations involve integration over the whole distribution functions, the modulated shape of the distribution functions is irrelevant. Secondly, the
    particles being ultrarelativistic, modulation due to F($\Psi$)
    will be very small. This is because the relative phase spread in a
    single cloud between the fastest and the slowest particles
    compared to the average velocity particles is very small $ (
    \gamma^2_\mathrm{fastest} - \gamma^2_\mathrm{slowest} )/
    \gamma^2_\mathrm{fastest} \gamma^2_\mathrm{slowest} \ll 1$. In this work, the shape of the gaussian distribution function remains unaltered at any $r/R_\mathrm{NS}$. Below we present a more
    generic way to construct EDF numerically.

{\textbf{Simulating EDF for {\em C}3 } : }
Let the gap closing timescale be `$\tau$'. The time required
  to form a single cloud is `$T= 30 \tau $' such that a fully formed
  spark corresponds to a cloud of electrons and positrons of length
  $L_\mathrm{s} = cT$. Let us consider two successive discharges
  giving rise to a leading cloud (labelled by index `1') and a
  trailing cloud (labelled by index `2'). Let the distribution
  function of each secondary plasma have a maximum and minimum
  dimensionless momenta cut-offs characterized by $p_\mathrm{max}$ and
  $p_\mathrm{min}$ respectively. The velocity corresponding to any $p$
  in the distribution function is given by the transformation $v = pc/
  \sqrt{1 + p^2}$ such that the corresponding cut-off velocities are
  given by $v_\mathrm{max}$ and $v_\mathrm{min}$ respectively. Let the
  particles with arbitrary velocity in cloud 1 and cloud 2 be labelled
  by $v^\mathrm{(1)}_\mathrm{arb}$ and $v^\mathrm{(2)}_\mathrm{arb}$
  respectively. We define the overlap region between the position of
  $v^\mathrm{(2)}_\mathrm{max}$ and the position of
  $v^\mathrm{(1)}_\mathrm{min}$ and give a description for the
  construction of the EDF below.\par

The time $t$ in which  $v^\mathrm{(2)}_\mathrm{max}$  overlaps with $v^\mathrm{(1)}_\mathrm{arb}$ is
\begin{align*}
 t_\mathrm{Ov} =
\frac{v^\mathrm{(1)}_\mathrm{arb}\tau -\left[ v^\mathrm{(2)}_\mathrm{max} -
    v^\mathrm{(1)}_\mathrm{arb}\right] \;T}
  {v^\mathrm{(2)}_\mathrm{max} -
    v^\mathrm{(1)}_\mathrm{arb} } 
\end{align*} for $ v^\mathrm{(1)}_\mathrm{upper} = v^\mathrm{(1)}_\mathrm{arb} \geq v^\mathrm{(1)}_\mathrm{min}$.

The position of overlap is given by
\begin{align*}
x_\mathrm{Ov} = \frac{v^\mathrm{(2)}_\mathrm{max} \;
  v^\mathrm{(1)}_\mathrm{arb}\tau } 
{v^\mathrm{(2)}_\mathrm{max} - v^\mathrm{(1)}_\mathrm{arb} } 
\end{align*} which can be represented as a function of $r/R_\mathrm{NS}$.

 The position of the minimum velocity particles of cloud
`1' at time $t_\mathrm{Ov}$ is given by 
\begin{align*}
x^\mathrm{(1)}_\mathrm{min} = \frac{v^\mathrm{(1)}_\mathrm{min} \;
  v^\mathrm{(2)}_\mathrm{max}\tau } 
{v^\mathrm{(2)}_\mathrm{max} - v^\mathrm{(1)}_\mathrm{min} }
   =  v^\mathrm{(2)}_\mathrm{lower}(T + t_\mathrm{Ov})  
\end{align*}The equality is used to solve for $ v^\mathrm{(2)}_\mathrm{lower}$ and the solution transformed to dimensionless momenta $ p^\mathrm{(2)}_\mathrm{lower}$ via the transformation $p = \beta/ \sqrt{1 - \beta^2}$.   \par 

The EDF $f^\mathrm{Ov}$ in the overlapped region is given by 
\begin{align*}
f^\mathrm{Ov} = f_{1} \; [p^\mathrm{(1)}_\mathrm{min}:p^\mathrm{(1)}_\mathrm{upper}] +   f_{2} \; [p^\mathrm{(2)}_\mathrm{lower}:p^\mathrm{(2)}_\mathrm{max}] 
\end{align*} where the notation $f_\mathrm{n}[a:b]$ refers to the portion of distribution function $f_\mathrm{n}$ from $a$ to $b$ for cloud with index `n'.

The spatial extent of the overlapped region is given by
\begin{align*}
\Delta_\mathrm{Ov}= x^\mathrm{(2)}_\mathrm{max} -
x^\mathrm{(1)}_\mathrm{min} = \frac{\left[
    v^\mathrm{(2)}_\mathrm{max}\;v^\mathrm{(1)}_\mathrm{min} -
    v^\mathrm{(1)}_\mathrm{min}
    v^\mathrm{(2)}_\mathrm{max}\right]\tau} 
  {v^\mathrm{(2)}_\mathrm{max} - v^\mathrm{(1)}_\mathrm{min}} 
\end{align*}  
  
The dispersion relation in the overlapped region is given
by the expression
\begin{align*}
 1-  \chi  \int_{-\infty}^{+\infty} dp \; \frac{ f^\mathrm{\Delta_{Ov}}}{\gamma^3} \;\frac{1}{\left(\Omega - \beta K\right)^2} = 0 \numberthis
\end{align*}

\begin{figure*}
\begin{tabular}{cc}
  \includegraphics[width=70mm]{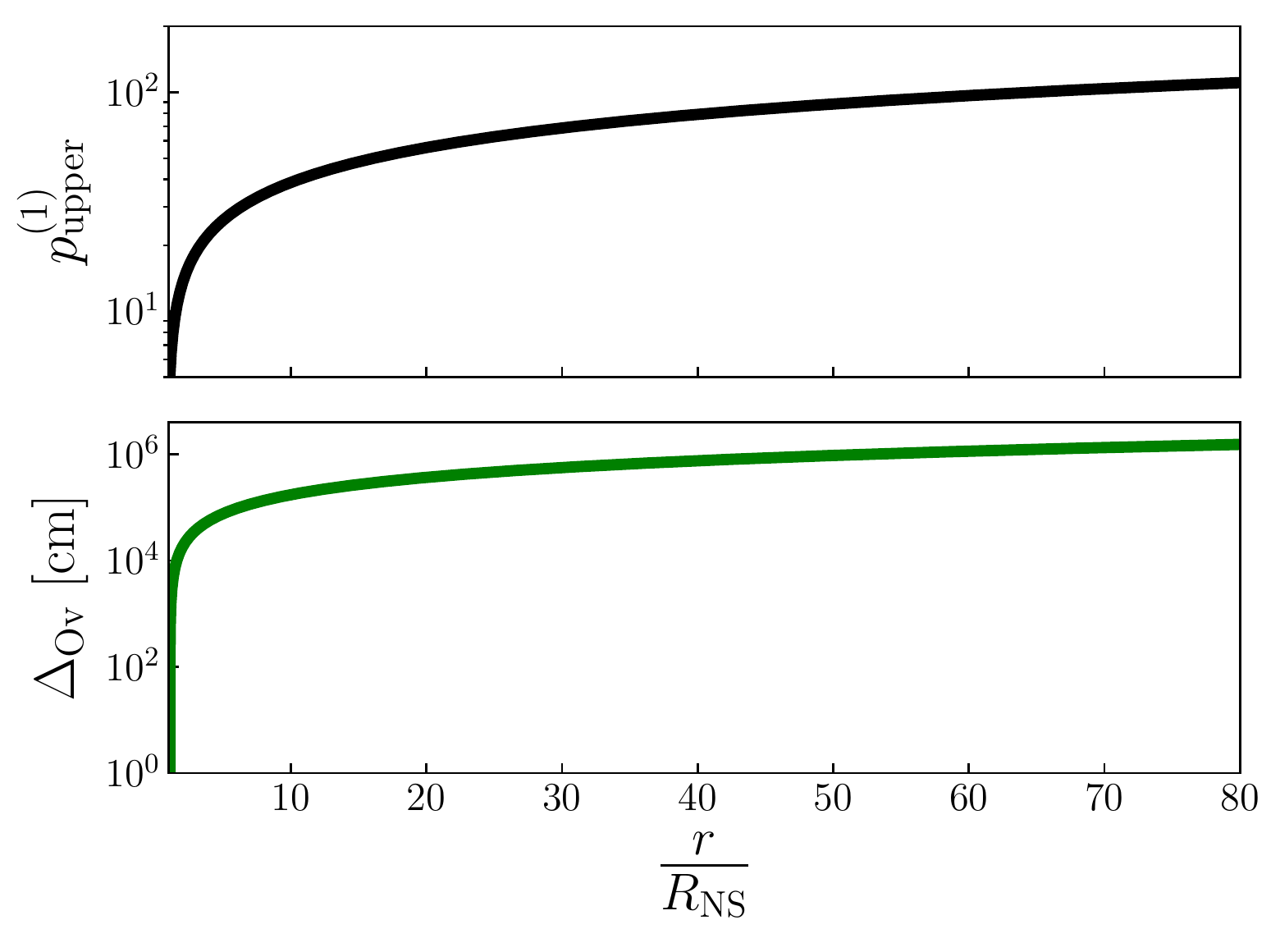} & \includegraphics[width=70mm]{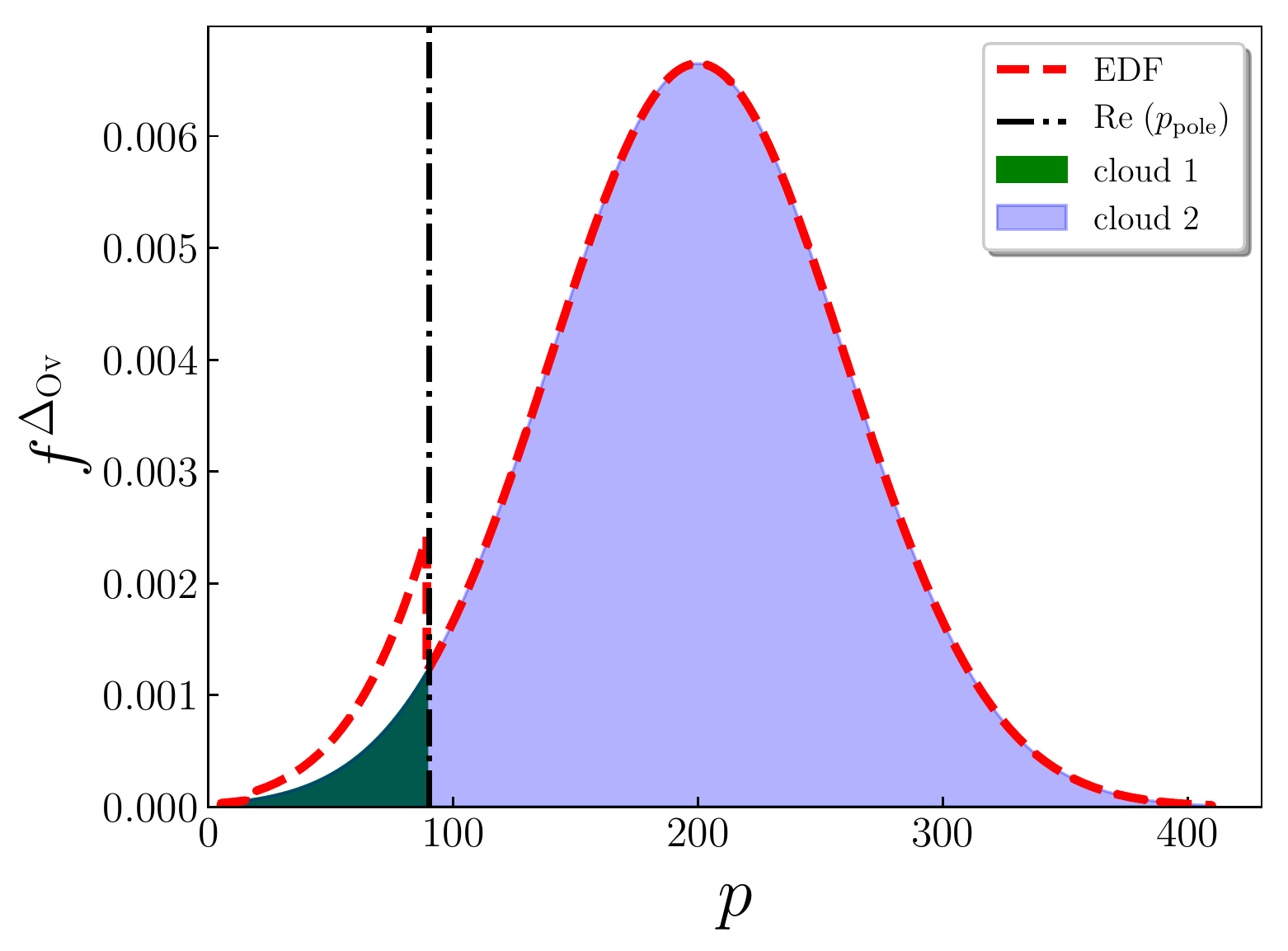} \\
(A) Particle ovelap and extent of overlap region as a function of $r/R_\mathrm{NS}$.  & (B) EDF at $r/R_\mathrm{NS} = $ 50   \\[4pt]
 \includegraphics[width=70mm]{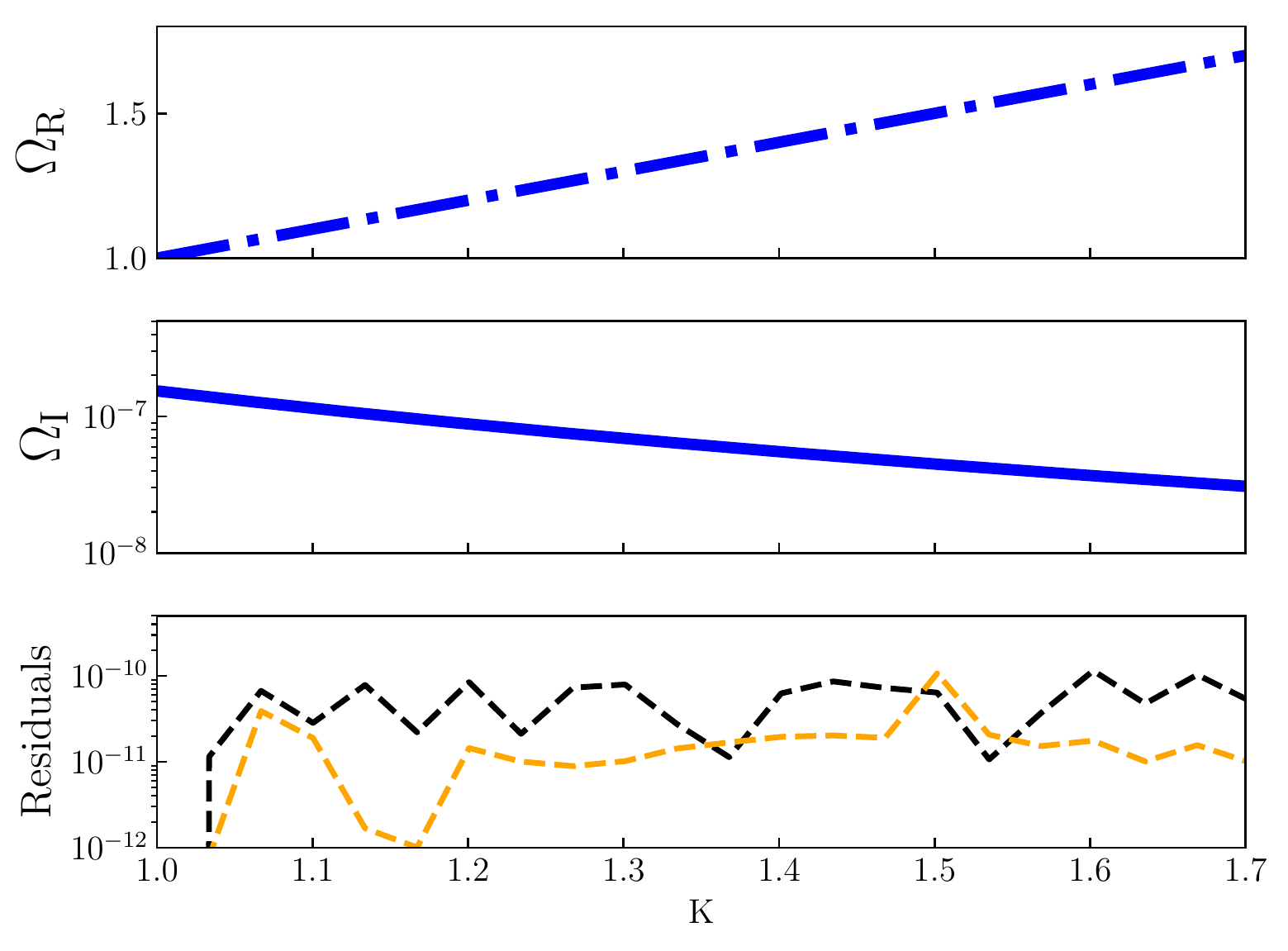} & \includegraphics[width=70mm]{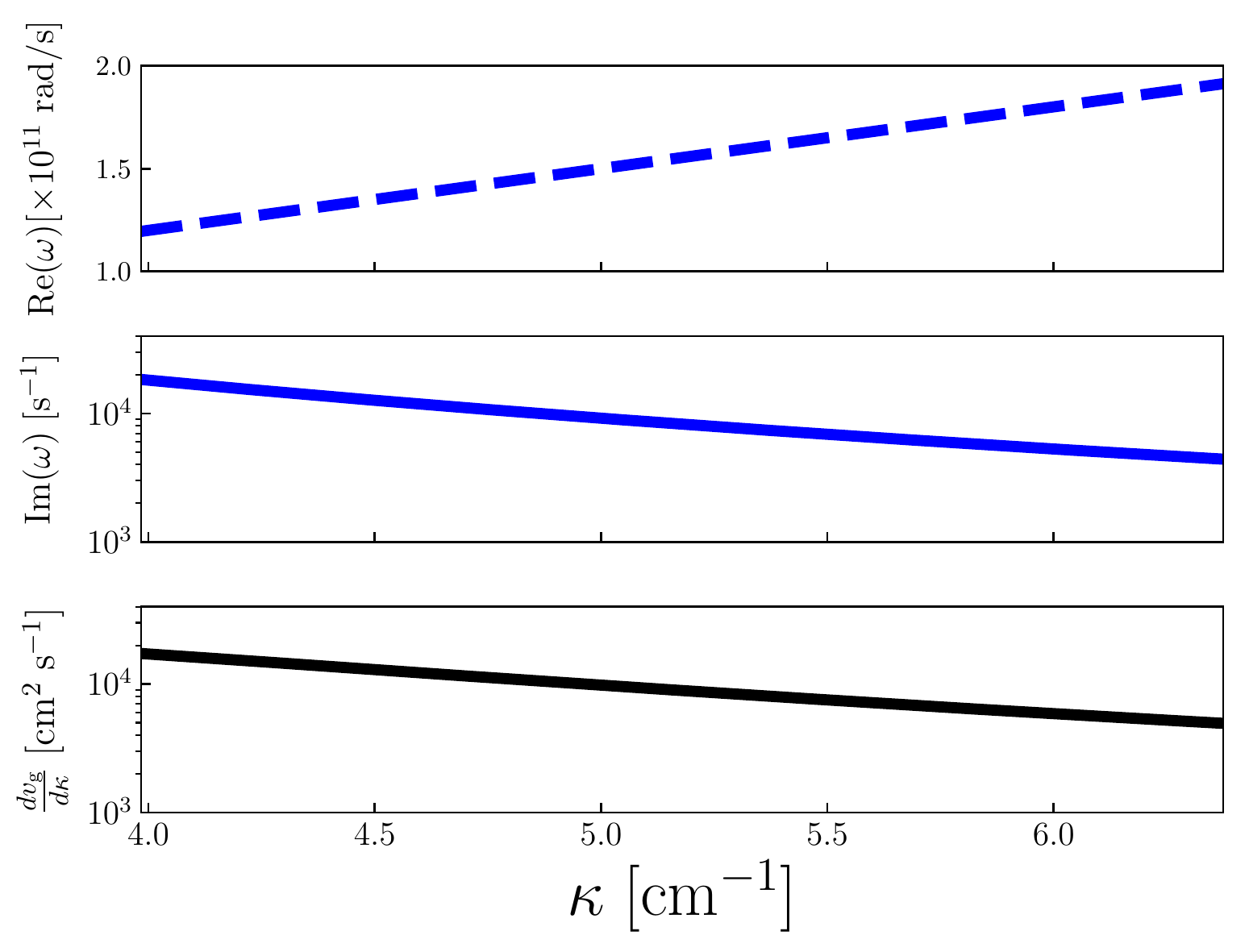} \\
(C) Dimensionless dispersion relation at $r/R_\mathrm{NS} = $ 50.   & (D) Dimensional dispersion relation at $r/R_\mathrm{NS} = $ 50 for $\kappa_\mathrm{GJ} = 10^4$.   \\[4pt]
\includegraphics[width=70mm]{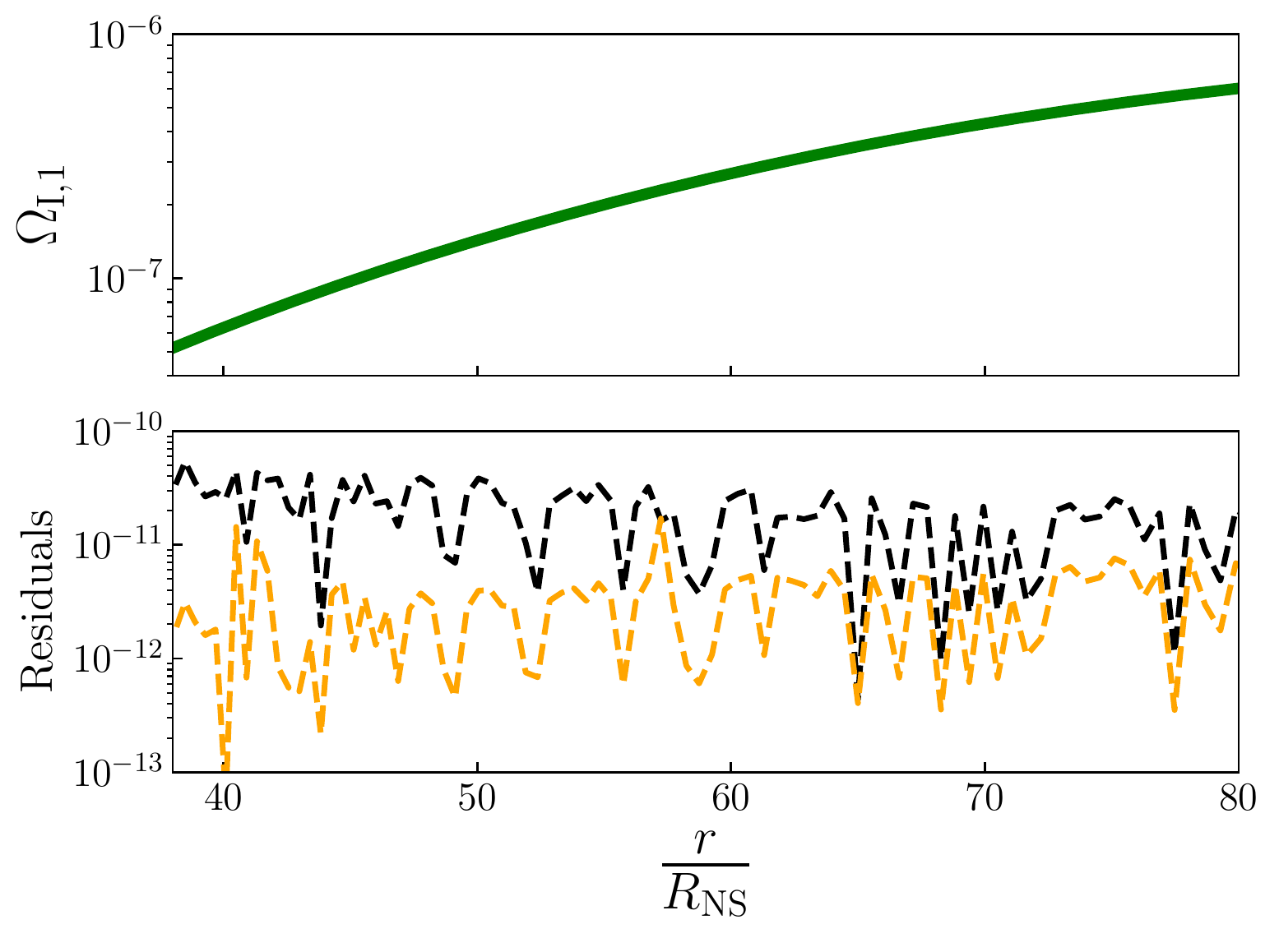} & \includegraphics[width=70mm]{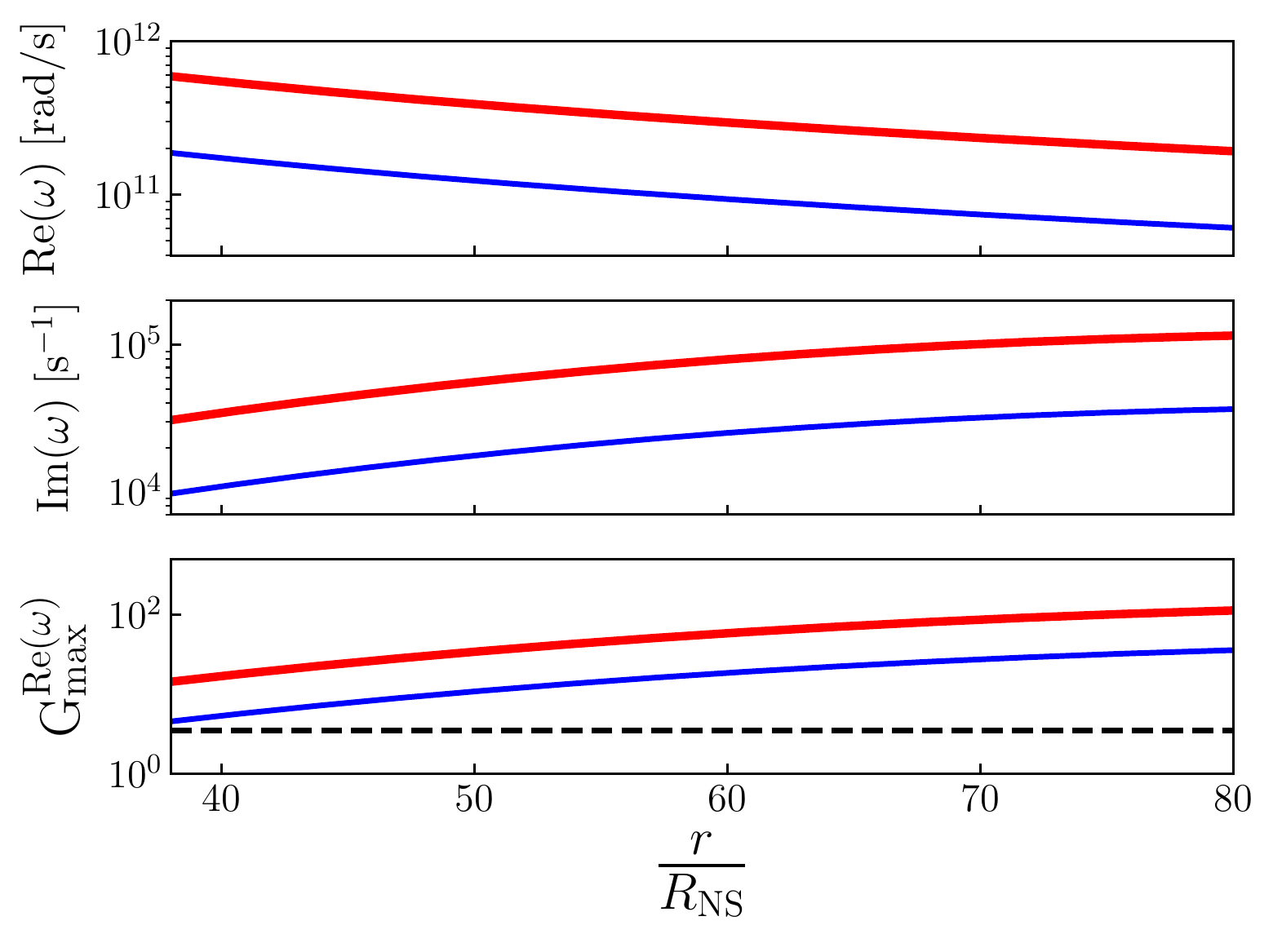} \\
(E) Dimensionless growth rate as a function of $r/R_\mathrm{NS}$.   & (F) Dimensional growth rate as a function of $r/R_\mathrm{NS}$.   \\[4pt]
\end{tabular}
\caption{Plots for {\em C}3 for $\tau = 100$
  nanoseconds and $ T = 30 \;\tau$ as discussed in section \ref{cloud-cloud overlap}. In the top left panel (A)
  the upper and the lower subplot shows the momenta of the particles in cloud 1 to be overtaken by the maximum velocity particles in cloud 2 and the extent of the overlapped region as a
  function of $r/R_\mathrm{NS}$. The top right panel (B) shows the EDF at $r \;=\; 50 \; R_\mathrm{NS}$ . The vertical black dot-dashed line shows Re$(p_\mathrm{pole})$. In the middle left panel (C) the first and second subplot shows the dimensionless real(in dot-dashed blue line) and imaginary parts(in solid blue line) of the dispersion relation as a function of the dimensionless wavenumber $K$   for the EDF shown in (B). The third subplot shows the residuals of the real (in dashed black line) and imaginary(in dashed orange line) parts of the dispersion relation as a function of the dimensionless wavenumber $K$. In
  the middle right panel (D) the first and second subplots show the corresponding dimensional dispersion relation to (C) for $\kappa_\mathrm{GJ} \;=\;
  10^{4} $ . The third subplot shows the group velocity dispersion as a
  function of wavenumber k. In the lower left panel (E) the first
  subplot shows the maximum dimensionless growth rate while the second subplot shows
  the residuals of the real(in dashed black line) and the imaginary part(in orange line) of the dispersion
  relation as a function of
  $r/R_\mathrm{NS}$. In the lower right panel (F) the first and the
  second subplot shows the solution of the dispersion relation for
  the maximum growth rate for $\kappa_\mathrm{GJ} = 10^4$ and
  $\kappa_\mathrm{GJ} = 10^{5}$ shown as blue and red solid line
  respectively as a function of $r/R_\mathrm{NS}$ . The third subplot shows the maximum gain calculated from Eq. \ref{Maximum
      gain } with the dashed black line showing the threshold Eq.
  \ref{Threshold for non-linear regime}}
\label{cloud-cloud overlap}
\end{figure*}

In this case, the EDF is being determined by
    the lower and higher momenta cut-off $p_\mathrm{min}$ and
    $p_\mathrm{max}$ of the secondary plasma distribution function and
    the gap closing time $\tau$. Here we assume a gaussian distribution function with $\mu = 200, \sigma
  = 60 , p_\mathrm{min} = 5, p_\mathrm{max}= 400$ for the secondary plasma clouds.After getting the EDF we follow the steps outlined in Appendix D to solve the hydrodynamic equations. The results of our numerical solution are shown in
Fig. \ref{cloud-cloud overlap}. As seen from the third subplot of (F)
sufficient growth rates can be obtained exceeding the gain threshold
defined in section \ref{Criterion for breakdown }.

\subsubsection{Effect due to longitudinal drift } \label{cloud-cloud aided by LDM}
\begin{figure*}[H]
\begin{tabular}{cc}
  \includegraphics[width=70mm]{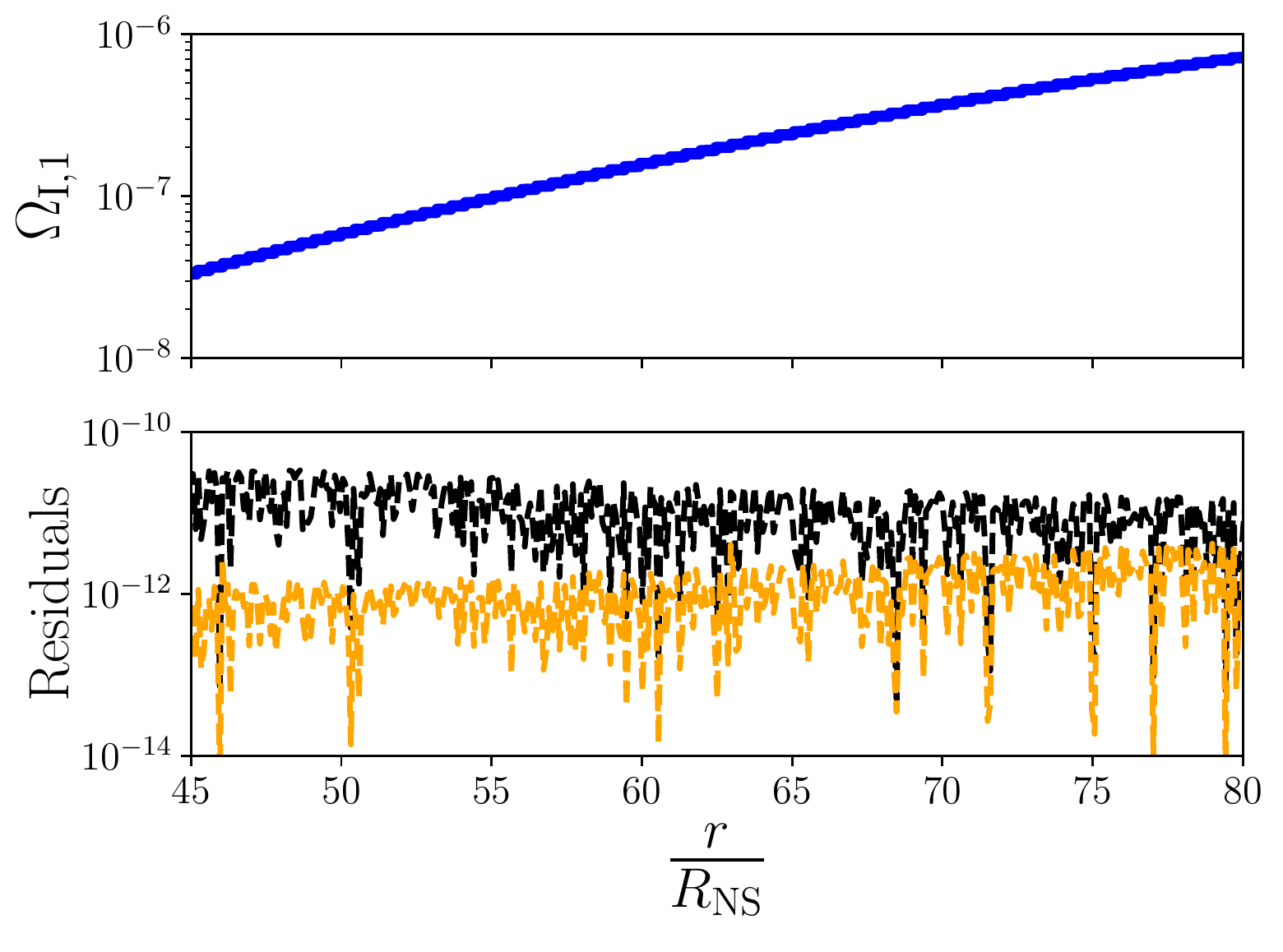} & \includegraphics[width=70mm]{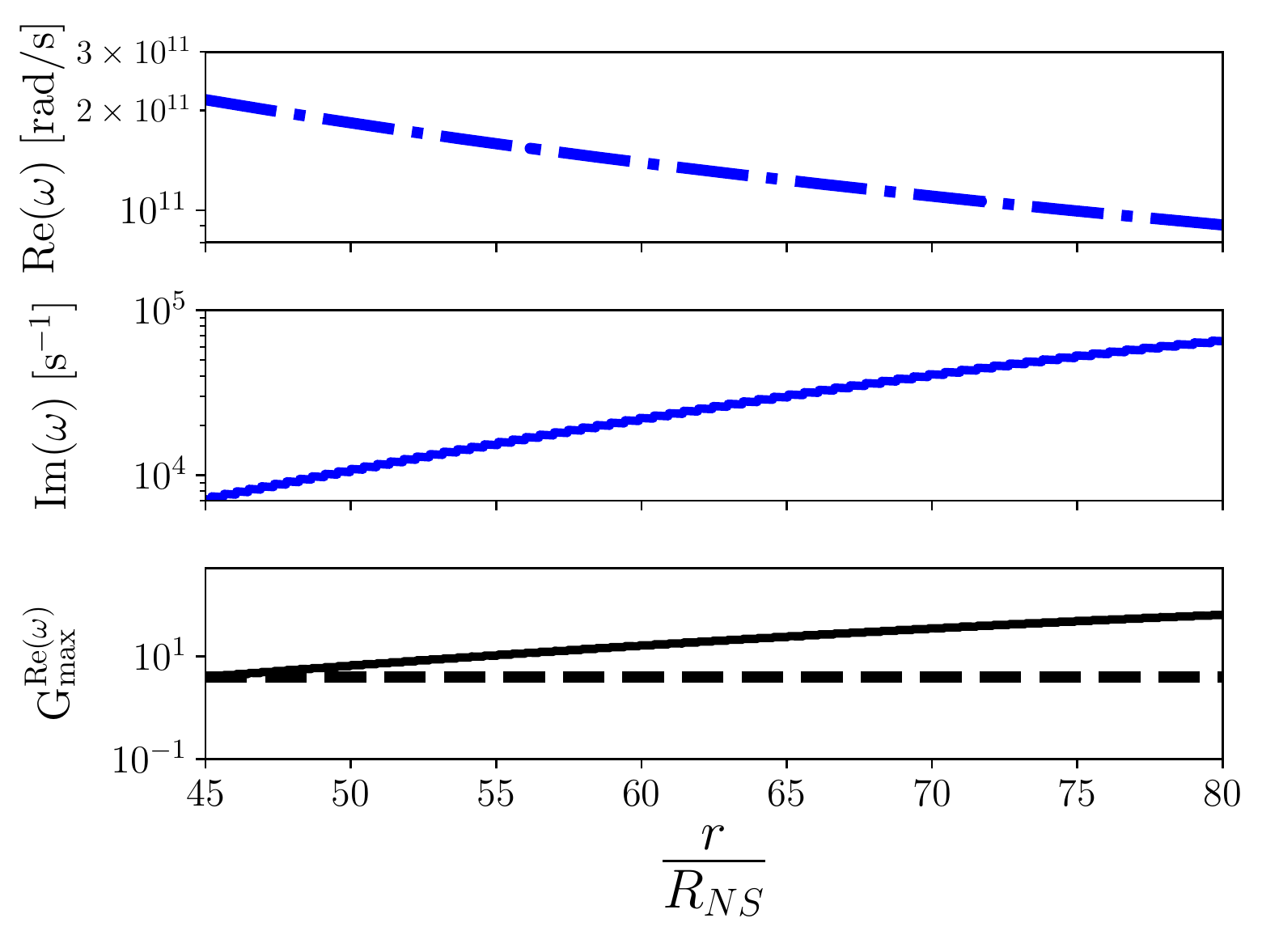} \\
(A) Dimensionless growth rate as a function of $r/R_\mathrm{NS}$.  & (B) Dimensional growth rate as a function of $r/R_\mathrm{NS}$. \\[4pt]
\end{tabular}
\caption{Plots for {\em C}3 aided by {\em C}2  along the last open field line for 
  multi-polar configuration as used in Fig. \ref{Comparison of dipolar
    and multipolar configuration} and discussed in section \ref{cloud-cloud aided by LDM} . Here  $\tau = 100$ nanoseconds and $ T
  = 30 \;\tau$. A gaussian with $\mu = 240, \; \sigma = 50 , \;
  p_\mathrm{min} = 25, \; p_\mathrm{max}= 400 $ has been assumed for
  the distribution function for the secondary plasma cloud at $r =
  1.02 \; R_\mathrm{NS}$. Multiplicity factor of $ \kappa_\mathrm{GJ}
  = 2 \times 10^{4}$ has been used. For left panel (A) the upper subplot shows the
  maximum dimensionless growth rate as a function of
  $r/R_\mathrm{NS}$. The lower
  subplot shows the residuals of the real(in dashed black line) and imaginary(in dashed orange line)
  parts of the dispersion relation respectively as a function of
  $r/R_\mathrm{NS}$. For the right panel (B) the upper and the middle subplot shows
  real and imaginary part of the dispersion relation in its
  dimensional form as a function of $r/R_\mathrm{NS}$. The lower subplot shows the
  maximum gain obtained using Eq. \ref{Maximum gain } as a function of
  $r/R_\mathrm{NS}$. The dashed black line in the second subplot of
  (B) represents the threshold Eq.\ref{Threshold for
      non-linear regime} }
\label{cloud-cloud overlap helped by Longitudinal drift}
\end{figure*}

   In the previous numerical simulation we have taken
    the value of $p_\mathrm{min} = 5$. However, if this value were to
    be higher the contribution to the EDF at a given height due to the
    leading cloud $f_{1} \;
    [p^\mathrm{(1)}_\mathrm{min}:p^\mathrm{(1)}_\mathrm{upper}]$
    becomes smaller which decreases the dimensionless growth rate
    drastically. We consider a situation where for some orientation of
    the crust-anchored dipole and high $\kappa_\mathrm{GJ}$, the
    longitudinal drift can lead to splitting in the electron-positron
    distribution function in the secondary plasma but does not produce
    minima in the EDF for {\em C}2. However when combined with {\em
      C}3 this separation lowers $p^\mathrm{(1)}_\mathrm{min}$ of the
    distribution functions as the cloud flows outward along the field
    line. We perform the next numerical simulation to study the effect
    of longitudinal drift on the cloud-cloud overlap for the
    aforementioned scenario. The results are shown in
    Fig. \ref{cloud-cloud overlap helped by Longitudinal drift}.  As
    seen from the third subplot of (F) even in this hybrid of cases
    {\em C}2 and {\em C}3 the maximum gain exceeds the gain threshold
    defined in section \ref{Criterion for breakdown }.\par

We find that in the absence of {\em C}2 the dimensionless growth rate
($\Omega_\mathrm{I} < 10^{-8}$ ) and comparable to the residuals of
the hydrodynamic equations.

\section{Discussion and comparisons with previous studies}\label{Discussions and results}
\begin{figure*}
\begin{tabular}{cc}
  \includegraphics[width=70mm]{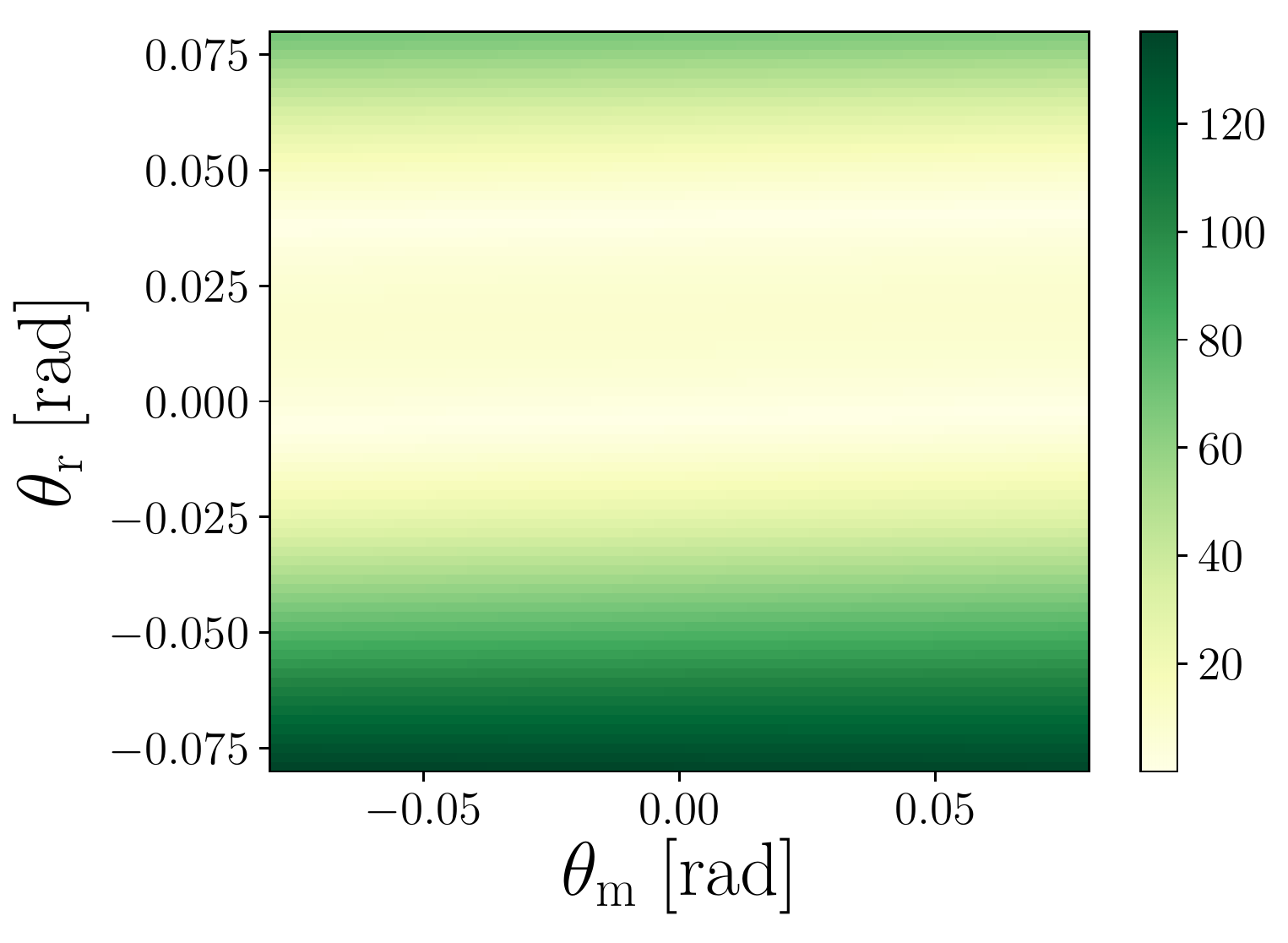} &
  \includegraphics[width=76mm]{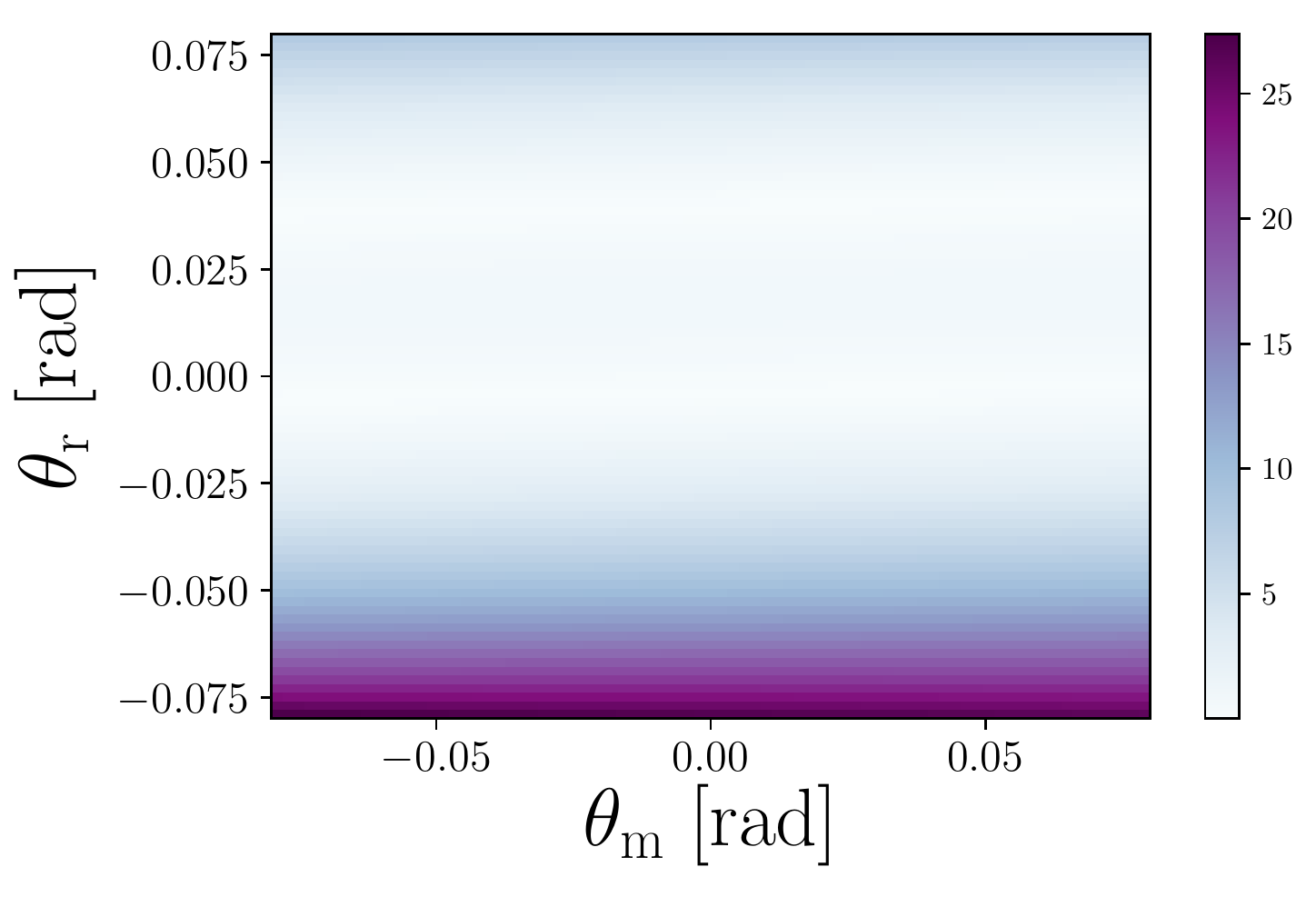} \\ (A) Color map for
  $\kappa_\mathrm{GJ} = 10^{4}$ at $r/R_\mathrm{NS} = $ 50 & (B) Color
  map for $\kappa_\mathrm{GJ} = 10^{5}$ at $r/R_\mathrm{NS} = $ 50
  \\[4pt]
\end{tabular}
\caption{Color map showing the separation ($\Delta \gamma_\mathrm{s}
  $) of the electron-positron distribution function due to
  longitudinal drift at $r/R_\mathrm{NS} = $ 50 due to various
  orientation ($\theta_\mathrm{m} , \theta_\mathrm{r} $) of the
  crust-anchored field ($b = 10, B_\mathrm{d} = 10^{12} $gauss$ $) as
  shown in fig. 1 of \citealt{2002A&A...388..235G}. The mean
  value of the gaussian is taken to be $\gamma_\mathrm{s} = 250$ and
  the ratio of dipole moments of the crust-anchored dipole to global dipole star centred dipole
  has been fixed to $|{m}/{d}| \;=\; 1.25 \times 10^{-3} $. For $\sigma =
  40$, the separation is said to be sufficient only if $\Delta
  \gamma_\mathrm{s} > = \mathrm{1.5} \sigma$.}
\label{Color map for multipolar field}
\end{figure*}

\begin{figure*}
\begin{tabular}{cc}
  \includegraphics[width=65mm]{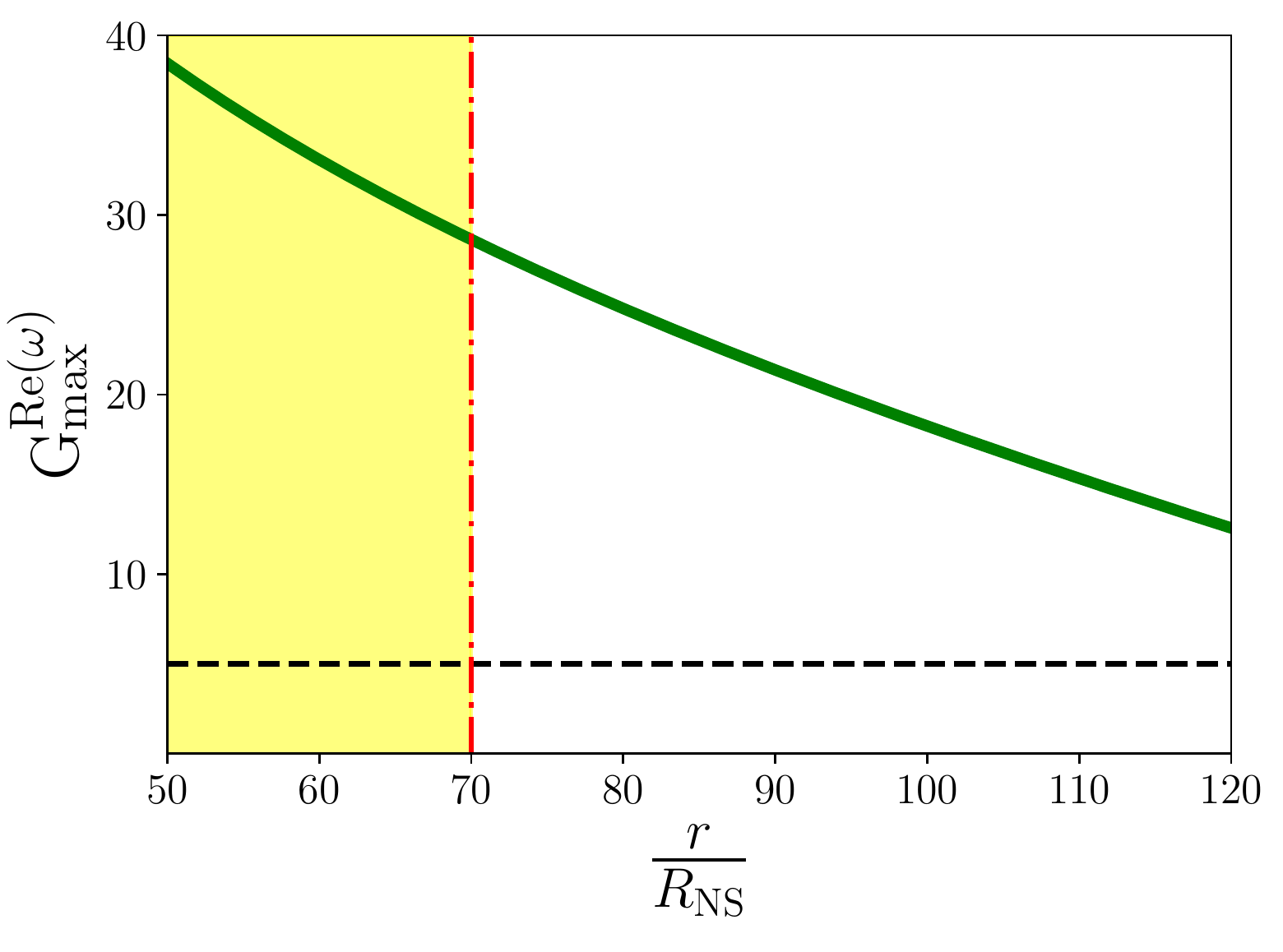} &
  \includegraphics[width=70mm]{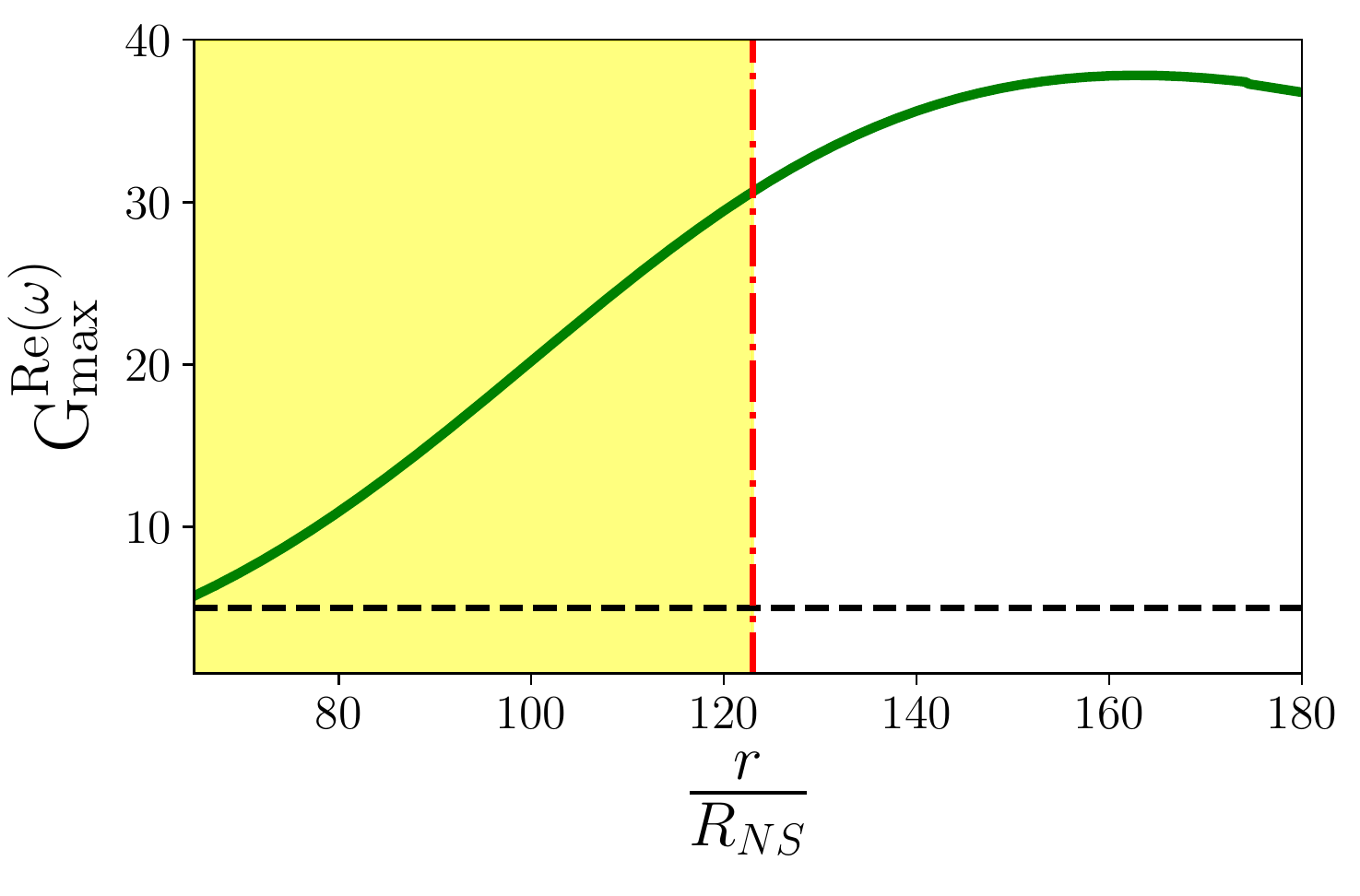} \\ (A)
  $\mathrm{G}_\mathrm{max}$ as a function of $r/R_\mathrm{NS}$ for
  longitudinal drift.  & (B) $\mathrm{G}_\mathrm{max}$ as a function
  of $r/R_\mathrm{NS}$ for cloud-cloud overlap.  \\[4pt]
  \includegraphics[width=70mm]{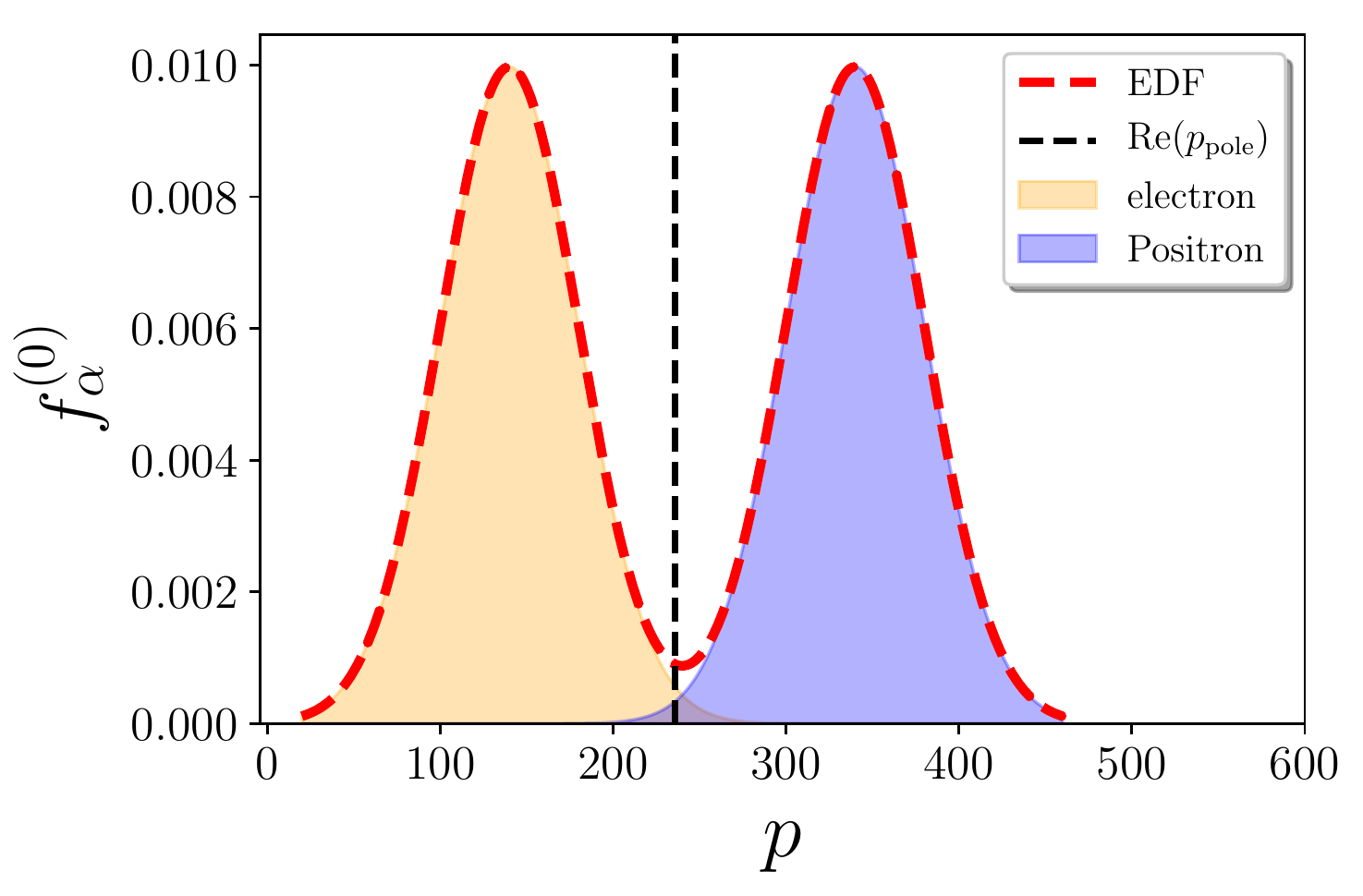}
  & \includegraphics[width=70mm]{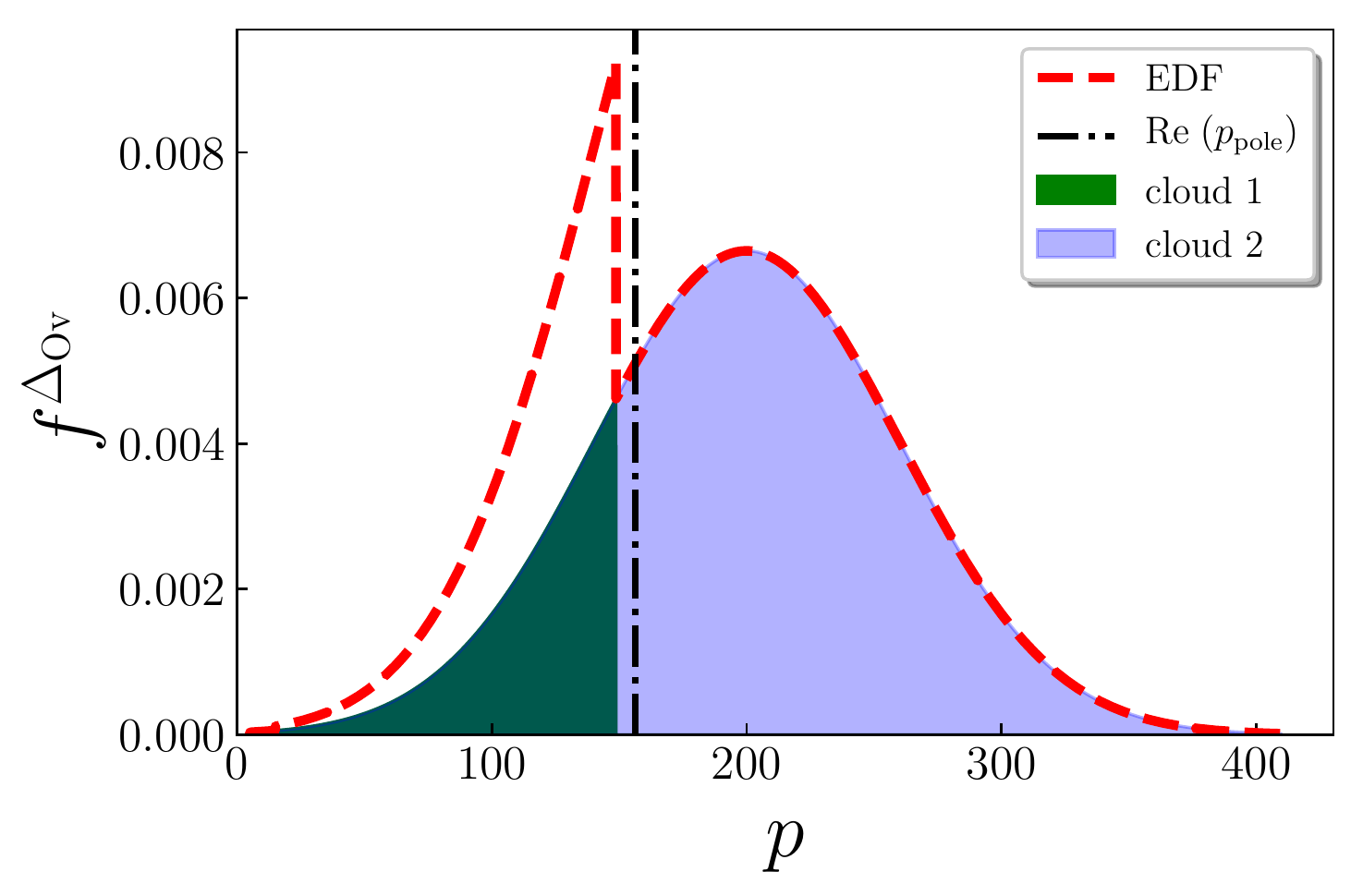} \\ (C) EDF for
  longitudinal drift in multipolar geometry at $r/R_\mathrm{NS} = $
  180.  & (D) EDF for cloud-cloud overlap at $r/R_\mathrm{NS} = $ 180.
  \\[4pt]
\end{tabular}
\caption{``Window of opportunity'' of Cherenk\'ov resonance. The plots
  for {\em C3} uses $\tau$ = 150 nanoseconds and
  $\kappa_\mathrm{GJ} \;=\; 10^{5}$ with gaussian distribution
  function parameters $\sigma = 60,\; p_\mathrm{min} = 5 ,\;
  p_\mathrm{max} = 400,\; p_\mathrm{mean} = 200$. The plots for
  {\em C2} uses multipolar configuration see  fig. 1
    of \citealt{2002A&A...388..235G} with $ b =10,
  \;\theta_\mathrm{r} = 0.08 \;\mathrm{rad} ,\; \theta_\mathrm{m} = -
  0.01 \; \mathrm{rad}, \;$ with $\kappa_\mathrm{GJ} \;=\; 10^{4} $
  and gaussian distribution function parameters $\sigma = 50,\;
  p_\mathrm{mean} = 240$. In the upper panels (A) and (B) the dashed horizontal black line in
  corresponds to gain threshold given in Eq
  \ref{Threshold for non-linear regime} . The dashed dot vertical red line corresponds to the density threshold given in  Eq.\ref{omega_1_threshold}. The
  shaded yellow region shows the ``Window of Opportunity'' where both these constraints are satisfied. The lower panels (C) and (D) shows the EDF for {\em C2} and {\em C3} at $r/R_\mathrm{NS}$ = 180. The vertical black dot-dashed line shows Re$(p_\mathrm{pole})$.  }
\label{Window of oppurtunity}
\end{figure*}

In sections ~\ref{Analysis of Langmuir instability} and 
\ref{Estimation of growth rates}
 we provided a hot plasma treatment of two-stream
instability and estimated growth rates of Langmuir mode for various models
of one-dimensional plasma flow in pulsars. Based on our analysis 
our final aim is to examine
under what conditions excitation of CCR is possible in pulsars. 
There are at least three conditions, namely,  (I), (II) and  (III) that need
to be fulfilled. The first condition (I) is that
for two-stream instability to occur in one-dimensional plasma flow, the EDF should not be single-humped (Gardner's theorem).
If condition (I) is satisfied, excitation
of CCR further requires the following two constraints to be satisfied
simultaneously viz., (II) The amplification criteria which requires the
maximum gain $G_\mathrm{max}$ to be greater than a gain threshold (see
Eq. \ref{Threshold for non-linear regime} ), and (III) The brightness
temperature criteria which requires a very dense plasma. Condition (III)  requires
the scaling factor $\omega_{1}$ to satisfy threshold criteria given by
Eq. \ref{omega_1_threshold}. \par

Note that the estimation of gain requires the description of EDF and the scaling
$\omega_{1}$. The solutions of the dimensionless hydrodynamic equation are determined 
by the EDF and subsequently one obtains the dimensionless growth rate
($\Omega_\mathrm{I}$) and the bandwidth of growing waves($\Delta
\Omega_\mathrm{R}$). The dimensional growth rate($\omega_\mathrm{I}$)
is a product of $\Omega_\mathrm{I}$ and $\omega_\mathrm{1}$. The
scaling factor varies as $\omega_{1} \propto \sqrt{ \kappa_\mathrm{GJ}
  \; n_\mathrm{GJ}}$ and falls monotonically with
distance. Thus to satisfy condition(III) high $\kappa_\mathrm{GJ}$ is 
necessary, which requires multi-polar surface magnetic field geometry. In
what follows we check how these conditions (II) and (III) are fulfilled
for cases {\em C}1, {\em C}2 and {\em C}3 respectively. We compare our results with
previous studies and discuss further implications.

\subsection{Results for {\em C}1} 

Observations suggest the presence of an ion component in pulsar plasma
(see \citealt{2003A&A...407..315G}) along with the $e^{+}/e^{+}$
beam. We assume the ion component to be
composed of iron and characterized by a bulk Lorentz factor of
$\gamma_\mathrm{ion} \approx 10^{3}$. While hot plasma treatment 
for $e^{+}/e^{+}$
beam exists in the literature, as far as we know that such treatment for an ion beam does not exist in the
literature. We analyze the ion component similarly as
\citet{1983Afz....19..753E} did for the high energy $e^{+}/e^{+}$
beam. We find that for width to mean ratio of 1$\%$ the kinetic regime
is completely suppressed. Using this width we estimate the maximum
gain for the ion beam in the hydrodynamic regime.  For the sake of
comparison, we also estimate the maximum gain for the high energy
$e^{+}/e^{-}$ beam. We find that although the gain for the ion beam is
$\sim$ 5 orders of magnitude higher than $e^{+}/e^{-}$ beam yet it
cannot satisfy condition (II). Since none of the beams can satisfy
condition (II) the beam driven Langmuir instabilities are excluded as
candidates for pulsar radio emission.

\subsection{Results for {\em C}2 and {\em C}3}  
AM98 found that growth rates in {\em C}3 exceed that of {\em C}2 for
the same $\kappa_\mathrm{GJ}$ at the radio emission region (see panels
(a) and (b) of fig. 6 in AM98). AM98 further asserts that {\em C}3
dominates {\em C}2 below $r/R_\mathrm{NS} =$ 50 and that the role
changes beyond this distance. In our work we find these assertions to
be inconsistent. We find the exact opposite result as can be seen from
the panel (F) in Fig. \ref{Longitudinal drift for multipolar geometry} and
\ref{cloud-cloud overlap} for $\kappa_\mathrm{GJ} = 10^{4}$. These
conflicting results can be understood by comparing the methodology for
the construction of EDF in this work and AM98.\par

\subsubsection{EDF for  C{\em 2}}
As discussed in section \ref{Longitudinal drift} the separation of the
bulk-velocities $\Delta \beta $ is equal to the product of the
geometrical term and the density term. However, both CR77 and AM98 assume the
geometrical term to be equal to unity. This assumption is not valid at
the radio emission region. For both dipolar and multipolar  surface magnetic field
geometries, the geometrical term is much less than unity ( See 
the upper subplot in panel A of Fig. \ref{Longitudinal drift for dipolar
  geometry} and Fig. \ref{Longitudinal drift for multipolar geometry}
).\par

Further CR77 assumed the radio emission region to be sufficiently far
from the surface.  Assuming a delta-function for $e^{+}-e^{-}$
distribution functions they estimated Im $(\omega)$ such that
Re($\omega$) $ \leq \omega_\mathrm{CCR}$. This is incompatible with
the condition (III).  Improving upon the cold plasma approximation of CR77,
AM98 presented a hot plasma treatment. But they incorrectly assumed
(in addition to neglecting the geometrical term), the same $\Delta
\beta$ for different $\kappa_\mathrm{GJ}$, whereas a self consistent
treatment requires different $\Delta \beta = 1/ \kappa_\mathrm{GJ}$.  However,
in their case $\Omega_\mathrm{I}$ is constant since the EDF has been
assumed to be the same for different
$\kappa_\mathrm{GJ}$. Consequently $\omega_\mathrm{I}$ at any height
scales only as $\omega_{1} \propto \sqrt{\kappa_{GJ}} $ (shown as
vertical shift of $\omega_\mathrm{I}$ in logarithmic scale in panel(a)
of fig.6 of AM98). Thus, AM98 made two inconsistent assumptions for
{\em C}2. Our work corrects these assumptions by incorporating
mutually consistent geometrical terms and $\kappa_\mathrm{GJ}$ for
different kinds of surface geometries.  As discussed in section
\ref{Evidence for multipolar field} a low/high $\kappa_\mathrm{GJ}$
should be associated with purely dipolar/multipolar surface magnetic
field geometry.  \par

 In section~\ref{Longitudinal drift} we considered a purely dipolar
 surface geometry in conjunction with $\kappa_\mathrm{GJ} = 50$ and
 $\kappa_\mathrm{GJ} = 500$ and found that condition(II) is
 satisfied. However, as seen from the first subplot of (F) in the lower panel of Fig. \ref{Longitudinal drift for dipolar geometry},
 condition (III) is not satisfied. Note that in reality
 $\kappa_\mathrm{GJ} \sim 10-100$ for the purely dipolar surface magnetic 
field case and
 hence under no circumstance can condition (III) be satisfied. We have
 shown this case only for the sake of illustration where condition (II)
 and (III) are not satisfied simultaneously.

Apart from being incompatible, using purely dipolar surface magnetic 
to get the geometrical
factor($\sim 10^{-3}$) with a very high $\kappa_{GJ} \sim 10^{4}$ is
unfruitful, as the separation $\Delta \beta \sim 10^{-7}$ is
negligible and condition (I) will not satisfied. However, even for
multi-polar surface magnetic field geometry, high $\kappa_\mathrm{GJ}$ necessarily
suppresses $\Delta \beta $ such that condition (I) cannot be
satisfied. An increase in $\Delta \beta$ can be achieved by only
increasing the geometrical term. This effect requires
$(\hat{\Omega}_\mathrm{Rot} \bm{.} \hat{B})_{O}$ to be lowered in the
geometrical term. To simulate the geometrical term for a multi-polar
geometry we use the prescription of \citet{2002A&A...388..235G} to
model the surface using two dipoles viz., the global star-centered
dipole and a local crust anchored dipole with orientation parameters
($\theta_\mathrm{m}, \theta_\mathrm{r}$) (see subsection
\ref{Multipolar geometry} for details). However, not all orientation
of the crust-anchored local dipole can lower this term. To explore
this effect we perform the following exercise: We change both
$\theta_\mathrm{m}$ and $\theta_\mathrm{r}$ in the range -0.08 radians
to + 0.08 radians for a fixed ratio of the dipole moments $ \lvert
{m}/{d} \rvert = 1.25 \times 10^{-3}$ and trace the last open magnetic
field line. Fig. \ref{Color map for multipolar field} shows the color
map of $\Delta \gamma$ for $\kappa_\mathrm{GJ} = 10^{4}$ and
$\kappa_\mathrm{GJ} = 10^{5}$ in sub-plots (A) and (B) at $r = 50
R_\mathrm{NS}$ for a gaussian distribution function with (mean $\mu$ =
250, width $\sigma$ = 40) at $r/R_\mathrm{NS} = 1.02$. The separation
is taken to be sufficient only if $ \Delta \gamma \geq 1.5
\sigma$. Within this parameter space, the separation is sufficient in
sub-plot (A) for a few special orientations while for sub-plot (B) the separation is
insufficient for any orientation. The results for one such orientation
in sub-plot (A) are shown in Fig. \ref{Longitudinal drift for multipolar
  geometry} where panel (F) shows both conditions (II) and (III)
are satisfied simultaneously in the radio emission region. The
geometrical factor for these special orientations quickly attains an
almost constant value beyond a few km from the surface (see the upper
panel of Fig. \ref{Comparison of dipolar and multipolar
  configuration}) which translates to a near steady EDF for
$r/R_\mathrm{NS} \geq 2$. A steady EDF translates to a steady
$\Omega_\mathrm{I}$. Thus $\omega_\mathrm{I}$ is completely dominated
by number density via the scaling $\omega_{1}$. This means for few
special orientations of the local crust anchored field, both
conditions (II) and (III) are satisfied simultaneously beyond few tens of
km from the surface. This is an essentially new result that has been
obtained using a very thorough analysis of the geometrical term. It
must also be pointed that for the same geometrical term the growth
rates and gain depend sensitively on $\kappa_\mathrm{GJ}$ as shown in
Fig. \ref{Longitudinal drift for multipolar geometry} for
$\kappa_\mathrm{GJ} = 10^{4}$ and $\kappa_\mathrm{GJ} = 8 \times
10^{3}$.\par

\subsubsection{EDF for C{\em 3}}
AM98 constructed the EDF at a few $r/R_{NS}$ (see Table 2 of AM98) for
which the shape of the EDF does not change and extrapolated the results in
between. In their analysis, $\Omega_\mathrm{I}$ remains fixed and the
growth rate ($\omega_\mathrm{I}$) falls monotonically as the scaling
$\omega_\mathrm{1}$ (see panel (b) of fig.6 in AM98). Panel (b) of fig.6
in AM98 also shows $\omega_\mathrm{I}$ to be high even for moderately
low $\kappa_\mathrm{GJ} \sim 10^{2} - 10^{3}$. However, both these
assertions are incomplete and invalid as is discussed below. Many
important features have been missed in AM98 due to the coarse
resolution of their numerical simulations.

We find that the variation of the growth rate $\omega_\mathrm{I}$ as a
function of $r/R_\mathrm{NS}$ in {\em C}2 is not monotonic. It can be
divided into two distinct spatial regions - the first part is
dominated by the EDF and the later part by the number density via
$\omega_{1}$. Although the overlap of the distribution functions begins
very close to the neutron star at $r \approx 2
p^\mathrm{2}_\mathrm{min} c \tau \sim $ 1.5 km, $\omega_\mathrm{I}$
remains very low until a few hundred km from the surface. It is in
this very regime that AM98 incorrectly asserts that {\em C}3 will
dominate {\em C}2. We find that $\Omega_\mathrm{I}$ remains very
low \footnote{see Appendix D for details of numerical simulations}($\ll
10^{-8}$) until a substantial contribution ( $f_{1} \;
[p^\mathrm{(1)}_\mathrm{min}:p^\mathrm{(1)}_\mathrm{upper}]$ ) from
the leading cloud ``$1$'' gives rise to a prominent low-momenta tail
in the EDF ( see panel B of Fig. \ref{cloud-cloud overlap} ). As seen
from panel (A) of Fig. \ref{cloud-cloud overlap},
$p^\mathrm{upper}_\mathrm{1}$ changes slowly beyond a few hundred
km. This implies the shape of the EDF changes rapidly closer to the
neutron star and vice versa. Thus as a prominent tail starts
developing $\Omega_{I}$ first increases rapidly and then attains a
steady value. Consequently for the first few hundred km
$\omega_\mathrm{I}$ remains very low, followed by a subsequent
increase and then a decline following a turnover. The panel (F) of
Fig. \ref{cloud-cloud overlap} shows $\omega_\mathrm{I}$ before the
turnover. However even after the development of a prominent tail
$\Omega_\mathrm{I}$ does not exceed $10^{-7}$(see panel E of
Fig. \ref{Section on cloud-cloud overlap}). This necessarily requires
very high $\kappa_\mathrm{GJ} = 10^{4} - 10^{5}$ via scaling
$\omega_{1}$ to give rise to high growth rates. This is again opposite
to what AM98 obtained. To conclude we find conditions  (II) and (III) are
satisfied simultaneously for {\em C}3 only beyond a few hundred km
from the surface and for very high $\kappa_\mathrm{GJ}$. This is an
essentially new result that has been obtained due to the finer
resolution of our numerical simulations.

To summarize, for {\em C}2 and {\em C}3 the coarseness of the
numerical resolution coupled with many simplifying assumptions led AM98
to general conclusions which are not valid for realistic pulsar
parameters. In this study, we have addressed the inconsistencies of
AM98.

\subsection{Window of Opportunity(WoU)}

 We are now interested to find the spatial region along the magnetic field 
where both conditions (II) and (III) are
 fulfilled simultaneously, and call this the ``Window of Opportunity"
 (hereafter WoU). WoU is shown as a shaded yellow region for {\em C}2
 and {\em C}3 in panels (A) and (B) of Fig.\ref{Window of
   oppurtunity}. It can be seen that for
 {\em C}2, the gain $G_\mathrm{max}$ decreases monotonically while for {\em C}3
 the gain shows a turnover.  As discussed previously the EDF for {\em C}2
 retains the same shape beyond few tens of km from the surface. This
 means the gain curve just reflects the scaling $\omega_\mathrm{1}$
 which decreases monotonically. On the other hand gain curve for {\em
   C}3 is dominated by EDF ($f^\mathrm{Ov}$) before turnover and by
 scaling $\omega_{1}$ beyond it.  \par

However there is an additional difference between the gain curves of
{\em C}2 and {\em C}3. The EDF for {\em C}2 follows condition (I) for
all $r/R_\mathrm{NS} \geq 2$, while EDF for {\em C}3 follows condition
(I) only till $p^\mathrm{(1)}_\mathrm{lower} <
p^\mathrm{(2)}_\mathrm{mean}$. For $p^\mathrm{(1)}_\mathrm{lower} \geq
p^\mathrm{(2)}_\mathrm{mean}$, the EDF is not single humped and
consequently $\omega_\mathrm{I}$ ceases. The gain for {\em C}3 can go
to zero abruptly. Lower the $p^\mathrm{(2)}_\mathrm{mean}$ , closer to
the neutron star surface can this termination occur. In this scenario
abrupt termination in gain can violate condition (II) much closer to
the neutron star. For our chosen parameters this termination occurs
beyond the distance where both conditions conditions (II) and (III) are
violated. Panels (C) and (D) in Fig. \ref{Window of oppurtunity} shows
EDF for {\em C}2 and {\em C}3 at $r/R_\mathrm{NS} = 180$, where it can
be seen ( compared to panel (B) for Fig. \ref{Longitudinal drift for multipolar  geometry} and Fig. \ref{cloud-cloud overlap} ) that far from the neutron star surface the dip in the EDF decreases for {\em C}3 and remains the
same for {\em C}2.

In a realistic scenario both {\em C}2 and {\em C}3 are at
work. However, in such a combination the EDF with the greater growth
rate dominates.  It is clear from the previous discussion for
Fig. \ref{Color map for multipolar field} that for high
$\kappa_\mathrm{GJ} = 10^{5}$ and a few orientations of the local
crust anchored dipole the EDF for {\em C}2 cannot satisfy the condition
(I). For these situations exclusively {\em C}3 can operate. As
discussed in section \ref{Section on cloud-cloud overlap} the EDF for
{\em C}3 depends on $\tau$ and the momenta cut-offs in the
distribution functions. If $p^\mathrm{(1)}_\mathrm{min}$ were to be
lower, the contribution $f_{1} \;
[p^\mathrm{(1)}_\mathrm{min}:p^\mathrm{(1)}_\mathrm{upper}]$ from the
leading cloud becomes smaller at a given height. This then decreases
$\Omega_\mathrm{I}$. However when combined with {\em C}2
$p^\mathrm{(1)}_\mathrm{min}$ decreases with height which then
increases the contribution due to $f_{1}$. The results for such a
hybrid scenario are shown in Fig. \ref{cloud-cloud overlap helped by
  Longitudinal drift} where it is seen that both conditions (II) and
(III) are satisfied.  Thus, a variety of combinations of {\em C}2 and
{\em C}3 can satisfy conditions (II) and (III) simultaneously, thereby
giving the ``WoU''. It must also be noted that our conclusions are
valid only for a strictly one-dimensional plasma flow where only the
Cherenk\'ov resonance condition operates. This assumption is not valid
very far away from the neutron star surface where curvature drift
(viz., the flow of particles perpendicular to the magnetic field plane $
u_\mathrm{d} \approx \sim{{v}_\mathrm{\parallel}^2 \gamma} /
{\omega_\mathrm{B,\alpha} \;\rho_\mathrm{c}}$ ; see
\citealt{1991MNRAS.253..377K}) will set in.\par

We suggest the presence of WoU can be applied to explain the
phenomenon of Radius-to-Frequency mapping where radio emission appears
to arise from a range of emission heights. The variation of the gain
curve can be used to explain time-variable features in pulsar radio
emission like moding and nulling. These aspects will be addressed in future work.

\subsection{Choice of the distribution functions}
We have adopted a semi-numerical approach in this work where we have
assumed the distribution functions of the species in plasma to be
gaussians. The mean values of secondary plasma distribution functions
reflect the order of magnitude estimate derived from considerations of
CCR. Recently the choice of relativistically streaming gaussians has
been criticized by \citet{2019JPlPh..85f9003R}. For {\em C}2 the
symmetric nature of the gaussian distribution function allowed the
bulk velocity to be associated with the mean. This allowed us to
construct EDF for {\em C}2 by only shifting the means without changing
the shape. For an asymmetric distribution function, no such
prescription exists and the shape of the $e^{+}-e^{-}$ distribution
functions can get distorted while separating. This can affect the dip
in the EDF and consequently the growth rate (see the first paragraph in
Appendix D). However, for {\em C}3 even for a non-gaussian
distribution function the most relevant parameters would still be the
location of the peak and $p_\mathrm{min}$. The generic features of the
amplification curve for {\em C}2 and {\em C}3 as shown in
Fig. \ref{Window of oppurtunity} would not change with the change in
the distribution function. The detailed aspects of WoU for different
distribution functions will be addressed in a future study.

\section{Conclusion}
\label{conclusion}
In this study, we find that both {\em C}2 and
    {\em C}3 can lead to excitation of large amplitude Langmuir waves,
    while the secondary plasma should be dense enough to account for high
    $T_\mathrm{b}$ in CCR. Contrary to the results obtained by AM98, 
     we find that for certain multipolar surface magnetic field configurations, the amplification gain for {\em C}2 vastly exceeds that of {\em C}3 for the same $\kappa_\mathrm{GJ}$. For these
    special configurations, very high amplification can be achieved
    very close to the neutron star and the spatial extent over which
    {\em C}2 operates vastly exceeds that due to {\em C}3. A generic feature for {\em C}3 is that the gain becomes high only for $\kappa_\mathrm{GJ} \geq 10^{4}$ after a few hundred km from the neutron star surface. For $\kappa_\mathrm{GJ} \sim 10^{5}$ {\em C}3
    operates exclusively as growth rates in {\em C}2 are suppressed completely. For surface field configurations and high
    $\kappa_\mathrm{GJ}$ wherein the EDF for {\em C}2 is single-humped, the separation nevertheless aids {\em C}3 by enhancing the low momenta tail in the EDF for {\em C}3. We find a
    window of opportunity (WoU) of Cherenk\'ov resonance around
    $r/R_\mathrm{NS} \sim $ 100 where any combination of {\em C}2 and
    {\em C}3 can account for CCR. 
    The presence of large amplitude Langmuir waves in WoU provides an impetus for a higher-order
    plasma theory.

\section*{Acknowledgement}
We thank the anonymous referee for critical comments and suggestions that 
helped to improve the manuscript significantly.
We thank Rahul Basu from IUCAA for discussions and suggestions.
We thank Marina Margishvili for careful reading of the manuscript, suggestion and support. Sk. MR
and DM acknowledge the support of the Department of Atomic Energy,
Government of India, under project no. 12-R$\&$D-TFR-5.02-0700. DM
acknowledges support and funding from the ``Indo-French Centre for the
Promotion of Advanced Research - CEFIPRA'' grant IFC/F5904-B/2018.

\section*{Data availability}

The data from the numerical simulations will be shared on reasonable request to the corresponding author, Sk. Minhajur Rahaman.

\bibliographystyle{mn2e} \bibliography{References}

\bsp	

\appendix 
The following appendix is being provided for an expanded discussion
and derivations that has been used extensively in the main paper.

\section{Making use of the Landau Prescription for electrostatic mode for a one-dimensional relativistic plasma}\label{Landau Prescription}

In the sub-Luminal region we make use of the Landau Prescription where
we allow $\omega$ to be complex and hence the integral in
the dispersion relation can be replaced by a contour
integral to give
\begin{equation}\label{sub-Luminal region}
 \kappa c + \sum_{\alpha} \omega_\mathrm{p,\alpha}^2 \oint_\mathcal{L} dp_{\alpha}\; \frac{\partial f_{\alpha}^{(0)}}{\partial p_{\alpha}} \;\frac{1}{\left(\omega - \beta_{\alpha}\kappa c\right)} = 0
\end{equation} 
where $\mathcal{L}$ stands for the Landau Contour.

Thus we have a pole at a value of particle three momenta
\begin{equation}\label{Pole}
p_\mathrm{pole} = \frac{\omega}{\sqrt{(\kappa c)^2 - \omega^2}}
\end{equation}

Since $\omega$ can now be complex we can write
\begin{equation}\label{Complex Omega}
\omega = \omega_\mathrm{R} \;+\; i\; \omega_\mathrm{I}
\end{equation}

Thus the pole can be written as 
\begin{equation}\label{Complex form of pole}
p_\mathrm{pole} \;= \; \frac{\left(\sqrt{\frac{|U|+ a}{2}} \pm i\;\sqrt{\frac{|U|- a}{2}}\right)}{Q}
\end{equation} 
where
\begin{align*}
&\ |U| = \sqrt{a^2 + b^2} \\
&\ a = \kappa^2 c^2 \left(\omega_R^2 - \omega_I^2\right)  - \left(\omega_\mathrm{R}^2 +\omega_\mathrm{I}^2\right)^2 \\
&\ b = 2\;\omega_\mathrm{R}\;\omega_\mathrm{I} \left(\kappa c\right)^2 \\
&\ Q = \sqrt{(\kappa c)^2 \left[(\kappa c)^2 + 2(\omega_\mathrm{R}^2 - \omega_\mathrm{I}^2)\right] + (\omega_\mathrm{R}^2 + \omega_\mathrm{I}^2)^2}   \numberthis
\end{align*}
which gives us
\begin{align*}
&\ \mathrm{Im}\;(p_\mathrm{pole}) < 0\; \textit{ if b $<$ 0} \Rightarrow \omega_\mathrm{I}<0\\
&\ \mathrm{Im}\;(p_\mathrm{pole}) > 0\; \textit{ if b $>$ 0} \Rightarrow \omega_\mathrm{I}>0\\ 
&\ \mathrm{Im}\;(p_\mathrm{pole}) = 0\; \textit{ if b $=$ 0} \Rightarrow \omega_\mathrm{I} \rightarrow 0\\
\end{align*}
Thus we can have three cases depending on the sign of $\omega_\mathrm{I}$. We are interested in the growth of waves. We explore the two regimes of growth viz, the kinetic and the hydrodynamic regime in the subsequent subsections.

\subsection{Resonant Landau/kinetic growth: $\omega_\mathrm{I} \ll \omega_\mathrm{R}$} \label{Kinetic regime}

First we consider the case  when $\omega_\mathrm{I} = 0$.\par 
\begin{figure}
\centering
\includegraphics[width = \columnwidth]{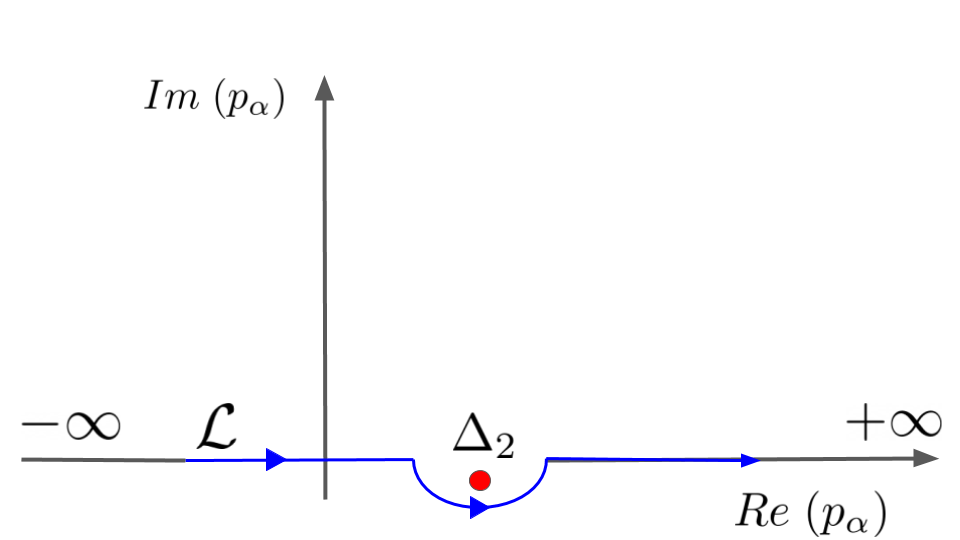}
\caption{\textbf{Contour diagram for $\omega_\mathrm{I} \ll \omega_\mathrm{R}$}}
\end{figure}

The pole is at 
\begin{equation}
p_\mathrm{pole} = \frac{\omega_\mathrm{R}}{\sqrt{(\kappa c)^2 - \omega_\mathrm{R}^2}} = \Delta_2
\end{equation} 

The dispersion relation \ref{sub-Luminal region} in this case given by
\begin{align*}
 \kappa c \;+\; \sum_{\alpha} \omega_\mathrm{p,\alpha}^2 & \;\mathcal{P}\int_{-\infty}^{+\infty} dp_{\alpha}\; \frac{\partial f_{\alpha}^{(0)}}{\partial p_{\alpha}} \;\frac{1}{\left(\omega - \beta_{\alpha}\kappa c\right)} \\ & + i\;\pi \sum_{\alpha}\; \omega_\mathrm{p,\alpha}^2\; \frac{\partial f_{\alpha}^{(0)}}{\partial p_{\alpha}}\Bigg\rvert_{p_{\alpha} \;=\; \Delta_2} \;=\; 0   \numberthis
\end{align*} 
where $\mathcal{P}$ stands for the Principal Value Integral.\par

The dimensionless dispersion relation is given by 
\begin{equation}\label{Normalised dispersion relation}
\epsilon(\Omega, K) = 1 + \sum_{\alpha} \frac{\chi_{\alpha}}{K} \oint_{\mathcal{L}} dp_{\alpha} \; \frac{\partial f_{\alpha}^{(0)}}{\partial p_{\alpha}} \frac{1}{\left(\Omega - \beta_{\alpha}K\right)} = 0
\end{equation}
where $\Omega = \frac{\omega}{\omega_1} \;,\; K = \frac{\kappa c}{\omega_1} \;,\; \chi_{\alpha} = \frac{\omega_\mathrm{p,\alpha}^2}{\omega_1^2} $.

\subsubsection{Derivation of growth rate in the Landau/Kinetic regime}
Let us consider the case when $\Omega = \Omega_\mathrm{R} \pm
i\;\Gamma_\mathrm{kin}$ such that $\Gamma_\mathrm{kin} <<
\Omega_\mathrm{R}$. The particle distribution function having the pole is labelled by subscript ``$b$" and the distribution function away from the pole is labelled by subscript ``$s$". 

Under this approximation the dispersion function can be written as 
\begin{align*}\label{Kinetic growth rate}
&\ \epsilon(\Omega_\mathrm{R} \pm i\; \Gamma_\mathrm{kin},K) = 0\\
&\ \Rightarrow \mathrm{Re} \;\epsilon(\Omega,K) + i\;\mathrm{Im} \; \epsilon(\Omega,K) \;\pm\; i\Gamma_\mathrm{kin} \;\frac{\partial\epsilon}{\partial\Omega}\Big\rvert_1 = 0\\
&\ \Rightarrow \Gamma_\mathrm{kin} \;=\; \mp\; \frac{\mathrm{Im} \; \epsilon(\Omega,K)}{\left(\frac{\partial \epsilon}{\partial \Omega}\right)\Big\rvert_1} \numberthis
\end{align*}

Again the dispersion function can be written as 
\begin{align*}\label{Kinetic dispersion function}
&\ \epsilon (\Omega,K) \\
&\ = Re(\epsilon) + i\; Im(\epsilon) \\
&\ = 1  +  \frac{\chi_{s}}{K}  \int_{-\infty}^{+\infty} dp_{s}\; \frac{\partial f_{s}^{(0)}}{\partial p_{s}} \frac{1}{(\Omega_R - \beta_{s}K)} \\ &\ + \frac{\chi_{b}}{K}  \mathcal{P}\int_{-\infty}^{+\infty} dp_{b}\; \frac{\partial f_{b}^{(0)}}{\partial p_{b}} \frac{1}{(\Omega_R - \beta_{b}K)} \\
&\ \mp i\pi  \frac{\chi_{b}}{K} \int_{-\infty}^{+\infty} dp_{b}\; \frac{\partial f_{b}^{(0)}}{\partial p_{b}} \delta(\Omega_R - \beta_{b}K)     \numberthis
\end{align*}

The root of the argument in the delta function is given by
\begin{align*}
&\ \Omega_\mathrm{R} - \beta_{b}^{res} K = 0 \Rightarrow p_{b,res} \;=\; \frac{\left(\frac{\Omega_\mathrm{R}}{K}\right)}{\sqrt{1-\left(\frac{\Omega_\mathrm{R}}{K} \right)^2}}
\end{align*}

We use the identity $ \delta(f(x)) = \sum_{i=1}^{n} \frac{\delta(x-a_\mathrm{i})}{|\frac{df}{dx}|_{x=a_\mathrm{i}}} $
where $a_i$ are the roots of $f(x)$.

Using the above identity we get 
\begin{align*}\label{Argument of delta function }
\delta(\Omega_\mathrm{R} - \beta_{b}K) = \frac{\delta(p_{b}- p_\mathrm{b,res})}{K \left(\frac{1}{\gamma^3}\right)\Big\rvert_\mathrm{p_{b} \;=\; p_\mathrm{b,res}}} \numberthis
\end{align*}

Equating the imaginary part we get
\begin{align*}\label{Imaginary part of DF}
\mathrm{Im}\; \epsilon(\Omega, K) \;=\; \mp \;\pi\;  \frac{\chi_{b}}{K^2}  \left(\frac{\partial f_{b}^{(0)}}{\partial p_{b}} \gamma^3\right)\Bigg\rvert_{p_{b} \;=\;p_\mathrm{b,res}}
\end{align*}

Now we evaluate the denominator of \ref{Kinetic growth rate}

\begin{align*}
&\ \frac{\partial \epsilon}{\partial \Omega} = -  \frac{\chi_{s}}{K} \int_{-\infty}^{+\infty} dp_{s}\; \frac{\partial f_{s}^{(0)}}{\partial p_{s}} \frac{1}{(\Omega - \beta_{s}K)^2} \\
&\ \Rightarrow \frac{\partial \epsilon}{\partial \Omega} \Big\rvert_1 = 2  \chi_{s} \int_{-\infty}^{+\infty} dp_{s} \; \frac{1}{\gamma^3} \frac{f_{s}^{(0)}}{(1-\beta_{s})^3}      \numberthis
\end{align*}\par 

Substituting \ref{Imaginary part of DF} into \ref{Kinetic growth rate} we get 
\begin{equation}\label{growth rate in kinetic regime}
\Gamma_\mathrm{kin} = \frac{\pi}{2K^2}\frac{ \chi_{b} \left(\frac{\partial f_{b}^{(0)}}{\partial p_{b}} \gamma^3\right)\Big\rvert_{p_{b} \;=\;p_\mathrm{b,res}}}{ \chi_{s} \left\langle \gamma^3(1+\beta_{\alpha})^3 \right\rangle_{s}}
\end{equation} 
where
\begin{align*}
\left\langle(...)\right\rangle_{\alpha} = \int_{-\infty}^{+\infty} dp_{\alpha}\; f_{\alpha}^{(0)} \;(...)
\end{align*}

\subsection{Non-resonant/Hydrodynamic Growth: $\omega_\mathrm{I} >0$} \label{Hydrodynamic regime}
\begin{figure}
\centering
\includegraphics[width = \columnwidth]{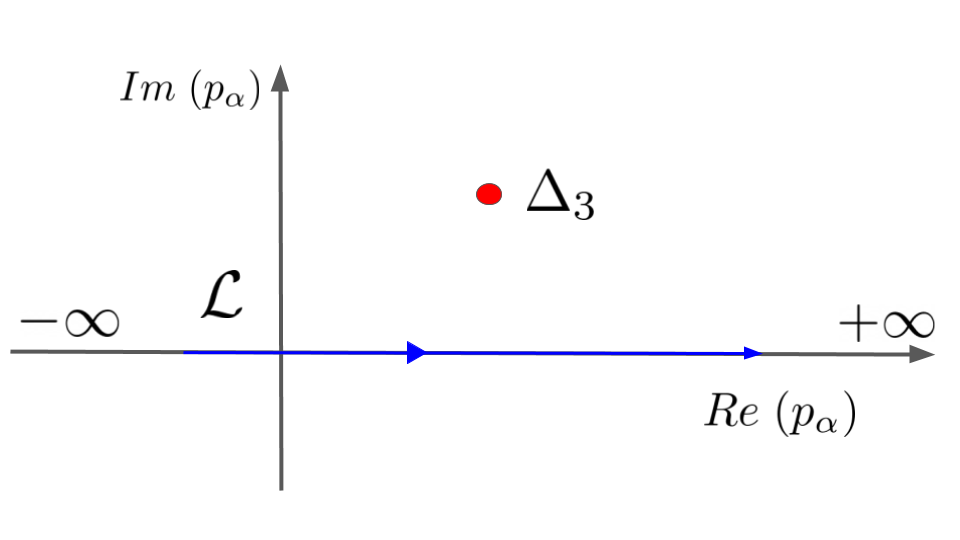}
\caption{\textbf{Contour diagram for $\omega_\mathrm{I} > 0$}}
\end{figure}

The pole is at
\begin{equation}
p_\mathrm{pole} \;= \; \frac{\left(\sqrt{\frac{|U|+ a}{2}} + i\;\sqrt{\frac{|U|- a}{2}}\right)}{Q} = \Delta_3
\end{equation}

The dispersion relation \ref{sub-Luminal region} in this case given by
\begin{equation}
 \kappa c \;+\; \sum_{\alpha} \omega_\mathrm{p,\alpha}^2 \int_{-\infty}^{+\infty} dp_{\alpha}\; \frac{\partial f_{\alpha}^{(0)}}{\partial p_{\alpha}} \;\frac{1}{\left(\omega - \beta_{\alpha}\kappa c\right)} \;=\; 0
\end{equation} 

The above equation can be  integrated by parts to get
\begin{equation}
 1 \;-\; \sum_{\alpha} \omega_\mathrm{p,\alpha}^2 \int_{-\infty}^{+\infty} dp_{\alpha}\; \frac{f_{\alpha}^{(0)}}{ \gamma^3} \;\frac{1}{\left(\omega - \beta_{\alpha}\kappa c\right)^2} \;=\; 0
\end{equation}

Normalising by $\omega_1$ we get
\begin{equation}
 1 \;-\; \sum_{\alpha} \chi_{\alpha} \int_{-\infty}^{+\infty} dp_{\alpha}\; \frac{f_{\alpha}^{(0)}}{ \gamma^3} \;\frac{1}{\left(\Omega - \beta_{\alpha}K\right)^2} \;=\; 0
\end{equation}
where $ \Omega = \frac{\omega}{\omega_1} , K = \frac{\kappa c}{\omega_1} , \chi_{\alpha} = \frac{\omega_\mathrm{p,\alpha}^2}{\omega_1^2}$

Substituting $\Omega = \Omega_\mathrm{R} + i\; \Omega_\mathrm{I}\;(\Omega_\mathrm{I} > 0)$ in the above equation 
and separating it into its real and imaginary part we get
\begin{align*}\label{Hydrodynamic equations}
&\  \epsilon_\mathrm{R} \;=\; 1 \;-\; \sum_{\alpha} \chi_{\alpha} \int_{-\infty}^{+\infty} dp_{\alpha}\; \frac{f_{\alpha}^{(0)}}{ \gamma^3} \;\frac{\left\lbrace\left(\Omega_\mathrm{R} - \beta_{\alpha}K\right)^2 - \Omega_\mathrm{I}^2\right\rbrace}{\left[\left(\Omega_\mathrm{R} - \beta_{\alpha}K\right)^2 + \Omega_\mathrm{I}^2\right]^2} \;=\; 0 \\
&\ \epsilon_\mathrm{I} \;=\; \;2\;\Omega_\mathrm{I}\;\; \sum_{\alpha} \chi_{\alpha} \int_{-\infty}^{+\infty} dp_{\alpha}\; \frac{f_{\alpha}^{(0)}}{ \gamma^3} \;\frac{(\Omega_\mathrm{R} - \beta_{\alpha}K)}{\left[\left(\Omega_\mathrm{R} - \beta_{\alpha}K\right)^2 + \Omega_\mathrm{I}^2\right]^2} \;=\; 0 \numberthis
\end{align*}

\section{Relativistic generalization of the Gardner's theorem}\label{Gardner's theorem}
We follow the methods outlined in ``Plasma Physics'' by Sturrock (1994). 

\begin{figure}
\centering
\includegraphics[width=\columnwidth]{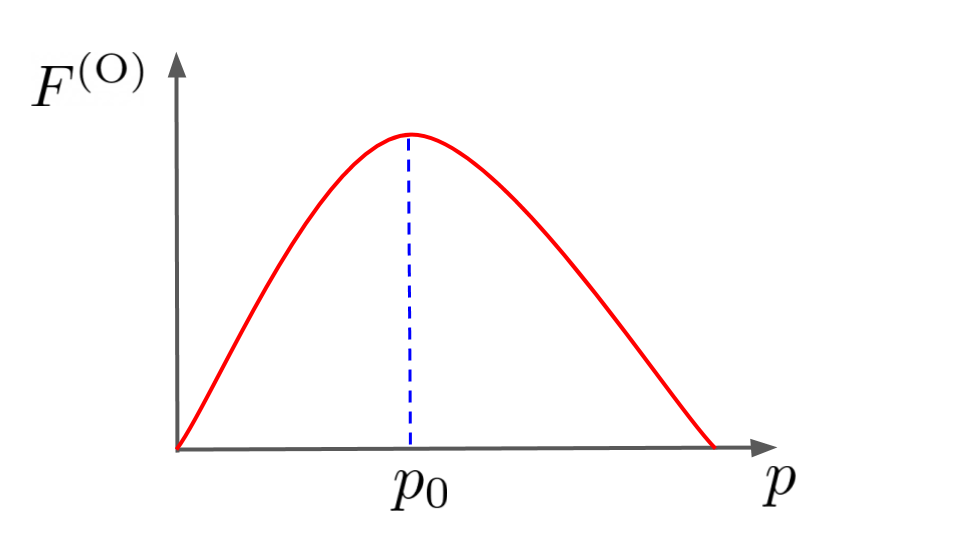}
\caption{\textbf{Schematic of the Equivalent distribution function}}
\label{Equivalent distribution function}
\end{figure}

We define equivalent distribution function(EDF) $F^{(0)}$ as the summation of  normalized distribution function weighted by their plasma frequency(squared) $f^{(0)}_{\alpha}$ of the species involved.
\begin{equation}
F^{(0)} \;=\; \sum_{\alpha} \; \eta_{\alpha} \; f^{(0)}_{\alpha}
\end{equation} where $\eta_{\alpha} \;=\; {\omega^2_\mathrm{p,\alpha}}/{\omega^2_\mathrm{p,ref}}$ where $\omega_\mathrm{p,ref}$ is the plasma frequency of a reference species.

\textbf{Gardner's theorem} states that a single humped equivalent distribution function(EDF) cannot support a growing set of waves.\par 

We adopt the method of contradiction. Let us assume on the contrary that there is a wave with $\omega_\mathrm{I} > 0$. 

The dispersion relation in terms of the EDF can be written as 
\begin{equation}
\kappa c \;+\; \omega^2_\mathrm{p,ref} \int^{+\infty}_{-\infty} \; dp \; \frac{\partial F^{(0)}}{\partial p}  \frac{1}{(\omega - \beta k c)} \;=\; 0
\end{equation}

Subsituting $\omega = \omega_\mathrm{R} + i\;\omega_\mathrm{I} $ we get
\begin{equation}
\epsilon \;=\; 1 - \frac{\omega^2_\mathrm{p,ref}}{(k c)^2}  \int^{+\infty}_{-\infty} dp\; \frac{\partial F^{(0)}}{\partial p} \; \frac{\left( \beta - \frac{\omega_\mathrm{R}}{k c} + i\frac{\omega_\mathrm{I}}{k c}\right)}{ \left(\beta - \frac{\omega_\mathrm{R}}{k c}\right)^2 + \left(\frac{\omega_\mathrm{I}}{k c}\right)^2 } \;=\; 0
\end{equation}

Separating the real and the imaginary part we get
\begin{align*}\label{Real and imaginary part of epsilon}
&\ \epsilon_\mathrm{R} \;=\; 1 - \frac{\omega^2_\mathrm{p,ref}}{(\kappa c)^2}  \int^{+\infty}_{-\infty} dp\; \frac{\partial F^{(0)}}{\partial p} \; \frac{\left( \beta - \frac{\omega_\mathrm{R}}{k c} \right)}{ \left(\beta - \frac{\omega_\mathrm{R}}{k c}\right)^2 + \left(\frac{\omega_\mathrm{I}}{k c}\right)^2 } \;=\; 0\\\\
&\ \epsilon_\mathrm{I} \;=\; - \frac{\omega_\mathrm{I}}{k c} \; \frac{\omega^2_\mathrm{p,ref}}{(k c)^2}  \int^{+\infty}_{-\infty} dp\; \frac{\partial F^{(0)}}{\partial p} \; \frac{1}{ \left(\beta - \frac{\omega_\mathrm{R}}{k c}\right)^2 + \left(\frac{\omega_\mathrm{I}}{k c}\right)^2 } \;=\; 0  \numberthis
\end{align*}

From Fig. \ref{Equivalent distribution function} of the EDF we have the maximum of the equivalent distribution function at $p \;=\; p_0$ which corresponds to $\beta_0 \;=\; \frac{p_0}{\sqrt{1 + p^2_0}}$  which gives us
\begin{align*}
&\ \frac{\partial F^{(0)}}{\partial p} > 0 \;\;\; \textit{for $p < p_0$ i.e, for $\beta < \beta_0$}\\\\ 
&\ \frac{\partial F^{(0)}}{\partial p} < 0 \;\;\; \textit{for $p > p_0$ i.e, for $\beta > \beta_0$}\ \numberthis
\end{align*}
so that
\begin{equation}\label{Condition on equivalent distribution function}
\left(\beta_0 - \beta \right) \frac{\partial F^{(0)}}{\partial p} \geq 0 \;\;\;\; \textit{for $-\infty < p < \infty$}
\end{equation}

Since $\epsilon_\mathrm{R} = 0$ and $\epsilon_\mathrm{I} = 0$ we have
\begin{equation}
\epsilon_\mathrm{R} + \left(\frac{\beta_0 k c - \omega_r}{\omega_\mathrm{I}}\right) \epsilon_\mathrm{I} \;=\; 0
\end{equation}

Using \ref{Real and imaginary part of epsilon} in the equation above we get ,
\begin{equation}
1 + \frac{\omega^2_\mathrm{p,ref}}{\left(\kappa c\right)^2} \int^{+\infty}_{-\infty} \frac{(\beta_0 - \beta) \frac{\partial F^{(0)}}{\partial p}}{\left(\beta - \frac{\omega_\mathrm{R}}{k c}\right)^2 + \left(\frac{\omega_\mathrm{I}}{\kappa c}\right)^2} \;=\; 0
\end{equation}

The integrand is non-negative for all values of $p$ as seen from \ref{Condition on equivalent distribution function}. Hence, the left hand side is \textit{always} greater than unity so that the equation cannot be satisfied. This proves that our original assumption must be incorrect.

\section{Useful derivations for the Langmuir mode}

\subsection{Energy distribution among Langmuir waves and particles in the plasma}\label{Energy distribution among waves and particles}

Following ``Introduction to Plasma Physics" by Nicholson (1983) we have 
\begin{align*}
\underbrace{\frac{1}{16\pi} \frac{d|E|^2}{dt}}_{\text{``Field energy''}} - \underbrace{\frac{1}{4} \frac{d\sigma_\mathrm{I}}{d\omega}\Big\rvert_{\omega_\mathrm{R}} \frac{d|E|^2}{dt} }_{\text{``Reactive part''}} = - \underbrace{\frac{1}{2} |E|^2 \sigma_\mathrm{R}(\omega_\mathrm{R})}_{\text{``Resistive part''}}        \numberthis
\end{align*}

The above expression can be further simplified as
\begin{align*}
&\ \underbrace{\left( 1 - 4\pi\frac{d\sigma_I}{d\omega}\Big\rvert_{\omega_\mathrm{R} }\right)\frac{1}{16\pi}\frac{d|E|^2}{dt}}_{ \;=\; \frac{dW_\mathrm{total}}{dt}} = - \frac{1}{2} |E|^2 \sigma_\mathrm{R}(\omega_\mathrm{R}) \\
\Rightarrow &\ \frac{dW_\mathrm{total}}{dt} \;=\; \left( 1 - 4\pi\frac{d\sigma_\mathrm{I}}{d\omega}\Big\rvert_{\omega_\mathrm{R}}\right) \frac{dW_\mathrm{Field}}{dt} \\ 
\Rightarrow &\ W_\mathrm{total} = \left( 1 - 4\pi\frac{d\sigma_\mathrm{I}}{d\omega}\Big\rvert_{\omega_\mathrm{R}}\right) W_\mathrm{Field}
\end{align*}

Using \ref{Real and imaginary part of epsilon} the above expression can be further simplified to finally obtain
\begin{equation}
W_\mathrm{total} \;=\; \frac{d}{d\omega} \left[\omega \epsilon_\mathrm{R}(\omega)\right]\Bigg\rvert_{\omega_\mathrm{R} } W_\mathrm{field}
\end{equation}

For our case we will evaluate the above expression at $\omega = \omega_1$ where the Langmuir mode touches the $\omega = \kappa c$ line. Please note the quantities for the Langmuir mode has been normalized at this frequency.

The energy distribution between particles and electric field for Langmuir mode for the relativistic pulsar plasma evaluated at the point where the mode touches the $\omega = \kappa c$ line is given as 
\begin{equation}
W_{total} = \frac{d}{d\Omega} \left[\Omega \;\epsilon(\Omega,K)\right]\Bigg\rvert_1 W_{Field}
\end{equation}

Now, we have
\begin{align*}
& \Omega\;\epsilon(\Omega,K)  \\
& \;=\; \Omega \left[ 1 + \sum_{\alpha} \frac{\chi_{\alpha}}{K} \int^{+\infty}_{-\infty} dp_{\alpha} \;\frac{\partial f^{(0)}_{\alpha}}{\partial p_{\alpha}} \frac{1}{(\Omega - \beta_{\alpha} K)}\right] \\\
\Rightarrow  & \frac{d}{d\Omega}[\Omega\;\epsilon(\Omega,K)] \\ 
& \;=\; \epsilon(\Omega,K) + \Omega \left(- \sum_{\alpha} \frac{\chi_{\alpha}}{K} \int^{+\infty}_{-\infty} dp_{\alpha} \;\frac{\partial f^{(0)}_{\alpha}}{\partial p_{\alpha}} \frac{1}{(\Omega - \beta_{\alpha} K)^2}\right)\\\
\Rightarrow  & \frac{d}{d\Omega}[\Omega\;\epsilon(\Omega,K)]\Bigg\rvert_1  \;=\;- \sum_{\alpha} \chi_{\alpha} \int^{+\infty}_{-\infty} dp_{\alpha} \;\frac{\partial f^{(0)}_{\alpha}}{\partial p_{\alpha}} \frac{1}{(1 - \beta_{\alpha})^2}\\\
& = - \sum_{\alpha} \chi_{\alpha} \left[ \frac{f^{(0)}_{\alpha}}{(1-\beta_{\alpha})^2}\Bigg\rvert_{-\infty}^{+\infty} - \int^{+\infty}_{-\infty} dp_{\alpha} \; f^{(0)}_{\alpha} \frac{(-2)}{(1 - \beta_{\alpha})^3} \left(- \frac{d\beta_{\alpha}}{dp_{\alpha}}\right)\right] \\\
& = 2 \sum_{\alpha} \chi_{\alpha} \left\langle \gamma^3 \; (1+\beta_{\alpha})^3 \right\rangle_{\alpha}     \numberthis
\end{align*}
 
Now, $\omega_1$ can be simplified as
\begin{align*}
\omega^2_{1} = \sum_{\alpha} \omega^2_\mathrm{p,\alpha} \left\langle \gamma (1+ \beta_{\alpha})^2 \right\rangle_{\alpha}  \numberthis
\end{align*}

For the ultra-relativistic case as is true for pulsar plasma $\beta_{\alpha} \approx 1$ which further simplifies $\omega_1$ to
\begin{align*}
\omega^2_{1} \;\approx \; 4 \sum_{\alpha} \omega^2_\mathrm{p,\alpha} \left\langle \gamma \right\rangle_{\alpha}  \numberthis
\end{align*} 

Thus, $\chi_{\alpha}$ can be simplified to
\begin{align*}
\chi_{\alpha} &\ \;=\;\frac{1}{4} \frac{\omega^2_\mathrm{p,\alpha}}{\sum_{\alpha} \omega^2_\mathrm{p,\alpha} \left\langle \gamma \right\rangle}_{\alpha} \\
&\ \;=\;\frac{1}{4} \frac{\eta_{\alpha}}{\sum_{\alpha} \eta_{\alpha} \left\langle \gamma \right\rangle}_{\alpha}  \numberthis
\end{align*}
where  $\eta_{\alpha} = \frac{\omega^2_\mathrm{p,\alpha}}{\omega^2_\mathrm{p,ref}}$ where $\omega_\mathrm{p,ref}$ is the plasma frequency of some reference species.

Then we have 
\begin{equation}
\frac{d}{d\Omega}[\Omega\;\epsilon(\Omega,K)]\Bigg\rvert_1  = \frac{1}{2} \sum_{\alpha} \left(\frac{\eta_{\alpha}}{\sum_{\alpha} \eta_{\alpha} \left\langle \gamma \right\rangle}_{\alpha} \right) \left\langle \gamma^3 \; (1+\beta_{\alpha})^3 \right\rangle_{\alpha} 
\end{equation} 

Making use of the ultra-relativistic approximation again we have
\begin{equation}
\frac{d}{d\Omega}[\Omega\;\epsilon(\Omega,K)]\Bigg\rvert_1 \; \approx \; 4 \sum_{\alpha} \left(\frac{\eta_{\alpha}}{\sum_{\alpha} \eta_{\alpha} \left\langle \gamma \right\rangle}_{\alpha} \right) \left\langle \gamma^3 \right\rangle_{\alpha} 
\end{equation} 
 
For a one-component plasma $\alpha = 1$ we have 
\begin{align*}
&\ W_\mathrm{total} = 4 \;\frac{\left\langle \gamma^3 \right\rangle}{\left\langle \gamma\right\rangle}\; W_\mathrm{field} \\
\Rightarrow &\ W_\mathrm{Field} \approx \frac{\left\langle \gamma \right\rangle}{4\left\langle \gamma^3 \right\rangle}\;W_\mathrm{total}
\end{align*} 

Thus, in an ultra-relativistic plasma the electric field energy density of an electrostatic wave is much lower than the energy of oscillation of the particles. As the Lorentz factor of the particles becomes very high ,the electric field energy tends to be very very small.

\subsection{Bandwidth of the growing waves}\label{Bandwidth of growing waves}

We have
\begin{align*}
\frac{\partial \epsilon}{\partial K} &\ = \frac{\partial}{\partial K}\left(1 + \sum_{\alpha} \frac{\chi_{\alpha}}{K} \int^{+\infty}_{-\infty} \; dp_{\alpha}\; \frac{\partial f_{\alpha}^{(0)}}{\partial p_{\alpha}} \frac{1}{(\Omega -\beta_{\alpha}K)}  \right) \\
&\ = - \sum_{\alpha} \chi_{\alpha} \left[ \left\langle \gamma^3 (1+\beta_{\alpha})^{4}\right\rangle_{\alpha} - \left\langle \gamma (1 + \beta_{\alpha})^2\right\rangle_{\alpha} \right] \numberthis
\end{align*}

Similarly we have
\begin{align*}
\frac{\partial \epsilon}{\partial \Omega} \Bigg\rvert_1 &\ = 2 \;\sum_{\alpha} \chi_{\alpha}\int^{+\infty}_{-\infty} dp_{\alpha} \; f_{\alpha}^{(0)}\frac{1}{\gamma^3}  \frac{1}{(1-\beta_{\alpha})^3} \\
&\ = 2 \sum_{\alpha} \chi_{\alpha} \left\langle \gamma^3 (1 + \beta_{\alpha})^3 \right\rangle_{\alpha} \numberthis
\end{align*} 

Using the ultra-relativistic approximation $\beta_{\alpha} \sim 1$ we can write
\begin{align*}
\frac{\partial \epsilon}{\partial K}\Bigg\rvert_1  = - \sum_{\alpha} \chi_{\alpha}\; 2\;\left\langle \gamma^3 (1+\beta_{\alpha})^{3}\right\rangle_{\alpha} \left[ 1 - \Lambda \right]   \numberthis
\end{align*}
where
\begin{align*}
\Lambda  = \frac{1}{2}\;\frac{\sum_{\alpha} \chi_{\alpha} \left\langle \gamma (1+\beta_{\alpha})^{2}\right\rangle_{\alpha}}{\sum_{\alpha} \chi_{\alpha} \left\langle \gamma^3 (1+\beta_{\alpha})^{3}\right\rangle_{\alpha}}
\end{align*}

Comparing both the expressions we get
\begin{align*}
\frac{\partial \epsilon}{\partial K} \Bigg\rvert_1 \;=\; -(1-\Lambda) \; \frac{\partial \epsilon}{\partial \Omega} \Bigg\rvert_1  \numberthis
\end{align*}

Again we have, from the normalised dispersion function
\begin{align*}
&\ \epsilon(\Omega,K) = 0\\
\Rightarrow &\ \frac{\Omega_\mathrm{R}}{K} \;=\; 1 - \Lambda \;\left[ 1 - \frac{1}{K}\right] \numberthis
\end{align*}

The bandwidth of the growing waves can be obtained using the constraint 
\begin{align*}
&\ |\Omega_\mathrm{R} - K\beta_p| \leq \frac{p_{T_{b}}}{\bar{\gamma_b}^3} \\
\Rightarrow &\ \Delta K \leq \frac{p_{T_{b}}}{\bar{\gamma_b}^3} \frac{1}{\Lambda} \numberthis
\end{align*}

We have
\begin{align*}
\Delta \Omega_\mathrm{R} \;=\; \Delta K\;(1 - \Lambda) \approx \Delta K  \numberthis
\end{align*}

Thus we have
\begin{align*}
\Delta \Omega_\mathrm{R} \leq \frac{p_{T_{b}}}{\bar{\gamma_b}^3} \frac{1}{\Lambda} \numberthis
\end{align*}

where
\begin{align*}
\Lambda  = \frac{1}{2}\;\frac{\sum_{\alpha} \chi_{\alpha} \left\langle \gamma (1+\beta_{\alpha})^{2}\right\rangle_{\alpha}}{\sum_{\alpha} \chi_{\alpha} \left\langle \gamma^3 (1+\beta_{\alpha})^{3}\right\rangle_{\alpha}}
\end{align*}

\section{ {Algorithm for numerically solving for the hydrodynamical equations and $G_\mathrm{max}$ using python packages}}
{ \textbf{Location of the hydrodynamic pole and the EDF:}} The imaginary part of the hydrodynamic dispersion relation $\epsilon_\mathrm{I}$ is a corollary of the Gardner's theorem discussed in section \ref{Gardner's theorem}. The condition $\epsilon_\mathrm{I} \;=\; 0$ can be satisfied only if $\mathrm{Re}\;(p_\mathrm{pole})$ of the hydrodynamic pole lies in the dip of the EDF. As discussed in \ref{Hydrodynamic regime} the hydrodynamic pole $p_\mathrm{pole}$ 
is necessarily complex with $ \mathrm{Im}\; (p_\mathrm{pole})$ > 0. As the dip in the  EDF decreases $\mathrm{Im}\;(p_\mathrm{pole})$ and the growth rate $\omega_\mathrm{I}$ also decreases until it vanishes altogether for a single humped EDF. Following this generic ideas we have carried out the following steps to solve the hydrodynamic equations.

\begin{itemize}
\setlength\itemsep{1em}

\item[\textsc{{\textbf{Step 1:}}  }] At a given  $r/R_\mathrm{NS}$ we first check  whether EDF is single-humped or not as shown in plot B of Fig.2,4 and 5 for $r/R_\mathrm{NS} = 50$; and plots (C) and (D) for $r/R_\mathrm{NS} \;=\; 180$ of Fig. 8 of the main paper. If it is not, the three-momenta values $p$ between the two peaks of the EDF/ dip of the EDF is divided uniformly into (n = 100) grid points to be used as guess values for $\mathrm{Re}\;(p_\mathrm{pole,guess})$. Using these we estimate guess values for  $\beta_\mathrm{pole,guess} = \mathrm{Re}\;(p_\mathrm{pole,guess}) / \sqrt{1 + ( \mathrm{Re}\;(p_\mathrm{pole,guess} ) )^2  } $.

\item[\textsc{{\textbf{Step 2:}}  }] We first need the the solution ($\Omega_\mathrm{R}, \Omega_\mathrm{I}$) for K = 1. The guess values for $Re(\Omega)$ is taken as $\Omega_\mathrm{R,guess} \;=\; \beta_\mathrm{pole,guess} \times $K$  $. The guess values for $\Omega_\mathrm{I}$ is taken from $10^{-8}$ to $10^{-6}$ uniformly divided into m = 1000 points. This gives a 2D grid with n $\times$ m points  such that each grid represents a guess value ($\Omega_\mathrm{R,guess}, \Omega_\mathrm{I,guess}$) for the hydrodynamic dispersion relations for K = 1. Both $\epsilon_\mathrm{R}$ and $\epsilon_\mathrm{I}$ are estimated for each grid point. The grid points for which $\epsilon_\mathrm{R} \; \& \; \epsilon_\mathrm{I} \leq 10^{-10}$ were filtered to be used as guess values for the python package $\texttt{fsolve}$. 

\item[\textsc{ {\textbf{Step 3:}} }] The python package $\texttt{fsolve}$ takes the filtered guess values ($\Omega_\mathrm{R,filtered}, \Omega_\mathrm{I,filtered}$) with arguments $( $K$ , \chi_{\alpha} , $EDF$ ) $ and a tolerance value $xtol \;= \; 10^{-12}$. It then solves for the real and imaginary part of the hydrodynamic dispersion relation $\epsilon_\mathrm{R} \;=\; 0$ and $  \epsilon_\mathrm{I} \; = \; 0$ simultaneously. It converges to a  ($\Omega_\mathrm{R}, \Omega_\mathrm{I})$. The  solution so obtained is then inserted into the expression for $\epsilon_\mathrm{R}$ and $\epsilon_\mathrm{I}$ to get the residuals. We take a conservative approach wherein the solution is taken as valid only if the residuals are atleast 3 orders of magnitude lower than $\Omega_\mathrm{I}$. 

After getting a solution we can estimate $\mathrm{Re} (p_\mathrm{pole})$  by following these steps $\beta_\mathrm{pole} = \Omega_\mathrm{R} \times K \rightarrow \mathrm{Re} (p_\mathrm{pole}) = \beta_\mathrm{pole} / \sqrt{1 - \beta^2_\mathrm{pole} } $. The $\mathrm{Re} (p_\mathrm{pole})$ so obtained for K = 1 is shown as a black dashed line in plot(B) of Fig. 2, 4 and 5 ; and plot (C) and (D) of Fig.8 of the main paper.

\item[\textsc{ {\textbf{Step 4}}  }] After getting the solution for K = 1, the wavenumber is changed in steps of $\Delta K_\mathrm{grid} = 10^{-3}$ and the previous step is repeated with the difference that from now on only one guess value needs to be provided. The guess values ($\Omega_\mathrm{R,guess}, \Omega_\mathrm{I,guess}$) for the $l-$th iteration is the solution for $(l-1)$th iteration. After a solution converges for wavenumber K = 1 + $l \times \Delta K_\mathrm{grid}$ , the residuals for $\epsilon_\mathrm{R}$ and $\epsilon_\mathrm{I}$ are estimated. The process is terminated at a wavenumber $K_\mathrm{cut-off}$ where either $\Omega_\mathrm{I} \leq 10^{-8}$ or the residuals do not satisfy the criteria mentioned above, whichever occurs first.

At the end of this stage we have obtained the dimensionless dispersion relation. The first, second and third  subplot of (C) in Fig. 2, 4 and 5 shows $\Omega_\mathrm{R}$, $\Omega_\mathrm{I}$ and the residuals (numerical errors) of $\epsilon_\mathrm{R}$ and $\epsilon_\mathrm{I}$ as a function of K.

\item[\textsc{ {\textbf{Step 5}}  }]:  To get the dispersion relation in the dimensionless form we estimate the scaling factor $\omega_{1}$ [in rad/s] via Eq. 3 of the main paper. The dispersion relation in the dimensional form is obtained by the following steps: $\omega_\mathrm{R} \;=\; \Omega_\mathrm{R} \times \omega_{1}  $ [in rad/s] , $\omega_\mathrm{I} \;=\; \Omega_\mathrm{I} \times \omega_{1}  $ [in s$^{-1}$] and k = K $\times \; \omega_{1} / c$ [in cm$^{-1}$]. The first and second subplot of (D) in Fig. 2, 4 and 5 of the main paper shows the dimensional dispersion relation ( also referred to as spectrum of the growing set of waves). The third subplot of (D) in the Fig. 2, 4 and 5 shows the group velocity dispersion $dv_\mathrm{g}/dk$ [in cm$^{2}$ s$^{-1}$] as a function of the wavenumber k. In all three figures group velocity dispersion is positive for the growing set of waves.

\item[\textsc{ {\textbf{Step 6}} }]: At a given $r/R_\mathrm{NS}$ the normalized bandwidth of the growing waves is given by $\Delta \Omega_\mathrm{R} = (\Omega_\mathrm{R}\lvert_{K_\mathrm{cut-off}} - \Omega_\mathrm{R}\lvert_{K \;=\; 1}) $. Using $r$, $\omega_\mathrm{1}$, $\Omega_\mathrm{I,1}$ and $\Delta \Omega_\mathrm{R}$ in Eq. 9 and Eq. 10 of the main paper we obtain the maximum gain $G_\mathrm{max}$ that can be associated with Re$(\omega) = \Omega_\mathrm{R,1} \times \omega_\mathrm{1}$. The Re($\omega_{R,1}$), Re($\omega_{I,1}$) and the gain curve are shown as upper, middle and lower subplots of panel (F) in Fig. 2, 4 and 5.  

\end{itemize}

\section*{References}
\noindent \textbf{Nicholson D. R, 1983}
 
\noindent \textbf{Sturrock P.A , 1994}

\label{lastpage}
\end{document}